\documentclass[fleqn,usenatbib]{mnras}

\usepackage{amsmath}
\usepackage{newtxtext,newtxmath}
\usepackage[T1]{fontenc}
\DeclareMathAlphabet{\mathcal}{OMS}{cmsy}{m}{n}  

\DeclareRobustCommand{\VAN}[3]{#2}
\let\VANthebibliography\thebibliography
\def\thebibliography{\DeclareRobustCommand{\VAN}[3]{##3}\VANthebibliography}

\usepackage{bm}
\usepackage{graphicx}
\usepackage{float}
\usepackage{natbib}
\usepackage{algorithm}
\usepackage{algpseudocode}
\usepackage{comment}
\usepackage{caption}
\usepackage{orcidlink}

\DeclareMathOperator*{\argmin}{arg\,min}

\newcommand{\bigparen}[1]{\left ( #1 \right )}

\newcommand{\bigbracket}[1]{\left [ #1 \right ]}

\newcommand{\Accelunits}{\,m\,s$^{-2}$}

\newcommand{\soft}[1]{\texttt{#1}}
\newcommand{\func}[1]{\textsc{#1}}

\newcommand{\SNR}{\mathcal{Z}}
\newcommand{\vlambda}{\bm{\lambda}} 
\newcommand{\Omegaorb}{\Omega_{\mathrm{orb}}}
\newcommand{\Porb}{P_{\mathrm{orb}}}
\newcommand{\Tobs}{T_{\mathrm{obs}}}
\newcommand{\kmax}{k_{\mathrm{max}}}
\newcommand{\tref}{t_{\mathrm{ref}}}
\newcommand{\tmid}{t_{\mathrm{mid}}}
\newcommand{\nrun}{n_{\mathrm{run}}}
\newcommand{\Ngrid}{N_{\mathrm{grid}}} 
\newcommand{\Tseg}{T_{\mathrm{seg}}} 
\newcommand{\Lc}{\mathbf{\Lambda}_{\mathrm{circ}}}
\newcommand{\Ld}{\mathbf{\Lambda}_{\mathrm{d}}}
\newcommand{\Lcart}{\mathbf{\Lambda}_{\mathrm{cart}}} 
\newcommand{\Ttrans}{T_{\mathrm{trans}}} 

\algrenewcommand\algorithmicrequire{\textbf{Input:}}
\algrenewcommand\algorithmicensure{\textbf{Output:}}

\title[Finding Short-Period Binary Pulsars with Pruning]{Coherent Signal Detection with Pruning -- I. Finding Short-Period Binary Pulsars in Circular Orbits}

\author[Kumar \& Zackay]{
Pravir Kumar\orcidlink{0000-0003-1913-3092}$^{1}$
\thanks{E-mail: \href{mailto:pravirka@gmail.com}{pravirka@gmail.com}}
and
Barak Zackay\orcidlink{0000-0001-5162-9501}$^{1}$
\thanks{E-mail: \href{mailto:barak.zackay@weizmann.ac.il}{barak.zackay@weizmann.ac.il}}
\\
$^{1}$Department of Particle Physics and Astrophysics, Weizmann Institute of Science, Rehovot 7610001, Israel}

\date{Accepted XXX. Received YYY; in original form ZZZ}

\pubyear{2026}

\begin{document}
\label{firstpage}
\pagerange{\pageref{firstpage}--\pageref{lastpage}}
\maketitle

\begin{abstract}
Detecting pulsars in short-period binary systems, which are unparalleled laboratories for fundamental physics and tests of general relativity, is a prime objective of radio astronomy. Their rapid orbital motion, however, presents a formidable computational challenge. Conventional searches are therefore limited to simplified signal models (e.g., constant acceleration) that remain valid for only short integrations ($\lesssim 4$--$10$\% of an orbital period). This fundamental limitation severely degrades search sensitivity, placing much of the faint, relativistic pulsar population beyond the reach of current surveys. We present a novel hierarchical search framework based on extreme pruning that overcomes these limitations by progressively eliminating improbable regions of parameter space across successive coherent integration stages. The algorithm achieves $>90$\% detection probability at the sensitivity threshold, with near-unity recovery for stronger signals, while reducing the computational complexity of full circular-orbit searches by up to 10 orders of magnitude relative to an unpruned hierarchical baseline. The resulting efficiency enables, for the first time, fully coherent integration over an entire orbital period and beyond. Compared to conventional acceleration searches, the proposed method delivers a 3- to 5-fold improvement in sensitivity, dramatically increasing the discovery potential for high-value targets such as pulsar--black hole binaries.
\end{abstract}


\section{Introduction} \label{sec:intro}
The search for binary pulsars has been a cornerstone of modern radio astronomy since the discovery of PSR\,B1913+16 \citep{Hulse:1975}. Binary pulsars, particularly compact, relativistic, short-period systems, are unique cosmic laboratories for fundamental physics: they probe the equation of state of nuclear matter at supra-nuclear densities and provide stringent tests of General Relativity (GR). They are also powerful probes of late-stage stellar evolution and neutron star mass measurements \citep{Ozel:2016}. PSR\,B1913+16 yielded the first indirect evidence for gravitational radiation, while the Double Pulsar system PSR~J0737$-$3039A/B has enabled some of the most precise tests of GR in the strong-field regime to date \citep{Kramer:2006, Wex:2014}. Discovering pulsars in compact binaries, in the dense cores of Globular Clusters (GCs), and ultimately in orbit around black holes remains a primary science goal in the modern era \citep{Bagchi:2025, Keane:2025}.

Detecting these systems, however, remains a difficult signal-processing problem. Pulsars are intrinsically faint radio sources that often require long integrations, $\Tobs$, to accumulate sufficient signal-to-noise ratio (S/N). For an isolated pulsar this integration is straightforward: the spin period is constant, so signal power is coherently combined by folding the time series at a single period. In a binary system, orbital motion imposes a time-dependent Doppler shift on the apparent spin frequency. Left uncorrected, this phase evolution smears the signal power across Fourier bins, reducing the coherent S/N and sharply degrading detectability \citep{Johnston:1991}. Recovering this lost coherence requires demodulation: the time series must be resampled to undo the orbital modulation and, because the orbit is \emph{a priori} unknown, this resampling must be performed over a dense grid of trial orbital parameters. What is a one-dimensional periodicity search for an isolated pulsar becomes, for a binary, a high-dimensional template-enumeration problem. The computational cost scales steeply with both $\Tobs$ and the orbital parameter-space volume. In the compact-binary regime where $\Tobs \sim \Porb$, fully coherent searches over the Keplerian parameters remain computationally intractable, even when the search volume is restricted and templates are placed optimally via a parameter-space metric \citep{Balakrishnan:2022}. Blind, exhaustive demodulation of wide-field survey data is therefore unfeasible with current computing resources.

Standard pulsar search pipelines therefore adopt approximate, lower-dimensional phase models. The dominant approach is the constant-acceleration search, which models the apparent spin evolution with a constant frequency derivative, sometimes augmented by a constant jerk term \citep{Johnston:1991, Ransom:2002, Andersen:2018}. This approximation holds only while the orbital phase changes modestly during the observation, but degrades once the integration spans a non-negligible fraction of the orbital period ($\Tobs \gtrsim 0.1\,\Porb$). In practice, this imposes a familiar trade-off: keep $\Tobs$ short to preserve the approximation, or integrate longer and lose sensitivity to the most compact systems \citep{Bagchi:2013}. At the opposite extreme, when the observation spans many orbits ($\Tobs \gg \Porb$), phase-modulation (sideband) searches exploit the regular comb of orbital sidebands around each spin harmonic \citep{Ransom:2003, Jouteux:2002}, while stack or segmented searches divide the data into short coherent blocks and combine them incoherently \citep{Wood:1991}. Each of these approaches targets a specific corner of the $\Tobs/\Porb$ plane, and the compact regime ($\Tobs \sim \Porb$), where the scientific payoff is richest, falls in the gap between them and is poorly served by existing methods.

Current state-of-the-art search pipelines fall into three broad classes. In the Fourier domain, acceleration and jerk searches (FDAS/FJAS), as implemented in \soft{PRESTO}, transform the entire time series once and perform a matched-filter correlation of the complex Fourier spectrum against a grid of constant-acceleration or constant-jerk templates \citep{Ransom:2002, Ransom:2011}. These methods are computationally efficient and widely used, but their sensitivity is reported to degrade rapidly once $\Tobs \gtrsim 0.1\,\Porb$ for acceleration searches and $\Tobs \gtrsim 0.15\,\Porb$ for jerk searches, where the low-order phase model breaks down \citep{Andersen:2018}. The second class, time-domain acceleration or jerk searches (TDAS/TJAS), instead resamples the time series over a grid of trial accelerations or jerks before performing a standard FFT-based periodicity search, as in \soft{Peasoup}-based pipelines \citep{Eatough:2013, Morello:2019, Barr:2020}. The template-bank methods construct a metric-based grid directly over the full circular or Keplerian orbital parameter space \citep{Allen:2013, Knispel:2013, Nieder:2020}. Although these template approaches achieve the broadest orbital coverage to date, the required template density scales steeply with spin frequency and the method necessarily accepts a non-negligible mismatch, both of which drive the computational cost beyond what large-scale blind surveys can sustain \citep{Balakrishnan:2022}.

A distinct third class comprises Fast Folding Algorithm (FFA)-based searches \citep{Staelin:1969}. Unlike FFT-based methods, which perform coherent integration only up to the Fourier transform stage and subsequently combine harmonic power incoherently, the FFA implements a fully phase-coherent matched filter directly in the time domain. Consequently, the FFA retains sensitivity across the full range of pulse duty cycles and approaches the theoretical optimum for periodic signals \citep{Morello:2020}. The recent revival of the FFA has established it as the preferred method for long-period, narrow-duty-cycle pulsars \citep{Cameron:2017, Parent:2018}; however, in its present form, it remains restricted to isolated-periodicity searches \citep{Morello:2020}. Despite this variety of existing techniques, no current method is simultaneously fully phase-coherent, sensitive to compact binaries, and computationally scalable beyond the low-order polynomial regime.

The need for such algorithmic capabilities is becoming more urgent. Current and next-generation radio facilities, including FAST, MeerKAT, Murriyang cryoPAF, DSA-2000 and ultimately the Square Kilometre Array (SKA), are delivering unprecedented raw sensitivity through wider bandwidths, longer integrations, and larger instantaneous sky coverage \citep{Stappers:2016, Hallinan:2019, Han:2021, Padmanabh:2023, Dunning:2023, Keane:2025}. As these surveys scale up, the discovery bottleneck increasingly shifts from raw collecting area to managing unprecedented data volumes and maintaining computational tractability at the instrument's full sensitivity. In the compact-binary regime, where the scientific payoff is highest, the gap between observational capability and search algorithm performance is widest, and it cannot be resolved by raw computing power alone. A practical, scalable route to fully coherent binary searches is therefore not a future desideratum; it is a present necessity \citep{Smith:2016}.

In this work, we introduce \emph{Extreme Pruning} (EP), a new framework for fully coherent searches over high-order polynomial phase models. The central idea is to organize the search hierarchically and to prune regions of parameter space that are statistically inconsistent with a coherent signal. Full-length coherent integration is applied only to a small subset of promising trajectories. In the pulsar-search context, this strategy enables fully coherent searches for circular binaries at a computational cost far below brute-force demodulation, while remaining highly competitive with established approximate methods. Although our first application is binary pulsar searching, the underlying pruning principle is general and applicable to other inference problems involving structured phase models.

We focus on binary pulsars in circular orbits, the dominant configuration in the observed binary pulsar population. This population is heavily skewed toward Millisecond Pulsars (MSPs) recycled via mass transfer from low-mass companions, the majority of which have evolved into white dwarfs \citep{Lorimer:2008}. In particular, the compact ``spider'' systems, comprising black widows ($\Porb \sim 1.5$--$10$\,h) and redbacks ($\Porb \sim 4$--$24$\,h) are characterized by circular or near-circular orbits and represents one of the most promising discovery spaces for compact-binary searches, precisely in the regime where $\Tobs/\Porb \sim 0.1$--$1$ \citep{Roberts:2013}.
For fully recycled MSPs with low-mass white-dwarf companions, prolonged tidal dissipation during the preceding X-ray binary phase efficiently circularises the orbit, suppressing eccentricities to $e \sim 10^{-6}$--$10^{-3}$ \citep{Phinney:1992, Tauris:2023}. Circular-orbit searches therefore offer both strong astrophysical motivation and provide an algorithmically well-posed setting in which to develop and validate the EP method via a clean analytical treatment \citep{Jouteux:2002, Ransom:2003}. For systems with low but non-zero  eccentricity, a circular-orbit search can still yield useful sensitivity. Extensions to substantially eccentric systems are natural but require a broader phase model and are deferred to future work.

This work is the first in a series. In Paper~I (this work), we present the algorithmic foundations, implementation details, and validation using simulated data. In a forthcoming paper (Paper~II), we will focus on the end-to-end search pipeline, its performance on real telescope data, and a comparative benchmark against contemporary search methods via robust injection-recovery tests.

This paper is structured as follows. In Section~\ref{sec:pruning_overview}, we provide a high-level overview of the pruning concept. Section~\ref{sec:phase_prelim} presents the phase model and other preliminaries. Section~\ref{sec:ffa_poly} generalizes the standard FFA to polynomial phase models, a prerequisite for EP method. Section~\ref{sec:pruning_alg} details the EP algorithm, its hierarchical structure, and its computational complexity analysis. In Section~\ref{sec:pruning_circular}, we discuss the specific application of EP to the search for circular binary orbits. We present the software implementation and performance benchmarks on simulated data in Section~\ref{sec:alg_implement}, and conclude with the implications for archival and ongoing pulsar surveys in Section~\ref{sec:discussion}.

\section{Pruning Concept: An Overview}\label{sec:pruning_overview}
The fundamental challenge in detecting relativistic binary pulsars is the steep polynomial scaling of the number of search templates with observation duration. For a phase-coherent search over time $\Tobs$, the number of orbital templates scales as $\Ngrid \propto \Tobs^{\kappa}$, where the exponent $\kappa$ is determined by the highest phase derivatives included in the orbital model: for instance, $\kappa = 1$ for isolated periodicity searches, $\kappa = 3$ for constant-acceleration searches, and $\kappa = 6$ for constant-jerk searches \citep{Smith:2016}. In traditional brute-force approaches, this scaling rapidly renders long integration times computationally infeasible, restricting searches to short data segments where sensitivity is severely suboptimal.

This enumeration problem is not unique to pulsar searches. Comparable challenges arise in other fields requiring exhaustive search over large parameter spaces, such as lattice enumeration for the shortest vector problem in lattice-based cryptography \citep{He:2024}. In such settings, probabilistic pruning techniques are widely employed to dramatically reduce enumeration costs by accepting a controllable probability of missing the optimal solution in exchange for substantial computational savings \citep{Gama:2010}.

Drawing on this principle, we introduce a \emph{pruning} algorithm specifically designed for the pulsar search problem. Our approach fundamentally alters the computational scaling by restructuring the search using a hierarchical tree in which depth corresponds to integration time and width spans the search parameter space. Rather than evaluating the full parameter tree at the complete observation time $\Tobs$, we employ a multi-stage sequential elimination strategy that progressively discards statistically implausible branches.

The intuition behind pruning can be framed in terms of statistical recoverability: only candidates achieving sufficient partial S/N at intermediate stages remain statistically capable of reaching the final detection threshold. Consider a hierarchical divide-and-conquer search (structured as a binary tree) targeting a final significance of S/N\,$= 10$ for a stationary signal. Since coherent integration in white Gaussian noise causes signal power to accumulate additively, a true astrophysical signal must produce detectable power in sub-segments of the data. For instance, to reach S/N\,$= 10$ in the full dataset, a candidate must yield S/N\,$\approx 7$ in each half. The implication is immediate: if a template yields S/N\,$< 5$ in the first half, consistent with noise, it becomes statistically implausible for the remaining half of the data to contain sufficient signal strength to reach the detection threshold. We can thus ``predict'' the final result will be a null detection without computing the second half. 

This logic extends recursively to earlier stages. Two levels prior, where each quarter-segment should exhibit S/N\,$\approx 5$ for a genuine detection, we can prune candidates with S/N\,$< 3$ (illustrative thresholds). Crucially, by discarding a candidate early, we eliminate not only the computation for that specific segment but the entire descendant sub-tree of finer parameters that would have branched from it. Thus, a set of appropriately calibrated thresholds forms a pruning funnel that yields a large (often exponential) reduction in the total number of template evaluations.

\begin{figure}
\includegraphics[width=\columnwidth]{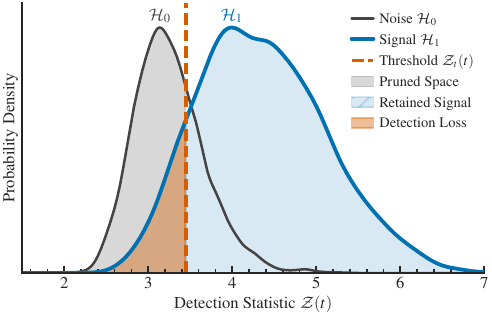}
\caption{Statistical snapshot of the pruning process at an intermediate stage. The noise distribution ($\mathcal{H}_0$, gray) is shifted from zero due to the maximization over search parameters (look-elsewhere effect). The signal distribution ($\mathcal{H}_1$, blue) separates to higher values as integration time increases. The pruning threshold $\SNR_{t}$ (red dashed line) acts as a filter: the gray shaded region represents pruned candidates (eliminated search volume), while the orange shaded region quantifies the false dismissal probability (risk of signal loss). \label{fig:score_dist_intermediate_lee}}
\end{figure}

The core principle of pruning relies on the distinct statistical evolution of noise ($\mathcal{H}_0$) and signal ($\mathcal{H}_1$) hypotheses as data is accumulated. Let $\SNR(t)$ denote the detection statistic evaluated on the data segment accumulated up to time $t$ for a given template. Under $\mathcal{H}_0$, the maximized statistic (after searching over a large bank of matched-filter templates) follows an extreme-value distribution in the asymptotic regime of many independent trials, due to the look-elsewhere effect. Under $\mathcal{H}_1$, assuming white Gaussian noise and a constant-amplitude signal, coherent integration leads to monotonic growth in the expected detection statistic:
\begin{equation}
    E[\SNR(t) \mid \mathcal{H}_1] \propto \sqrt{t}.
\end{equation}
This divergence permits the definition of a time-dependent pruning threshold $\SNR_{t}(t)$. At any intermediate stage $t$, we retain a candidate only if $\SNR(t) > \SNR_{t}(t)$. As illustrated in Figure~\ref{fig:score_dist_intermediate_lee}, the threshold is chosen to ensure that the cumulative probability of rejecting a true signal (Type II error) remains acceptably low (e.g., $<10\%$), while the probability of retaining a noise candidate (Type I error) drops exponentially with time.

The computational efficiency arises from the competition between the polynomial expansion of the search grid and the exponential contraction of the survival space. To derive an approximate scaling, we employ a Gaussian tail approximation for the noise distribution assuming a single template. The fraction of noise candidates surviving the threshold cut, $P_{\rm surv}$, drops exponentially with time. Setting the threshold to scale with expected signal growth, $\SNR_{t}(t) \propto \sqrt{t}$, the survival probability for noise behaves as:
\begin{equation}
    P_{\mathrm{surv}}(t \mid \mathcal{H}_0) \approx \exp\left(-\frac{\SNR_{t}(t)^2}{2}\right) \propto \exp\left(-\frac{t}{2\,T_0}\right),
\end{equation}
where $T_0$ is a characteristic timescale representing the pruning starting stage (i.e., the time at which $S/N \sim 1$). The total computational cost $C_{\rm total}$ is then the integral of the active search volume over the observation time:
\begin{equation}
    C_{\rm total} \propto \int_0^{\Tobs} t^\kappa \cdot \exp\left(-\frac{t}{2\,T_0}\right) dt.
\end{equation}
This integral takes the form of a Gamma function. Crucially, in the regime $\Tobs/T_0 \gg 1$, the integral converges to a finite bound:
\begin{equation}
    C_{\rm total} \propto (2\,T_0)^{\kappa+1} \Gamma(\kappa+1) \approx C(2T_0) \cdot \kappa!\,,
\end{equation}
where $C(2T_0)$ aggregates the proportionality constants and represents the base enumeration volume required to coherently integrate a data segment of length $2T_0$.

The key insight is that once signal accumulation sufficiently separates the $\mathcal{H}_0$ and $\mathcal{H}_1$ distributions, the pruning mechanism eliminates false candidates faster than the parameter space expands. Consequently, for a search with fixed $\kappa$, the integrated computational cost approaches a finite asymptotic value as $\Tobs/T_0 \gg 1$, rather than growing polynomially with $\Tobs$. Increasing the search order raises the asymptotic cost and shifts the peak computational load to later stages of the hierarchy, but the exponential contraction of the survival space continues to dominate. This bounded complexity enables coherent integrations over durations that would otherwise be computationally prohibitive. We apply this framework to pulsar searches in Section~\ref{sec:pruning_alg}. Sections~\ref{sec:phase_prelim}–\ref{sec:ffa_poly} provide the algorithmic foundations and implementation details necessary for the pruning strategy.

\section{Phase-model preliminaries}\label{sec:phase_prelim}
In this paper, we focus on detecting pulsars in circular binary orbits. This serves as the algorithmic foundation for our pruning framework, while the necessary extensions to handle significant orbital eccentricity are deferred to future work. However, for typical observation durations ($\Tobs \lesssim$ a few hours), we can safely neglect other deviations such as post-Keplerian relativistic effects and intrinsic pulsar spin-down. We further assume that the sky position is known to within the telescope's beam uncertainty, allowing for the correction of Doppler shifts due to Earth's motion relative to the solar system barycentre \citep{Lorimer:2004}. The input data are assumed to be frequency-averaged and de-dispersed, producing a time series $\mathcal{T}(t)$ in the barycentric frame. This is uniformly sampled at interval $t_s$, giving discrete samples $\mathcal{T}_n = \mathcal{T}(n t_s)$ for $n = 0,\ldots,N_{s}-1$, where $N_{s}$ is the total number of samples.

\subsection{Coherent phase model}\label{subsec:coherent_phase_model}
The detection of pulsars in binary systems requires modelling the periodic Doppler modulation of their observed pulse arrival times, which arises from orbital motion. For a pulsar with an intrinsic spin frequency $f_{\mathrm{int}}$ (defined in its rest frame) and moving with instantaneous line-of-sight velocity $v(t)$, the observed frequency $f(t)$ in the non-relativistic limit is
\begin{equation}\label{eq:doppler}
    f(t) = f_{\mathrm{int}} \bigbracket{1 - \frac{v(t)}{c}} + \mathcal{O}\bigparen{v^2/c^2},
\end{equation}
where $v(t) = \dot{d}(t)$ is the line-of-sight velocity component, defined such that $v(t) > 0$ when the distance to the pulsar is increasing (source receding). The $\mathcal{O}\bigparen{v^2/c^2}$ terms (e.g., relativistic time dilation and higher-order Doppler corrections) are omitted here for simplicity, as we focus on the leading-order Doppler modulation relevant for search modelling. For a general orbit, the observed signal phase $\Phi(t)$, measured in cycles, as a function of arrival time $t$ can be expressed as:
\begin{equation}\label{eq:phase_generic}
    \Phi(t) = \Phi_{\mathrm{ref}} + f_{\mathrm{int}} \bigbracket{(t - \tref) - \frac{d(t)}{c}},
\end{equation}
where $\Phi_{\mathrm{ref}} \equiv \Phi(\tref)$ is the reference phase at epoch $\tref$, $d(t)$ is the varying distance between the pulsar and the observer, and $c$ is the speed of light. 

\subsection{Polynomial phase model}\label{subsec:poly_phase_model}
When the observation span $\Tobs$ is much shorter than the orbital period $\Porb$ (i.e., $\Tobs \ll \Porb$), the orbital motion causes only a small change in the observed frequency. In this regime, the physical phase model can be efficiently approximated by a Taylor series expansion around a reference epoch $\tref$. This polynomial approach simplifies the search but may lose accuracy if the observation span becomes a significant fraction of $\Porb$. The phase evolution is expanded as:
\begin{equation}\label{eq:phase_taylor}
    \Phi(t) = \Phi_{\mathrm{ref}} + \sum_{k=0}^{\kmax} \frac{f_{k}}{(k+1)!}(t - \tref)^{k+1}.
\end{equation}
Here, $\kmax$ is the highest derivative order required to accurately model the phase evolution over $\Tobs$, and the coefficients $f_k$ represent the observed spin frequency and its time derivatives evaluated at $\tref$:
\begin{align}\label{eq:f1}
    f_{k} \equiv f^{(k)}(t= \tref) = \left.\frac{d^k f(t)}{dt^k}\right|_{t = \tref},
\end{align}
where $f(t) = \dot{\Phi}(t)$. The reference epoch is commonly taken as the mid-point of the observation (i.e., $\tref=t_{C}$) to minimize both the maximum phase error and correlation between polynomial terms. In this formalism, the search is performed over a grid in the parameter space defined by the coefficients $\mathbf{\Lambda}_f = \{f_0, f_1, \dots, f_{k_{\max}}\}$.

These frequency derivatives $f_k$ are directly related to the pulsar's intrinsic spin frequency $f_{\mathrm{int}}$ and the line-of-sight kinematic derivatives of the pulsar's motion. We can expand $d(t)$ in a Taylor series about $\tref$:
\begin{equation}\label{eq:dist1}
    d(t) = \sum_{k=0}^{\kmax} \frac{d_{k}}{k!}(t - \tref)^k,
\end{equation}
where $d_{k} \equiv d^{(k)}(t = \tref)$ represents the $k$-th time derivative of the line-of-sight distance variation evaluated at $\tref$. By taking successive time derivatives of the Doppler-shifted frequency given in equation~\eqref{eq:doppler}, we obtain
\begin{equation} \label{eq:fderiv}
    f_k = -\frac{f_{\mathrm{int}}}{c} d_{k+1}, \quad \text{for} \quad k \geq 1.
\end{equation}
For $k=0$, $f_0=f(\tref)=f_{\mathrm{int}}\bigl[1-d_1(\tref)/c\bigr]$, corresponding to the observed spin frequency at the reference epoch. In practice, searches operate directly on $f_0$, which absorbs any bulk Doppler shift. Here, $d_1$, $d_2$, $d_3$, $d_4$, $d_5$ denote the line-of-sight velocity, acceleration, jerk, snap and crackle respectively. Thus the parameter vector $\Ld = \{f_0, d_2, \dots, d_{\kmax+1}\}$ also defines the polynomial search space. All subsequent search algorithms are formulated as searches over $\Ld$.

\subsection{Phase-Coherent Folding}\label{subsec:brute_folding}
To enhance periodic signals buried in noisy data, a standard technique is to fold the time series according to a predictive phase model, producing an integrated phase-resolved profile \citep{Lorimer:2004}. In the time-domain, this process coherently combines signal intensity over the entire observation duration $\Tobs$, amplifying the pulsar's periodic signature relative to the noise. For a given phase bin $b$, the folded profile $\mathcal{P}(b)$ is computed as:
\begin{align}\label{eq:time_folding}
    \mathcal{P}(b) &= \sum_{n=0}^{N_{s}-1} \mathcal{T}_n \cdot \delta_{b, \hat{b}(t_n)}, \\
    \hat{b}(t_n) &= \lfloor \Phi(t_n) N_b \rceil \pmod{N_b},
\end{align}
where $\delta_{i,j}$ is the Kronecker delta, $\Phi(t_n) \in [0, 1)$ is the instantaneous rotational phase of the pulsar (derived from the ephemeris), $N_b$ is the number of phase bins, and $b \in \{0, \ldots, N_b - 1\}$. This formulation is equivalent to histogramming the samples according to their instantaneous phase.

The choice of $N_b$ is constrained by the temporal resolution of the data. The maximum meaningful resolution is bounded such that $N_{b,\max} = \lfloor (f_{\mathrm{int}} t_s)^{-1} \rfloor$. Conversely, selecting $N_b < N_{b,\max}$ effectively down-samples the folded profile. While this downsampling is desirable for computational efficiency, particularly when searching slow pulsars ($f_{\mathrm{int}} \lesssim 1$\,Hz) where high phase resolution is not critical, it acts as a low-pass boxcar filter. This effectively smooths sharp pulse features and introduces ``bin straddling'' losses, where the energy of a narrow duty-cycle pulse is split between adjacent bins due to phase quantization.

\subsubsection{Fourier-Domain Folding}\label{subsec:fourier_folding}
To mitigate the phase quantization errors inherent in time-domain folding, we compute the harmonic content of the folded profile directly in the Fourier domain. Instead of mapping time-series samples to discrete phase bins, we compute the complex Fourier coefficients of the folded profile using the exact floating-point phase $\Phi(t_n)$. This approach yields the precise harmonic amplitudes of the signal's profile in rotational phase, without the information loss associated with binning. The complex coefficient for the $m$-th harmonic, $\hat{\mathcal{P}}(m)$, is given by:
\begin{equation}\label{eq:fourier_folding}
    \hat{\mathcal{P}}(m) = \sum_{n=0}^{N_{s}-1} \mathcal{T}_n \exp\bigbracket{-2\pi i m \Phi(t_n)}
\end{equation}
where $m \in \{0, 1, \ldots, \lfloor N_b/2 \rfloor\}$. The DC term ($m=0$) corresponds to the sum of all samples, while higher harmonics ($m \geq 1$) capture the shape and structure of the pulse profile.

This method avoids the discretization noise associated with time-domain binning and preserves the harmonic content of the pulse profile up to the Nyquist limit set by the chosen resolution $N_b$. The time-domain folded profile can be reconstructed by performing an inverse discrete Fourier transform (IDFT):
\begin{equation}\label{eq:ifft_recovery}
    \mathcal{P}(b) = \frac{1}{N_b} \left[ \hat{\mathcal{P}}(0) + 2\,\text{Re} \left\{ \sum_{m=1}^{N_h} \hat{\mathcal{P}}(m) \exp\left(2\pi i\, m\, \frac{b}{N_b} \right) \right\} \right],
\end{equation}
where $N_h = \lfloor N_b / 2 \rfloor$ is the highest retained harmonic. For brevity, equation~\eqref{eq:ifft_recovery} is written in its standard compact form; the Nyquist-frequency term requires separate treatment when $N_b$ is even. Because no phase quantization is applied, this Fourier-domain approach preserves sharp features and pulse centroids with high fidelity, yielding superior sensitivity for narrow duty-cycle pulses where bin-straddling losses would otherwise be significant. The primary drawback of direct Fourier-domain folding is its steep computational cost, scaling as $\mathcal{O}(N_{s}\, N_h)$, though this can be somewhat mitigated using Nyquist interpolation.

\subsubsection{Optimal Detection Statistics}
The traditional folding approach sums unweighted samples, implicitly assuming uniform noise properties across the entire observation. However, real observational data frequently exhibit time-varying noise variance (\emph{heteroscedasticity}) due to factors such as instrumental gain fluctuations, intermittent RFI, or changing observing conditions. To optimize detection sensitivity under these non-stationary conditions, we employ an optimal weighting strategy derived from maximum likelihood principles under a multiplicative signal model in which the pulse amplitude scales with the local mean level (see Appendix~\ref{app:weighted_fold} for the full derivation).

This approach requires maintaining two auxiliary arrays for the primary time-series data that encode both the measurements and their uncertainties:
\begin{align}
    \label{eq:weighted_data}
    \mathcal{T}_{w,n} &= (\mathcal{T}_n - \mu_n) \cdot \frac{\mu_n}{\sigma^2_n}, \\
    \label{eq:weights}
    \mathcal{T}_{s,n} &= \frac{\mu^2_n}{\sigma^2_n},
\end{align}
where $\mu_n$ and $\sigma^2_n$ are the local mean and variance, respectively, estimated robustly from the data (e.g., using a sliding window or median filtering to mitigate outliers). Here, $\mathcal{T}_{w,n}$ represents the weighted signal contribution, while $\mathcal{T}_{s,n}$ encodes the corresponding statistical weight. Together, these quantities form a pair of sufficient statistics that propagate both signal and variance information through the folding process. The specific weighting adopted here follows from the multiplicative signal model assumed above; alternative noise models may lead to different definitions of $\mathcal{T}_{w,n}$ and $\mathcal{T}_{s,n}$ without altering the subsequent folding formalism. 

These arrays are folded independently into phase bins using the chosen phase model. In the time domain, this yields the weighted profile $\mathcal{P}_w(b)$ and the weight profile $\mathcal{P}_s(b)$ via equation~\eqref{eq:time_folding}. In the Fourier domain, we compute the complex coefficients $\hat{\mathcal{P}}_w(m)$ and $\hat{\mathcal{P}}_s(m)$ via equation~\eqref{eq:fourier_folding}, which are subsequently transformed via IFFT to recover $\mathcal{P}_w(b)$ and $\mathcal{P}_s(b)$. This Fourier-based path preserves the optimal statistical properties while eliminating the phase quantization errors discussed in Section~\ref{subsec:brute_folding}. The resulting $\mathcal{P}_s(b)$ correctly accounts for both inverse-variance weighting and bin occupancy effects when downsampling.

A straightforward approach to construct a detection statistic is to first normalize the profile bin-wise:
\begin{equation} \label{eq:norm_profile}
    \mathcal{P}_{\mathrm{norm}}(b) = \frac{\mathcal{P}_w(b)}{\sqrt{\mathcal{P}_s(b)}},
\end{equation}
and then apply a matched filter using a normalized template profile $T(b)$ (where $\sum_b T(b)^2 = 1$):
\begin{equation} \label{eq:folded_snr}
    \SNR_\alpha = \sum_{b=0}^{N_b-1} \mathcal{P}_{\mathrm{norm}}(b) \, T(b).
\end{equation}
While convenient because the normalized profile $\mathcal{P}_{\mathrm{norm}}(b)$ can be reused across multiple template evaluations,  $\SNR_\alpha$ is statistically suboptimal when the accumulated weights $\mathcal{P}_s(b)$ vary significantly across pulse phase. The statistically optimal detection statistic, derived from maximum likelihood principles, is:
\begin{equation} \label{eq:optimal_snr}
    \SNR_\beta = \frac{\sum_{b} \mathcal{P}_w(b) \, T(b)}{\sqrt{\sum_{b} \mathcal{P}_s(b) \, T(b)^2}}.
\end{equation}
Here, the numerator represents the projection of the weighted data onto the template, while the denominator provides the correct normalization accounting for both the template shape and the per-bin variance. This formulation ensures that bins with higher accumulated weight $\mathcal{P}_s(b)$ contribute appropriately to the final statistic, maximizing overall sensitivity. The choice between $\SNR_\alpha$ and $\SNR_\beta$ involves a trade-off between computational efficiency and statistical optimality.

For periodic signals with known period and assumed pulse shape, the optimal detection procedure involves forming a phase-coherent folded profile and correlating it against a family of zero-mean, unit-energy templates \citep{Morello:2020}. Common choices include boxcar or single-Gaussian templates. In this work, we adopt a time-domain matched filtering approach using boxcar templates as a computationally efficient baseline and defer more sophisticated template families and Fourier-domain detection statistics to future work. Specifically, we implement the $\SNR_\alpha$ statistic and search over a set of boxcar widths $w \in \mathcal{W}$ spanning the target duty cycle range. The final detection statistic is the maximum over all trial widths:
\begin{equation}
    \SNR_{\max} = \max_{w \in \mathcal{W}} \SNR_\alpha(w).
\end{equation}
To compute boxcar correlations efficiently across multiple widths, we employ a circular prefix-sum algorithm that reduces computational complexity from $\mathcal{O}(w_{\max}\,N_b)$ for an incremental running-sum implementation to $\mathcal{O}(N_w\,N_b)$ after an initial $\mathcal{O}(N_b + w_{\max})$ setup cost, where $N_w=|\mathcal{W}|$ and $w_{\max}=\max(w\in\mathcal{W})$. The improvement is most significant when the trial widths are sparsely sampled (e.g., logarithmically spaced), such that $N_w \ll w_{\max}$ \citep{Morello:2020}.

As a faster alternative to the exhaustive width search, we also implement an approximate scoring method based on Kadane's maximum subarray algorithm \citep{Kadane:2023}, which reduces the per-profile complexity to $\mathcal{O}(K\,N_b)$ with $K\sim3$ fixed linear bias passes over the profile (D. Gazith et al.\ 2026, in preparation). The approximation incurs a small false-dismissal rate of $\lesssim 5\%$ in the low-S/N regime ($\mathrm{S/N}\lesssim 5$), with the rate dropping rapidly at higher $\mathrm{S/N}$.

\subsection{Search Grid Design}\label{sec:search_grid_design}
A coherent search over a multidimensional parameter space necessitates discretizing it with a grid fine enough to prevent significant signal loss between neighbouring grid points \citep{Allen:2013}. The optimal grid resolution in each parameter dimension is derived by bounding the phase error incurred due to a mismatch between the true signal parameters and the nearest grid point over the observation interval $t \in [0, \Tobs]$. 

Grid construction strategies vary depending on the trade-off one wishes to make between detection sensitivity and computational cost. In this work, we adopt a conservative gridding criterion based on limiting the cumulative phase drift. Specifically, we require that any mismatch from the true parameters induces no more than $\eta$ fold bins of cumulative phase offset for a folded profile with $N_b$ bins. This leads to the constraint:
\begin{equation}\label{eq:grid_criteria}
    |\Delta \Phi(t)| \lesssim \frac{\eta}{N_b}, \quad \text{for all } t \in [0, \Tobs],
\end{equation}
which ensures minimal loss of coherent power. The tolerance parameter $\eta$ regulates the grid density, balancing sensitivity against search complexity.

A straightforward approach is to treat deviations in each Taylor coefficient $f_k$ independently. Since the maximum phase deviation for monomials occurs at the endpoints of the observation interval, evaluating $\Delta \Phi(t)$ at $t = \Tobs$ yields the ``naive" grid spacing:
\begin{equation}\label{eq:grid_params}
    \Delta f_k = \frac{\eta}{N_b} \frac{(k+1)!}{(\Tobs - \tref)^{k+1}}.
\end{equation}
The total number of grid points required to cover a search range $\Delta f_k^{\mathrm{range}}$ up to order $k_{\max}$ is
\begin{equation}\label{eq:grid_vol}
    N_{\mathrm{grid, brute}} = \prod_{k=0}^{k_{\max}} \left\lceil \frac{\Delta f_k^{\mathrm{range}}}{\Delta f_k} \right\rceil,
\end{equation}
where $\Delta f_k^{\mathrm{range}} = f_k^{\mathrm{max}} - f_k^{\mathrm{min}}$. In practice, search implementations often operate in terms of kinematic parameters $\Ld$, which are related to the phase model via frequency derivatives. For such cases, it is useful to translate the gridding criterion accordingly. Assuming a frequency search range $[f_{\min}, f_{\max}]$ and conservatively setting the intrinsic frequency $f_{\mathrm{int}}$ to $f_{\max}$, the required grid spacing in the $k$-th $\Ld$ parameter using equation~\eqref{eq:fderiv} is:
\begin{equation}
    \Delta d_k = \frac{c}{f_{\max}} \Delta f_{k-1}, \quad k \ge 2.
\end{equation}

However, this method is computationally inefficient. The monomial basis functions $\{ (t-t_c)^k \}$ underlying the Taylor expansion are not orthogonal over the observation span, leading to strong correlations between the model parameters. Geometrically, the valid parameter search volume is a highly elongated hyper-ellipsoid (a ``needle'') rather than a hyper-rectangle. A simple rectangular grid in the $\{d_k\}$ space is therefore highly redundant, as the grid axes do not align with the principal axes of the parameter metric. 

To address this, we employ a \emph{hybrid strategy} for Taylor-basis based search: we retain the physically intuitive Taylor coefficients for the search coordinates but define the grid density based on an orthogonal basis analysis. This procedure, detailed in Appendix~\ref{app:optimal_gridding}, utilizes Chebyshev polynomials to diagonalize the parameter metric. The analysis yields an optimally spaced grid for the kinematic parameters:
\begin{equation}\label{eq:dk_optimal}
    \Delta d_k^{\mathrm{opt}} = \frac{2^{2k-1} \eta c \, k!}{N_b f_{\max} \Tobs^k}, \quad k \geq 2.
\end{equation}
The orthogonalization approach thus allows for a coarser, more efficient grid by a coarsening factor of $2^{k-1}$ for each derivative order $k \ge 2$ compared to the naive method. This dramatically reduces the total number of grid points, $\Ngrid$, making higher-order searches computationally tractable.

\section{Polynomial Fast Folding Algorithm}\label{sec:ffa_poly}
The core idea behind the Fast Folding Algorithm (FFA) is to avoid redundant computation by exploiting the hierarchical structure of the folding process. Standard FFA implementations efficiently explore frequency or period parameter space by reusing partial folds \citep[see, e.g.,][]{Staelin:1969, Lovelace:1969, Cameron:2017, Parent:2018, Morello:2020, Pearlman:2021, Shahaf:2022}. Here, we generalize the FFA to efficiently search the multi-dimensional parameter space $\Ld$, where a vector $\vlambda = \{f_0, d_2, \dots, d_{k_{\max}+1}\} \in \Ld$ governs the polynomial phase model.

\subsection{Algorithm Description}
The polynomial FFA (P-FFA) employs a dynamic programming strategy. The input data, spanning a total duration $\Tobs$ are partitioned into $N_0$ non-overlapping base segments of equal duration $T_0$. For clarity of exposition, we assume $\Tobs = N_0\,T_0$; non-dyadic lengths can be handled by padding or by a final partial merge. The algorithm progressively combines folded profiles from shorter segments to construct profiles over longer durations, reusing computed fold sums across both parameter trials and time segments.

\begin{figure}
\includegraphics[width=\columnwidth]{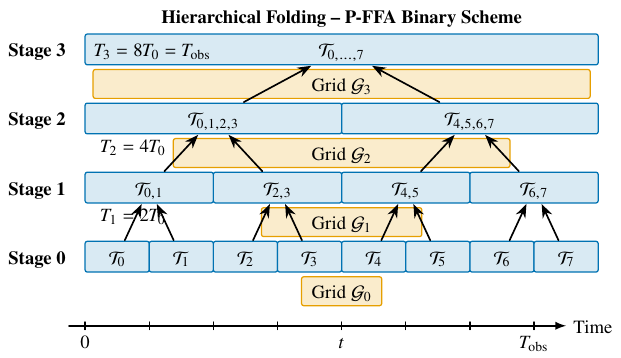}
\caption{Schematic of the bottom-up, breadth-first P-FFA merge. An observation of duration $\Tobs$ is partitioned into eight equal segments. Folding profiles are initialized at stage 0 on the coarsest parameter grid. Each subsequent stage coherently combines profiles from adjacent segments in a level-synchronous manner, halving the number of segments at each step until profiles spanning the full observation on the final grid are constructed.
\label{fig:folding_binary}}
\end{figure}

Conceptually, this hierarchical process is recursive. We define a \emph{profile state} as a tuple $\mathcal{V} = \{\mathcal{P}_w(b), \mathcal{P}_s(b)\}$, representing the weighted folded profile vector and its corresponding weight vector, respectively. Let $\mathcal{V}(t_{\mathrm{mid}}, T, \vlambda)$ denote the state for a segment of duration $T$ centred on $t_{\mathrm{mid}}$, folded with the phase model specified by $\vlambda$. The state for a duration $2T$ is constructed from the states of its two constituent halves,
\begin{equation} \label{eq:ffa_recursion_conceptual}
\mathcal{V}(0, 2T, \vlambda_{i+1})  =
\mathcal{V}\!\left(-\frac{T}{2}, T, \vlambda_{i,L}\right)
\oplus
\mathcal{V}\!\left(+\frac{T}{2}, T, \vlambda_{i,R}\right).
\end{equation}
Here, $\vlambda_{i+1}$ is a parameter vector on the stage-$(i+1)$ grid $\mathcal{G}_{i+1}$, defined for duration $2T$. The vectors $\vlambda_{i,L}$ and $\vlambda_{i,R}$ are the parameters on the coarser stage-$i$ grid $\mathcal{G}_i$, defined for duration $T$, that best approximate the phase evolution of $\vlambda_{i+1}$ over the left and right sub-intervals. The operation $\oplus$ denotes coherent combination of two pre-computed states. It involves retrieving the two sub-states, applying the phase shifts ($\Delta\phi_L$, $\Delta\phi_R$) required to align them to the target model $\vlambda_{i+1}$, and summing the shifted states component wise. The recursion terminates at a base duration $T_0$, where states are initialized by direct brute-force folding. On this shortest time-scale, a constant-frequency approximation is often sufficient for initialization. While the recursion provides the conceptual framework, we employ an iterative bottom-up implementation (Algorithm~\ref{alg:ffa_poly}), as it is generally more efficient. Figure~\ref{fig:folding_binary} shows a schematic of the hierarchical merge.

The algorithm begins with base segments of duration $T_0$. At each stage $i$, the integration time doubles ($T_i = 2^i T_0$), so that under the dyadic assumption, the total number of merge stages is
\begin{equation}
    N_{\mathrm{stages}} = \left\lfloor \log_2 \left( \frac{\Tobs}{T_0} \right) \right\rfloor .
\end{equation}
The core operations in Algorithm~\ref{alg:ffa_poly} are: 
\begin{itemize}
    \item \func{DetermineGrid}: Generates the stage-dependent parameter grid $\mathcal{G}_i \subset \Ld$ for the search bounds $\vlambda_{\mathrm{bounds}}$, tolerance $\eta$ and number of phase bins $N_b$, referenced to the current segment midpoint.
    \item \func{ComputeBruteFold}: Initializes the base profile states by direct folding of the time series over segments of duration $T_0$.
    \item \func{Resolve}: Projects a target parameter vector $\vlambda_{i+1} \in \mathcal{G}_{i+1}$ into the local frame of $\mathcal{G}_i$ accounting for time translation and computes the residual phase offset.
    \item \func{Shift}: Applies the required phase shift to a profile state, either as a cyclic shift in the time domain or as a complex phase rotation in the Fourier domain.
\end{itemize}
The final output is a set of coherently folded profiles corresponding to the final grid at $\Tobs$. A matched-filter search using boxcar templates is then performed on each profile to produce the detection statistic across the target range of pulse widths.

\begin{algorithm}[H]
\caption{Bottom-Up, Breadth-First Hierarchical P-FFA}
\label{alg:ffa_poly}
\begin{algorithmic}[1]
\Require Time series $\mathcal{T}(t)$, sampled at interval $t_s$
\Require Maximum polynomial order $\kmax$
\Require Search bounds $\vlambda_{\mathrm{bounds}}$ for $\Ld = \{f_0, d_2, \dots, d_{\kmax+1}\}$
\Require Coherence tolerance $\eta$ (bins), fold bins $N_b$
\Require Base segment duration $T_0$, with $\Tobs / T_0 \in \mathbb{Z}$

\Ensure Folded states $\mathcal{V}$ for duration $\Tobs$ on grid $\mathcal{G}_{N_{\mathrm{stages}}}$

\State \textbf{Initialize:}
\State $N_0 \gets \Tobs / T_0$ \Comment{Number of base segments}
\State $N_{\mathrm{stages}} \gets \lfloor \log_2(N_0) \rfloor$ \Comment{Number of full merge stages}
\State $\mathcal{G}_0 \gets \func{DetermineGrid}(T_0, k_{\max}, \vlambda_{\mathrm{bounds}}, \eta, N_b)$ 
\State Allocate ping-pong buffers: $\mathcal{B}_{\mathrm{curr}}$ and $\mathcal{B}_{\mathrm{next}}$

\State \textbf{Base Folding (Stage 0):}
\For{$j = 0$ to $N_0 - 1$} \Comment{Loop over base segments}
    \State $t_{\mathrm{start}} \gets j\,T_0$, \quad $t_{\mathrm{end}} \gets (j + 1)\,T_0$
    \State $t_{\mathrm{mid}} \gets \left(j + \tfrac{1}{2}\right) T_0$
    \State Extract segment: $\mathcal{T}_{\mathrm{seg}} = \mathcal{T}[t_{\mathrm{start}} : t_{\mathrm{end}})$
    \For{each $\vlambda \in \mathcal{G}_0$}
        \State $\mathcal{V}_{\vlambda} \gets \func{ComputeBruteFold}(\mathcal{T}_{\mathrm{seg}}, \vlambda, t_{\mathrm{mid}})$
        \State $\mathcal{B}_{\mathrm{curr}}[j][\vlambda] \gets \mathcal{V}_{\vlambda}$
    \EndFor
\EndFor

\State \textbf{Hierarchical Merging:}
\For{stage $i = 0$ to $N_{\mathrm{stages}} - 1$}
    \State $T_i \gets T_0 \cdot 2^i$, \quad $T_{i+1} \gets 2 \cdot T_i$
    \State $\mathcal{G}_{i+1} \gets \func{DetermineGrid}(T_{i+1}, k_{\max}, \vlambda_{\mathrm{bounds}}, \eta, N_b)$
    \State $N_{i+1} \gets N_0 / 2^{i+1}$ \Comment{Merged segments at next stage}
    
    \For{$j' = 0$ to $N_{i+1} - 1$}
        \For{each target $\vlambda_C \in \mathcal{G}_{i+1}$}
            \State $\vlambda_L, \Delta\phi_L \gets \func{Resolve}(\vlambda_C, \mathcal{G}_i, -T_i/2)$
            \State $\vlambda_R, \Delta\phi_R \gets \func{Resolve}(\vlambda_C, \mathcal{G}_i, +T_i/2)$

            \State $\mathcal{V}_L \gets \mathcal{B}_{\mathrm{curr}}[2j'][\vlambda_L]$
            \State $\mathcal{V}_R \gets \mathcal{B}_{\mathrm{curr}}[2j' + 1][\vlambda_R]$
            
            \State $\mathcal{V}_C \gets \func{Shift}(\mathcal{V}_L, \Delta\phi_L) + \func{Shift}(\mathcal{V}_R, \Delta\phi_R)$
            \State $\mathcal{B}_{\mathrm{next}}[j'][\vlambda_C] \gets \mathcal{V}_C$
        \EndFor
    \EndFor
    \State Swap buffers: $\mathcal{B}_{\mathrm{curr}} \leftrightarrow \mathcal{B}_{\mathrm{next}}$
\EndFor
\State \Return $\mathcal{B}_{\mathrm{curr}}[0]$  \Comment{All final states over $\mathcal{G}_{N_{\mathrm{stages}}}$}

\end{algorithmic}
\end{algorithm}

\subsection{Adaptive Parameter Grids and Phase Coherence}\label{sec:adaptive_grids}
A key feature of the P-FFA is the use of an adaptive parameter grid $\mathcal{G}_i$ at each hierarchical stage $i$. As the coherent integration time doubles, the grid resolution must increase, and the dimensionality of the active search space may also expand in order to maintain phase coherence.

\subsubsection{Grid Refinement}
To keep the phase error below a prescribed tolerance of $\eta/N_b$ cycles, the grid spacing for each parameter in $\vlambda$ must scale with the coherent integration time $T_i$ as $\Delta f_k \propto T_i^{-(k+1)}$ (see equation~\eqref{eq:grid_params}). Consequently, the stage-dependent grid $\mathcal{G}_i$ is refined as $T_i$ increases. Higher-order parameters $f_k$ (where $k \ge 1$) only require inclusion in the search once their maximum phase contribution over the accumulated segment duration $T_i$ exceeds the tolerance. We therefore define a characteristic activation time-scale $T_{c,k}$ for each parameter:
\begin{equation}\label{eq:threshold_time}
    T_{c,k} \approx 2 \left( \frac{\eta (k+1)!}{N_b |f_{k, \mathrm{max}}|} \right)^{\frac{1}{k+1}},
\end{equation}
where $|f_{k, \mathrm{max}}|$ is the maximum expected magnitude of the parameter. The frequency term $f_0$ is active by default, i.e. $T_{c,0}=0$. At stage $i$, the active set is $A_i = \{k : T_i \gtrsim T_{c,k}\}$, and only coefficients in $A_i$ are included in $\mathcal{G}_i$. This adaptive activation of parameters substantially reduces the search volume at early stages without sacrificing phase coherence.

The total number of grid points at stage $i$ is then:
\begin{equation} \label{eq:n_grid_adaptive}
    N_{\mathrm{grid, FFA}}(T_i) = \prod_{k \in A_i} \left\lceil \frac{\Delta f_k^{\mathrm{range}}}{\Delta f_k(T_i)} \right\rceil,
\end{equation}
where $\Delta f_k^{\mathrm{range}}$ is the search range in the $k$th active dimension and $\Delta f_k(T_i)$ is the spacing required at duration $T_i$. For inactive parameters ($T_i < T_{c,k}$), the grid size contribution is effectively unity.

\subsubsection{Phase alignment and Coherent Combination}\label{sec:phase_alignment}
The hierarchical merge depends critically on the precise alignment of the constituent profile states ($\mathcal{V}_L, \mathcal{V}_R$) to a common phase reference before summation. This ``stitching'' step ensures coherence with the target phase model $\Phi_C(t)$ defined by parameters $\vlambda_C \in \mathcal{G}_{i+1}$ for the merged segment of duration $T_{i+1}=2T_i$.

Let the merged segment be centred at reference epoch $t_C = 0$, spanning $[-T_i, T_i)$.  Its left and right halves are centred at $t_L = -T_i/2$ and $t_R = +T_i/2$, respectively. To combine them, we must determine the equivalent states of the target parameter $\vlambda_C$ in the reference frames of the halves. Because the constituent states were folded relative to their local reference epochs ($t_L, t_R$), they must be transported to a common phase convention $t_C$, before they can be combined. 

We use the Taylor basis transformation $\mathbf{T}(\Delta t)$, described in Appendix~\ref{app:taylor_basis}, to project the global parameters onto the local frames. The projected vectors for the two halves are:
\begin{equation}\label{eq:param_translation}
    \vlambda_{\{L,R\}} = \mathbf{T}(\Delta t_{\{L,R\}})\,\vlambda_C, \qquad \Delta t_{\{L,R\}} = \mp \frac{T_i}{2}.
\end{equation}
In general, this translation produces a non-zero zeroth-order delay term, $d_{0,\{L, R\}}$, in the local expansion. This term represents the integrated geometric path-length offset accumulated over the interval $\Delta t_{\{L, R\}}$. The required phase transport is the phase difference between the global model evaluated at the local reference epoch and the locally referenced model at its origin:
\begin{equation}\label{eq:phase_transport}
    \Delta \phi_{\{L, R\}} = f_{0, C} \left( \Delta t_{\{L, R\}} - \frac{d_{0,\{L, R\}}}{c} \right) \pmod 1,
\end{equation}
where $f_{0, C}$ is the frequency parameter defined in $\vlambda_C$. In practice, \func{Resolve} returns both the nearest stage-$i$ grid point and the corresponding residual phase offset $\Delta\phi_{\{L,R\}}$.

To form the merged profile state $\mathcal{V}_C$, we apply cyclic shifts to the constituent profiles and sum them element-wise:
\begin{equation}
    \mathcal{V}_C(b) = \mathcal{V}_L(b - \Delta b_L) + \mathcal{V}_R(b - \Delta b_R),
\end{equation}
where the shift operator is realized as a nearest-bin integer shifts with $\Delta b_{\{L,R\}} = \lfloor \Delta \phi_{\{L,R\}} \cdot N_b \rceil$. This operation effectively ``rewinds'' the local profiles to the common reference epoch of the merged segment and preserves coherence across the full observation span.

\subsection{Sensitivity Loss and Error Bounds}
Although the hierarchical construction substantially reduces computational cost, it introduces approximations that accumulate over the $N_{\mathrm{stages}}$ merge levels and can degrading sensitivity. The dominant contributions to this error budget are grid discretization and the numerical realisation of profile shifts.

\subsubsection{Grid Discretization}
The intrinsic mismatch of the hierarchical search arises from discretizing the local parameter grid $\mathcal{G}_i$. At stage $i$, the ideal parameter vector obtained by projecting a parent-grid point $\vlambda_C \in \mathcal{G}_{i+1}$ into a child segment will, in general, not lie exactly on $\mathcal{G}_i$ (see equation~\eqref{eq:param_translation}). The \func{Resolve} function therefore selects the nearest neighbour $\vlambda_{\{L, R\}} \in \mathcal{G}_i$, introducing a residual parameter offset $\delta \vlambda = \vlambda_{\mathrm{ideal}} - \vlambda_{\{L, R\}}$. 

For a polynomial phase model, the worst-case phase mismatch at a single merging stage occurs when the parameter residuals in all active dimensions contribute constructively. The magnitude of this single-stage error is bounded by the linear sum of the contributions from each dimension:
\begin{equation}
    |\delta \Phi_{\mathrm{stage}}| \approx \left| \sum_{k} \frac{\delta f^{(k)}}{(k+1)!} \left(\frac{T_i}{2}\right)^{k+1} \right| \lesssim \sum_{k} \frac{\eta}{2 N_b},
\end{equation}
where $\delta f^{(k)} \le \Delta f_k(T_i)/2$ is the maximum distance to the nearest grid point in the $k$-th dimension.

Crucially, this mismatch is not an isolated penalty but a cumulative phase error. Because each sequential merge operation aligns profiles based on these discrete nearest neighbours, the single-stage mismatches compound over the full integration duration. Assuming the nearest-neighbour offsets are effectively uncorrelated across levels, the total accumulated phase error can be approximated as $\approx \sqrt{N_{\mathrm{stages}}} |\delta \Phi_{\mathrm{stage}}|$ (behaving as a random walk), though it can scale linearly in the absolute worst-case scenario.

This accumulated parameter mismatch is controlled by the coherence tolerance $\eta$, which sets the grid resolution. A larger $\eta$ implies a coarser grid and reduced computational cost, but increases the risk of signal smearing, where the uncompensated phase drift causes the signal power to disperse across multiple phase bins. This trade-off is particularly critical for pulsar signals with narrow duty cycles ($\lesssim 5\%$), where even a sub-bin cumulative drift can lead to significant S/N degradation.

\begin{figure*}
\includegraphics[width=\textwidth]{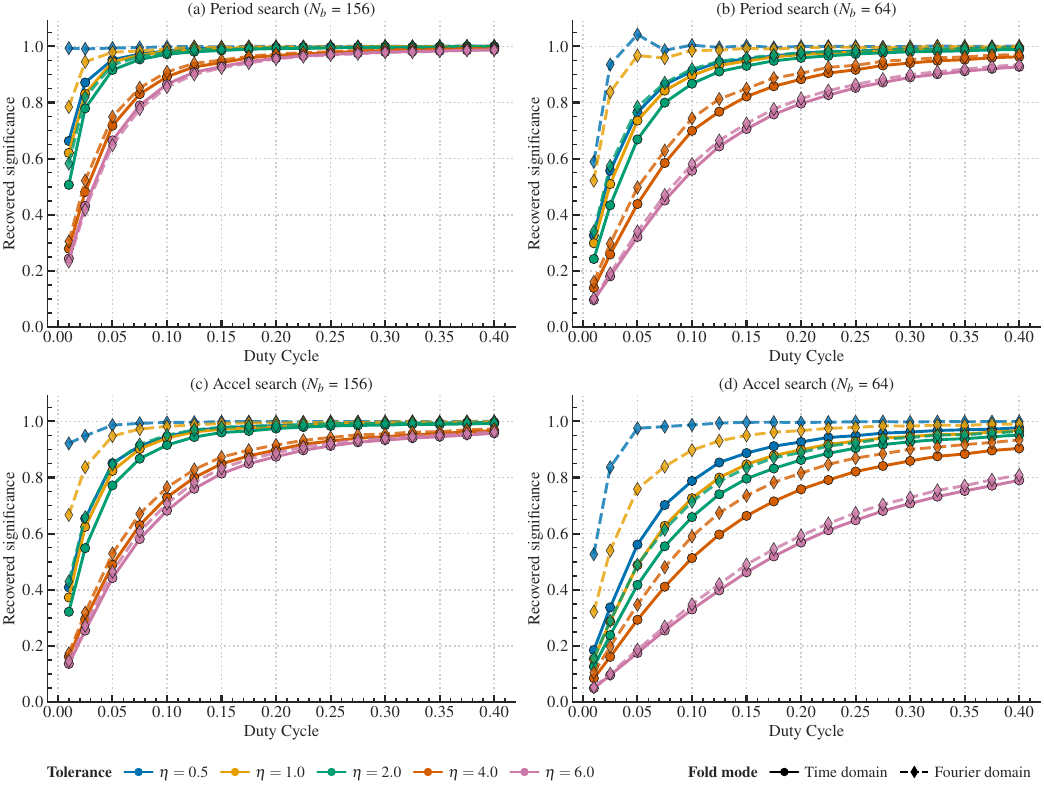}
\caption{Convergence analysis of P-FFA sensitivity for injected pulsar signals. Each panel shows the recovered significance or $(S/N)^2$ fraction $R = (\mathcal{Z}_{\mathrm{detected}}/\mathcal{Z}_{\mathrm{injected}})^2$, as a function of pulsar duty cycle, for two simulated test cases in white Gaussian noise. \textbf{Panel (a--b):} Pulsar with an intrinsic period of 10\,ms. \textbf{Panel (c--d):} Same period (10\,ms), but with a constant acceleration of 100\,m\,s$^{-2}$.
Coloured lines indicate different tolerance levels ($\eta$), ranging from 0.5 to 6 fold bins. Each case is shown for two folding resolutions: \textbf{Left:} Search with the Nyquist-limited maximum folding resolution ($N_b = 156$ bins), \textbf{Right:} Coarser folding with a custom choice of $N_b = 64$.
All simulations use 9-min observations with a sampling interval of 64\,$\upmu$s. The injected signal has a nominal $S/N = 1000$, chosen to isolate algorithmic recovery losses from statistical noise fluctuations. Solid lines represent time-domain P-FFA recovery; dashed lines show the corresponding Fourier-domain results.
\label{fig:ffa_sensitivity}}
\end{figure*}

\subsubsection{Phase Shift Quantization}
A second, implementation-dependent error source arises when the corrective phase shifts are applied to discretized folded profiles. In a purely time-domain implementation,  a shift by $\Delta\phi_{\{L,R\}}$ cycles is approximated by a nearest-integer bin rotation. The resulting rounding error is bounded by
\begin{equation}
    |\delta \phi_{\mathrm{round}}| \le \frac{1}{2N_b}
\end{equation}
for each shifted profile at each merge step. Like grid discretization, this error accumulates over $N_{\mathrm{stages}}$ merging steps, leading to decoherence. In the worst case these errors can add coherently over successive levels, although in practice these are expected to be only weakly correlated between stages, leading to partial cancellation and an approximately random-walk accumulation.

This error can be eliminated by performing the profile combination in the Fourier domain. By the Fourier shift theorem, a spatial shift in the time domain is equivalent to applying a linear phase ramp in the frequency domain. For a profile $\mathcal{P}(b)$ with $N_b$ bins, a fractional translation corresponding to $\Delta \phi$ cycles is achieved by multiplying its discrete Fourier Transform, $\hat{\mathcal{P}}(m)$, by a complex exponential:
\begin{equation}\label{eq:fft_shift}
    \mathcal{F}\left[ \mathcal{P}(b - \Delta b) \right] = \hat{\mathcal{P}}(m) \exp\bigparen{-2\pi i\frac{m\Delta b}{N_b}}.
\end{equation}
In our P-FFA implementation, we also provide a Fourier-domain folding mode. The profile states $\mathcal{V}$ are stored as a set of complex Fourier coefficients from the base folding stage onward (see Section~\ref{subsec:fourier_folding}). The fractional phase shifts $\Delta \phi_{\{L,R\}}$ are then applied at fractional-bin precision through equation~\eqref{eq:fft_shift}, and the two states are added directly in the Fourier domain. This technique is mathematically equivalent to perfect sinc interpolation for band-limited signals and entirely removes the rounding error $\delta \phi_{\mathrm{round}}$.

With the elimination of rounding errors via Fourier-domain P-FFA, the grid discretization error becomes the dominant residual source of sensitivity loss. The coherence tolerance $\eta$ therefore sets the principal trade-off of the method: smaller values improve recovered $\mathrm{S/N}$, particularly for narrow duty-cycle signals, but require a larger search grid and correspondingly higher computational cost. Figure~\ref{fig:ffa_sensitivity} illustrates this trade-off, showing the convergence of retrieval efficiency towards the asymptotic limit as $\eta$ is reduced.

\subsection{Complexity analysis}\label{sec:ffa_complexity}
We now estimate the computational cost and memory requirements of the hierarchical P-FFA, using a brute-force (BF) coherent fold as the reference baseline. The BF method folds the full time series of $N_s = \Tobs/t_s$ samples for every point in the final parameter grid $\mathcal{G}_{\mathrm{final}}$. The cost per grid point is $\mathcal{O}(N_s)$, giving
\begin{equation}\label{eq:brute_cost}
    C_{\mathrm{BF}} = \mathcal{O}\!\left(N_s \cdot \Ngrid(\Tobs)\right).
\end{equation}
Its memory requirement is modest, consisting primarily of the time series plus a single profile state, i.e. $\mathcal{O}(N_s + N_b)$.

For the P-FFA, the total cost is composed of an initialization step and a series of merge steps. Initialization folds all  $N_0 = \Tobs/T_0$ base segments over the stage-0 grid $\mathcal{G}_0$, giving
\begin{equation}
    C_{\mathrm{init}} = \mathcal{O}\!\left(N_s \cdot N_{\mathrm{grid}}(T_0)\right).
\end{equation}

At merge stage $i$, the algorithm combines $N_0/2^{i+1}$ adjacent segment pairs for each point in the next-stage grid $\mathcal{G}_{i+1}$. Each merge requires retrieving two profile states, evaluating the phase transport, shifting both states, and summing them, for a cost of $\mathcal{O}(N_b)$ per merge (time-domain folds). The total merge cost is therefore
\begin{equation}
    C_{\mathrm{merge}} = \sum_{i=0}^{N_{\mathrm{stages}}-1} \frac{N_0}{2^{i+1}}\,N_{\mathrm{grid}}(T_{i+1})\,\mathcal{O}(N_b).
\end{equation}

For a constant-frequency search ($\kmax=0$), one has $N_{\mathrm{grid}}(T_i)\propto T_i$, and hence 
\begin{equation}
    C_{\mathrm{merge}} = \mathcal{O}\!\left(\Tobs N_b \log\!\frac{\Tobs}{T_0}\right),
\end{equation}
recovering the standard FFA scaling \citep{Morello:2020}. For polynomial searches with $\kmax \ge 1$, the parameter-space volume grows super-linearly with integration time (e.g., $\propto T_i^3$ for acceleration searches and even more steeply for higher-order models), causing the merge sum to be dominated by its final stages ($i = N_{\mathrm{stages}}-1$). The cost of this final stage is $\mathcal{O}(\Ngrid(\Tobs) \cdot N_b)$. The overall P-FFA cost is then
\begin{equation} \label{eq:ffa_cost_cpu}
    C_{\mathrm{FFA}} \approx \mathcal{O}(N_s \cdot N_{\mathrm{grid}}(T_0) + \Ngrid(\Tobs) \cdot N_b).
\end{equation}
Because the final grid is generally much larger than the initial grid, the second term dominates in the polynomial-search regime. In this limit, the asymptotic speedup relative to brute-force folding is
\begin{equation}
    \text{Speedup} = \frac{C_{\mathrm{BF}}}{C_{\mathrm{FFA}}} \approx \mathcal{O}\!\left(\frac{N_s}{N_b}\right).
\end{equation}

This speedup comes at the cost of substantially higher memory usage. Since the algorithm traverses the hierarchy in a breadth-first (level-synchronous) manner, it must retain the entire grid level in memory before proceeding to the next. Each stored state consists of a folded profile associated with a specific segment and grid point. With a \emph{ping-pong} buffer strategy, the memory footprint at stage $i$ is
\begin{equation}
M_i = \frac{N_0}{2^i}\,N_{\mathrm{grid}}(T_i)\,\mathcal{O}(N_b). 
\end{equation}
For polynomial searches with super-linear grid growth, the peak memory occurs in the final stages and scales as
\begin{equation}\label{eq:ffa_memory_peak}
    M_{\mathrm{FFA,peak}} = \mathcal{O}\!\left(\Ngrid(\Tobs)\cdot N_b\right).
\end{equation}
While this peak requirement might appear to impose a strict ceiling, potentially limiting practical searches to low polynomial orders such as constant-acceleration searches, it is easily mitigated using a standard time-memory trade-off. To evaluate higher-dimensional parameter spaces without exceeding hardware memory limits, we can truncate the dynamic programming hierarchy. Instead of a pure breadth-first traversal up to the final stage, memoization is halted at an intermediate stage $i_{\mathrm{trunc}}$ where the total state footprint remains within the available memory. The final folded profiles for the dense target grid $\mathcal{G}_{\mathrm{final}}$ are then constructed on-the-fly by querying and combining these stored intermediate states. This hybrid approach shifts the later layers from breadth-first memoization to depth-first computation. 

Although this strategy strictly bounds peak memory to a user-defined threshold, it introduces redundant fold-merge operations that increase the total computational cost $C_{\mathrm{FFA}}$. For higher-order polynomial searches ($\kmax \ge 2$), the combinatorial explosion of grid points at the terminal merge levels turns this time-memory trade-off into a severe computational bottleneck. Consequently, while memory is no longer a hard constraint, the sheer number of terminal-node evaluations naturally restricts the practical application of P-FFA to low polynomial orders, typically no higher than constant acceleration.

In summary, the hierarchical P-FFA achieves a dramatic reduction in computational complexity compared to brute-force methods by exploiting dynamic programming, making it a highly efficient engine for low-dimensional searches. However, overcoming the prohibitive enumeration costs of higher-order parameter spaces requires a fundamentally different strategy.

\section{Extreme Pruning: A Novel Algorithm}\label{sec:pruning_alg}
The central computational challenge in binary pulsar detection is the combinatorial explosion of the search space as the observation duration $\Tobs$ increases. Existing approaches are limited by the need to evaluate a number of trial templates that scales as a high-order polynomial in $\Tobs$ \citep{Balakrishnan:2022}. Here, we propose \emph{Extreme Pruning} (EP), an algorithm that couples hierarchical grid refinement to adaptive candidate elimination in order to reduce the effective search complexity while preserving coherent phase tracking.

\begin{figure}
\includegraphics[width=\columnwidth]{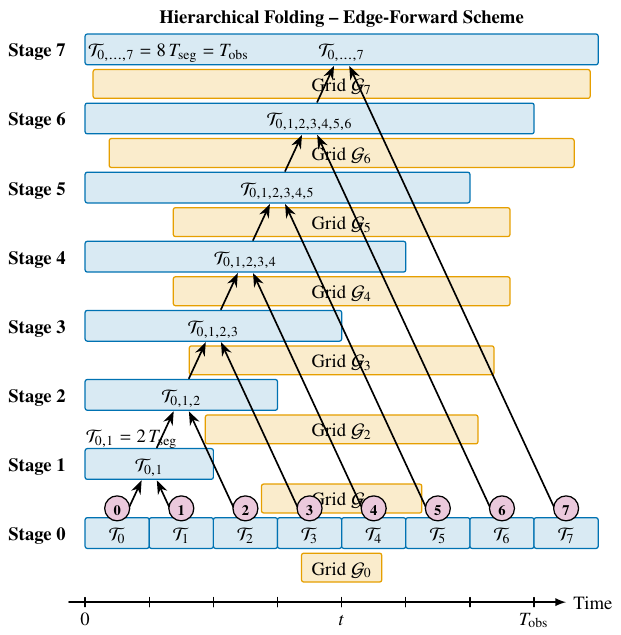}
\caption{Schematic of the hierarchical merging step of the EP algorithm for the edge-forward scheme ($q=0$). An observation of duration $\Tobs$ is partitioned into eight equal segments. Folding profiles are initialized at stage 0 on the base grid for each segment. Each subsequent stage coherently adds the next segment to the accumulated profile states, refining the grid as the coherent span increases, until the profiles covering the full observation on the final grid are constructed.
\label{fig:folding_edge_forward}}
\end{figure}

\subsection{Algorithm Description}\label{sec:pruning_algo_desc}
\begin{figure}
\includegraphics[width=\columnwidth]{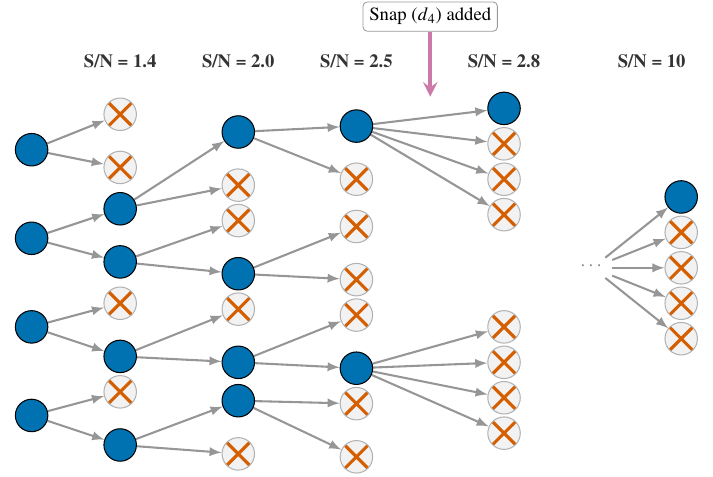}
\caption{Schematic of the pruning procedure used in the EP algorithm. Each node represents a surviving candidate after scoring at a given stage, and each branch corresponds to refinement onto the finer parameter grid of the next stage. Candidates falling below the stage-specific threshold are pruned, so that only high-significance branches are propagated forward.
\label{fig:pruning_schematic}}
\end{figure}

The EP algorithm partitions an input time series of total duration $\Tobs$ into $M$ non-overlapping base segments, each of duration $\Tseg=\Tobs/M$. Let $\mathcal{T}_j$ denote the data in segment $j$, spanning the interval $[j\Tseg, (j+1)\Tseg)$. The algorithm proceeds through $M$ accumulation stages indexed by $s = 0, \dots, M-1$, using a \emph{middle-out} traversal scheme anchored at a starting segment $q \in \{0, \dots, M-1\}$. At stage $s$, it incorporates the new segment indexed by $j = \mathcal{J}(s, q)$, so that the coherent integration time becomes $T_s = (s+1)\Tseg$. The mapping $\mathcal{J}$ is defined in Appendix~\ref{app:middle_out}. In the special case $q=0$, the traversal reduces to a simple \emph{edge-forward} accumulation, in which stage $s$ incorporates segments $0$ to $s$. At each stage, the profile state of the newly added segment is coherently accumulated into the surviving candidates from the previous stage. A matched-filter detection statistic (using boxcar templates) is then evaluated on the currently accumulated profile, and low-significance branches are pruned. Candidates that remain above threshold are propagated to the next stage and refined on a progressively finer parameter grid. Figure~\ref{fig:folding_edge_forward} illustrates the hierarchical merging process in the edge-forward case, while Figure~\ref{fig:pruning_schematic} summarizes the overall pruning logic with a schematic.

The EP algorithm operates on pre-computed profile states $\mathcal{V}(t_{C,j}, \Tseg, \vlambda_{0}) \in \mathbb{V} $ for each segment $\mathcal{T}_j$ and each parameter vector $\vlambda_0\in\mathcal{G}_0$, where $t_{C,j}$ is the segment midpoint and $\mathcal{G}_0$ is the base grid appropriate for the short duration $\Tseg$. These states are generated efficiently using the partial P-FFA described in Section~\ref{sec:ffa_poly}. The number of base segments $M$ is inherited from the partial P-FFA stage. Since the partial P-FFA produces profile states through a binary merging hierarchy, $M$ is typically power of two. In practice, $M$ is determined by the level at which the P-FFA hierarchy is terminated, such that profile states of duration $\Tseg$ remain computationally tractable on the base parameter grid $\mathcal{G}_0$.
We define a \emph{Search Candidate} as the tuple $\mathcal{U} = \{\vlambda, \mathcal{V}, \SNR\}$, containing the current parameter vector, the accumulated profile state, and the detection statistic. Here, $\vlambda$ is defined at the current reference epoch of the candidate, $\mathcal{V}$ denotes the accumulated folded state (weighted profile and weights), and $\SNR$ is the matched-filter S/N.

The algorithm begins by seeding a set of base candidates from the anchor segment $q$. It then advances through stages $s=1,\dots,M-1$. At each stage, the surviving candidates from duration $T_{s-1}$ are extended to duration $T_s$ by coherently combining them with the newly incorporated segment $j=\mathcal{J}(s,q)$. This extension consists of four key operations: refining each surviving parameter node into a set of finer \emph{leaf} points, resolving each leaf into the local frame of the new segment, coherently accumulating the resolved segment profile, and evaluating the computed detection statistic against a stage-dependent pruning threshold $\SNR_{t,s}$. 

A primary constraint in this pipeline is the memory required to store candidate states. We therefore impose a strict deterministic limit, $\mathcal{C}_{\max}$, on the number of candidates propagated between stages. If the number of surviving candidates in the output buffer exceeds $\mathcal{C}_{\max}$, an overload-pruning step is triggered: the detection threshold is dynamically raised to retain only the highest-scoring candidates (e.g., those above the median of the current buffer). This serves as a fail-safe mechanism to guarantee that memory usage remains within fixed bounds. Ideally, $\mathcal{C}_{\max}$ is chosen large enough that overload pruning is rarely (if ever) triggered, but the mechanism ensures robust and predictable behaviour under all conditions. The core operations detailed in Algorithm~\ref{alg:pruning_seq} are:
\begin{itemize}
    \item \func{Seed:} Generates the initial candidate tuple in the buffer $\mathcal{C}_{\mathrm{curr}}$ (denoted as \func{WorldTree}) from the anchor segment $q$. This involves computing scores for the stored fold profile states in $\mathbb{V}[q]$.
    \item \func{Branch:} For each surviving candidate from the previous stage, the parameter vector $\vlambda$ is refined. A local, high-resolution grid $\Lambda_{\mathrm{leaves}}$, bounded by the grid cell of the previous stage, is generated with resolution appropriate for the longer coherent span $T_s$. This step effectively activates the higher-order polynomial terms required for longer integration times.
    \item \func{Validate:} An optional step to filter out non-physical parameter vectors from the refined grid $\Lambda_{\mathrm{leaves}}$.
    \item \func{Resolve:} Projects a leaf parameter $\vlambda_{\mathrm{leaf}} \in \Lambda_{\mathrm{leaves}}$, defined at the accumulator reference epoch into the local epoch of the new segment $j$. It identifies the nearest pre-folded parameter vector on $\mathcal{G}_0$, and computes the required phase-alignment shift $\Delta \phi$.
    \item \func{Shift:} Retrieves the segment profile state $\mathcal{V}_j$ from $\mathbb{V}$, applies the phase shift $\Delta \phi$, and combines it to the accumulated candidate state.
    \item \func{Score:} Computes the detection statistic for the updated profile against the template bank (see equation~\eqref{eq:optimal_snr}). If the resulting $\SNR$ is below the stage-dependent threshold $\SNR_{t,s}$, the candidate is discarded.
    \item \func{Transform:} An optional step to updates the coordinate definition of the surviving leaf parameter $\vlambda_{\mathrm{leaf}}$ to the midpoint of the new duration $T_s$.
    \item \func{PruneOverload:} Enforces the buffer capacity constraint by retaining only the highest-scoring candidates whenever the WorldTree exceeds $\mathcal{C}_{\max}$.
    \item \func{Ascend:} Reconstructs the fold profile of a surviving candidate by re-integrating the pre-computed segment fold states stored in $\mathbb{V}$ along its inferred parameter trajectory and compute its score.
\end{itemize}

\begin{algorithm}[H]
\caption{Extreme Pruning (EP): Sequential Coherent Accumulation and Pruning}
\label{alg:pruning_seq}
\begin{algorithmic}[1]
\Require Number of base segments $M$, base segment duration $\Tseg$
\Require Pre-computed base profile states $\mathbb{V}[j][\vlambda_0]$ for all segments $j=0,\dots,M-1$ and base-grid points $\vlambda_0 \in \mathcal{G}_0$
\Require Stage threshold scheme $\{\SNR_{t,s}\}$ for $s=1,\dots,M-1$
\Require Anchor segment index $q \in \{0,\dots,M-1\}$
\Require Maximum candidate buffer size $\mathcal{C}_{\max}$
\Ensure Final surviving candidate list $\mathcal{B}_{\mathrm{final}}$

\State Allocate $\mathcal{B}_{\mathrm{curr}}$ and $\mathcal{B}_{\mathrm{next}}$ with capacity capacity $\mathcal{C}_{\max}$
\State \textbf{Initialization (Stage 0):}
\State $t_{\mathrm{init}} \gets \left(q+\frac{1}{2}\right)\Tseg$ \Comment{Anchor midpoint}
\State $\mathcal{B}_{\mathrm{curr}} \gets \func{Seed}(\mathbb{V}[q], \mathcal{G}_{0}, t_{\mathrm{init}})$  \Comment{Initialize WorldTree}

\State \textbf{Hierarchical Sequential Merging:}
\For{$s \gets 1$ to $M - 1$} \Comment{Iterate Stages}
    \State $T_{s} \gets (s+1)\Tseg$
    \State $j \gets \mathcal{J}(s, q)$ \Comment{Index of new segment to add}
    \State $t_{\mathrm{seg}} \gets \left(j+\frac{1}{2}\right)\Tseg$
    \State $t_{\mathrm{acc}} \gets \func{Midpoint}(s-1,q)$ \Comment{Current accumulator epoch}
    \State $t_{\mathrm{new}} \gets \func{Midpoint}(s,q)$ \Comment{Epoch after adding segment $j$}
    \State $\SNR_{t,\mathrm{cur}} \gets \SNR_{t,s}$ \Comment{Current effective pruning threshold}
    
    \For{each candidate $\mathcal{U} \in \mathcal{B}_{\mathrm{curr}}$}
        \State $\Lambda_{\mathrm{leaves}} \gets \func{Branch}(\mathcal{U}.\vlambda, t_{\mathrm{acc}}, T_{s})$ \Comment{Refine grid}
        \State $\Lambda_{\mathrm{leaves}} \gets \func{Validate}(\Lambda_{\mathrm{leaves}})$ \Comment{Drop unphysical nodes}
        
        \For{each $\vlambda_{\mathrm{leaf}} \in \Lambda_{\mathrm{leaves}}$}
            \State $(\vlambda_{0}, \Delta\phi) \gets \func{Resolve}(\vlambda_{\mathrm{leaf}}, \mathcal{G}_0, t_{\mathrm{init}}, t_{\mathrm{acc}}, t_{\mathrm{seg}})$
            \State $\mathcal{V}_{\mathrm{seg}} \gets \mathbb{V}[j][\vlambda_{0}]$
            \State $\mathcal{V}_{\mathrm{new}} \gets \mathcal{U}.\mathcal{V} + \func{Shift}(\mathcal{V}_{\mathrm{seg}}, \Delta\phi)$
            \State $\SNR \gets \func{Score}(\mathcal{V}_{\mathrm{new}})$
            
            \If{$\SNR  \ge \SNR_{t,\mathrm{cur}}$}
                \State $\vlambda_{\mathrm{store}} \gets \func{Transform}(\vlambda_{\mathrm{leaf}}, t_{\mathrm{acc}}, t_{\mathrm{new}})$
                \State $\mathcal{U}_{\mathrm{new}} \gets (\vlambda_{\mathrm{store}}, \mathcal{V}_{\mathrm{new}}, \SNR)$
                \If{$|\mathcal{B}_{\mathrm{next}}| < \mathcal{C}_{\max}$}
                    \State Add $\mathcal{U}_{\mathrm{new}}$ to $\mathcal{B}_{\mathrm{next}}$
                \Else
                    \State $\SNR_{t,\mathrm{cur}} \gets \func{PruneOverload}(\mathcal{B}_{\mathrm{next}}, \SNR_{t,\mathrm{cur}})$
                    \If{$\SNR \ge \SNR_{t,\mathrm{cur}}$} \Comment{Re-check}
                         \State Add $\mathcal{U}_{\mathrm{new}}$ to $\mathcal{B}_{\mathrm{next}}$
                    \EndIf
                \EndIf
            \EndIf
        \EndFor
    \EndFor

    \If{$|\mathcal{B}_{\mathrm{next}}| = 0$}
        \State \textbf{break}
    \EndIf
    \State Swap buffers: $\mathcal{B}_{\mathrm{curr}} \leftrightarrow \mathcal{B}_{\mathrm{next}}$
\EndFor
\State \textbf{Finalize:}
\State $\mathcal{B}_{\mathrm{curr}} \gets \func{Ascend}(\mathcal{B}_{\mathrm{curr}}, \mathbb{V}, t_{\mathrm{new}})$  \Comment{Re-integrate folds}
\State \Return $\mathcal{B}_{\mathrm{curr}}$
\end{algorithmic}
\end{algorithm}

The final output is the set of candidates (profile states, parameter vectors, and scores) that survive all pruning stages across the full dataset. The key innovation of this method is the integration of hierarchical grid refinement with adaptive pruning, creating a scalable search framework across increasingly longer observation spans while maintaining computational feasibility.

\subsection{Grid Refinement and Transformation}\label{sec:unpruned_growth}
Before analysing pruning efficiency and elimination strategies, we first establish the baseline complexity governed by the unpruned growth of the EP search tree. Specifically, we consider how the stage-dependent parameter grid $\mathcal{G}_s$ evolves as the coherent duration increases from $T_{s-1}$ to $T_s$. The linear hierarchical approach naturally creates a branching pattern as the parameter grid is progressively refined. Grid refinement is driven by two factors: finer grid spacing in the dimensions already active at stage $s-1$, and additional higher-order polynomial parameters become active once $T_s$ exceeds the corresponding activation scale $T_{c,k}$ (defined in equation~\eqref{eq:threshold_time}).

Let $A_s = \{k : T_s \ge T_{c,k}\}$ denote the set of parameters active at stage $s$. The stage-dependent branching factor is defined as the ratio of grid sizes between adjacent stages, 
\begin{equation}\label{eq:branching_factor_def}
    B(s) \approx \frac{\Ngrid(T_s)}{\Ngrid(T_{s-1})}, \qquad s\ge 1.
\end{equation}

Using the scaling $\Delta f_k(T)\propto T^{-(k+1)}$ for each active dimension, and incorporating an additional discrete expansion factor for the newly activated parameters, we obtain
\begin{align}\label{eq:branching_factor}
B(s) \approx \left(\frac{s+1}{s}\right)^{\kappa_{s-1}}\, \prod_{k=1}^{\kmax} \delta_k(s),
\end{align}
where
\begin{equation}
    \kappa_{s-1} = \sum_{k\in A_{s-1}} (k+1),
\end{equation}
is the cumulative exponent over parameters active at stage $s-1$, and
\begin{equation}
    \delta_k(s) =
    \begin{cases}
        \max\left(1,\left\lceil \dfrac{\Delta f_k^{\mathrm{range}}}{\Delta f_k(T_s)} \right\rceil\right),
        & \text{if } k\in A_s \setminus A_{s-1}, \\[8pt]
        1, & \text{otherwise}.
    \end{cases}
\end{equation}
The first factor captures refinement within the parameter subspace already active at stage $s-1$, while the product over $\delta_k(s)$ accounts for discrete growth when new polynomial dimensions enter the search space. Because refinement in the kinematic parameters $d_k$ is frequency-dependent, the branching factor $B(s)$ represents the average number of new grid regions spawned from each existing region during this transition. 

The total unpruned candidate count after the full linear traversal is the product of these stage-wise branching factors applied to the base-grid size,
\begin{equation}\label{eq:n_total_unpruned}
    N_{\mathrm{grid, EP}}(\Tobs) = N_{\mathrm{grid},0}\prod_{s=1}^{M-1} B(s),
\end{equation}
where $N_{\mathrm{grid},0}=\Ngrid(\Tseg)$. The eventual computational and memory cost of the EP depends on the effectiveness of the thresholding strategy $\SNR_{t,s}$ in controlling this theoretical candidate volume.

\subsubsection{The Reference Frame Dilemma}\label{sec:grid_issues}
The scaling law for the branching factor $B(s)$ in equation~\eqref{eq:branching_factor} assumes ideal geometric scaling, which is strictly valid only for a coordinate grid anchored to a fixed reference epoch. More generally, the spacing required in the $k$-th polynomial coefficient is controlled by the maximum excursion from the chosen reference epoch,
\begin{equation}
    \Delta f_{k,s} \propto L_s^{-(k+1)}, \qquad L_s = \max_{t\in I_s}|t-\tref|,
\end{equation}
where $I_s$ is the time interval accumulated at stage $s$. The computational efficiency of the search therefore depends critically on the choice of the reference epoch $\tref$.

In a linear traversal, the centre of the accumulated interval shifts at every stage. If the reference epoch is fixed at the start of the run ($\tref = t_{C,q}$), the maximum temporal excursion from the reference grows asymmetrically as additional segments are added. This makes the branching pattern (and the required grid density) strongly dependent on the anchor segment $q$. Specifically, for a centrally anchored run ($q=M/2$), the polynomial expansions remain well centred, whereas for an edge-anchored run ($q=0$ or $M-1$) the far boundary of the interval lies at a distance of order $T_s$ from the reference epoch. The resulting over-resolution is a purely coordinate effect: the same physical phase model must be sampled much more densely simply because the expansion is evaluated far from its origin.

This produces a large disparity in branching behaviour across different choices of $q$. The optimal configuration occurs when the reference epoch lies near the midpoint of the accumulated interval, while the worst-case scenario occurs for an edge-anchored run ($q=0$). Stage by stage, the penalty is approximately bounded by a factor of $2^{\kappa_{s-1}}$, corresponding to the doubling of the maximum excursion relative to a centred frame. Although a fixed midpoint frame ($q=M/2$) is the most efficient static choice, it still leaves half of the accumulation stages appreciably off-centre. More importantly, the multi-run pruning strategy introduced in Section~\ref{sec:ep_multi} requires the branching to be independent of the anchor segment. Enforcing a fixed reference frame would therefore require adopting the worst-case edge-anchored ($q=0$) grid density globally to guarantee coverage, leading to intractable over-gridding.

\subsubsection{Moving Grid and Axis Misalignment}\label{sec:tiling_issue}
To ensure the branching factor remains independent of the anchor segment $q$, the reference epoch $\tref$ must track the accumulated interval. We therefore adopt a \emph{moving reference frame} strategy: at stage $s$, the candidate grid is represented in the frame of the current accumulator epoch $t_{C, s-1}$. At the end of the stage, a linear \textsc{Transform} operation shifts all surviving candidates to the midpoint of the newly accumulated interval $t_{C,s}$. This keeps the polynomial domain approximately symmetric within $[-T_s/2, +T_s/2]$, thereby maintaining stable near-optimal grid density throughout the search.

However, while shifting $\tref$ ensures a consistent branching pattern $B(s)$, it alters the orientation and shape of the grid cells with respect to the new coordinate axes. Although the underlying metric is invariant, the validity regions become sheared in the transformed Taylor frame, creating a non-trivial tiling problem for axis-aligned search grids. A parameter vector $\vlambda$ representing a specific phase evolution maps to a local hyper-rectangular validity region (a ``tile'') in the frame of $t_{C,s}$. When transformed to $t_{C,s+1} = t_{C,s} + \Delta t$, the Taylor coefficients mix via the transformation matrix $\mathbf{T}(\Delta t)$ (see Appendix~\ref{app:taylor_basis}). Geometrically, the transformation acts as a shear: a validity region that is orthogonal in the frame of $t_{C,s}$ becomes a sheared hyper-parallelepiped in the frame of $t_{C,s+1}$, even though the underlying physical phase model remains unchanged.

\begin{figure}
\includegraphics[width=\columnwidth]{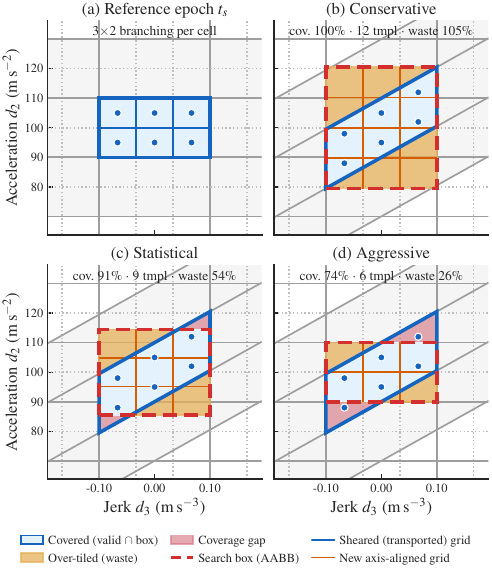}
\caption{Grid tiling problem in the Taylor basis, illustrated as a 2D projection in the acceleration--jerk ($d_2$--$d_3$) plane; the actual search space could be higher-dimensional, so tiling inefficiencies compound with dimension.
\textbf{Panel (a):} At reference epoch $t_s$, an axis-aligned search tile is subdivided into finer orthogonal child cells. 
\textbf{Panel (b)--(d):} Advancing the epoch to $t_s + \Delta t$ introduces off-diagonal coupling in the parameter transformation matrix, shearing the validity region into a parallelogram.
\textbf{Panel (b):} Conservative tiling forms the minimal AABB enclosing the sheared region, guaranteeing full coverage at the cost of a substantial over-tiled volume.
\textbf{Panel (c):} Statistical tiling propagates extents in quadrature, trading a modest corner coverage gap for reduced template count.
\textbf{Panel (d):} Aggressive tiling retains uncoupled diagonal extents only, minimising wasted volume but leaving pronounced coverage gaps near the corners of the sheared tile.}
\label{fig:tiling}
\end{figure}

Consequently, restricting the search to efficient axis-aligned bounding boxes (AABBs) requires a compromise among three sub-optimal strategies for constructing the refined grid $\Lambda_{\mathrm{leaves}}$:
\begin{itemize}
    \item Conservative Tiling (absolute-matrix propagation): Grid extents are propagated using the element-wise absolute value of the transformation matrix,
    \begin{equation}
        e'_a = \sum_b |T_{ab}| \, e_b.
    \end{equation}
    This constructs the smallest AABB guaranteed to enclose the full sheared validity region. It ensures complete parameter-space coverage but introduces substantial geometric redundancy outside the true valid volume (Figure~\ref{fig:tiling}b).

    \item Quadrature Tiling (root-sum-square propagation): Grid extents are propagated component-wise in quadrature,
    \begin{equation}
        e'_a = \sqrt{\sum_b (T_{ab} e_b)^2}.
    \end{equation}
    This effectively approximates the sheared region as an ellipsoid and provides a pragmatic compromise by limiting template growth at the expense of minor sensitivity gaps at the tile corners (Figure~\ref{fig:tiling}c).

    \item Aggressive Tiling (diagonal-only): Only the diagonal elements of $\mathbf{T}$ ($T_{a,a}=1$) are retained, ignoring off-diagonal coupling terms. This produces the most compact inner AABB and minimizes template volume, but leaves significant sensitivity gaps near the true boundaries of the sheared region (Figure~\ref{fig:tiling}d).
\end{itemize}

Because the reference frame is updated at every stage, these alignment mismatches accumulate throughout the EP traversal. The resulting effect is not a uniform loss of sensitivity, but rather progressively larger regions of local under-coverage. These gaps are partially mitigated in practice by natural overlap between neighbouring templates. Although Figure~\ref{fig:tiling} illustrates the effect using a 2D projection for clarity, the redundant volume and under-covered corner regions grow rapidly with parameter-space dimensionality.

\subsubsection{Orthogonal Basis and Stability}\label{sec:ortho_basis}
The tiling problem is further aggravated by the strong covariance inherent in the Taylor monomial basis functions $\{ (t-t_c)^k \}$. Even before any frame translation, the constant-phase-error region in Taylor space is typically highly anisotropic. Subsequent shear transformations $\mathbf{T}(\Delta t)$ therefore act on a parameter cell that is already elongated and poorly aligned with axis-aligned template placement.

To mitigate this covariance, we consider a Chebyshev polynomial representation obtained by mapping the time coordinate onto the normalized interval $x \in [-1, 1]$. Writing the half-span of the current observation window as $h_s=T_s/2$, the line-of-sight distance becomes
\begin{equation}\label{eq:cheby_d}
    d(t)=\sum_{k=0}^{\kmax}\alpha_k\,T_k\!\left(\frac{t-t_c}{h_s}\right),
\end{equation}
where $\{\alpha_k\}$ are the Chebyshev coefficients and $T_k$ are the Chebyshev polynomials of the first kind (see Appendix~\ref{app:chebyshev_basis} for the basis transformations).

The principal advantage of the Chebyshev basis is improved geometric isotropy. Because the Chebyshev polynomials form a near-orthogonal and well-conditioned basis over the normalized interval, the local parameter covariance is significantly reduced compared to the monomial basis, and the constant-mismatch region is substantially less elongated. However, unlike Taylor coefficients which are invariant to domain size, Chebyshev coefficients $\{\alpha_k\}$ are explicitly tied to the domain scale $h_s$ ($\propto h_s^k$). As the coherent duration grows, the coefficients undergo both shear mixing (due to $t_c$ shifting) and volumetric scaling (due to $h_s$ expansion). For conservative and quadrature AABB tiling, this conflates physical refinement with coordinate rescaling, causing the transported bounding boxes to inflate even more rapidly than in the Taylor basis.

\begin{figure}
\includegraphics[width=\columnwidth]{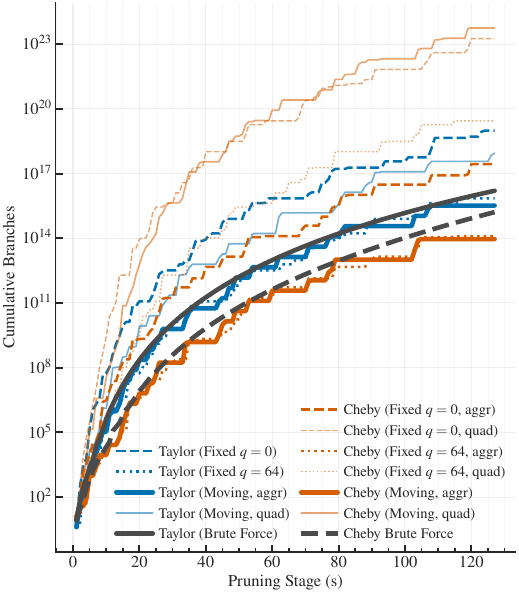}
\caption{Cumulative grid size, expressed as the product of branching factors $B(s)$, for different reference-frame, basis, and tiling strategies. The example assumes a circular-orbit search ($\Porb \geq \Tobs$) over an 18-min observation divided into 128 segments ($\eta = 1.0$, $N_b = 64$). Fixed grid strategies are shown for both the best-case midpoint reference ($q=64$; dotted lines) and the worst-case edge reference ($q=0$; dashed lines) baselines. Moving grid strategies demonstrate the low-overhead scaling of aggressive tiling in both Taylor and Chebyshev bases, contrasted with the much larger growth produced by quadrature tiling.
\label{fig:moving_reference}}
\end{figure}

\subsubsection{Branching Strategy}\label{sec:strategy_recommendation}
An alternative approach to avoiding this dilemma is transporting a non-orthogonal lattice that exactly tracks the coordinate shear, perfectly conserving the local parameter-space volume from stage to stage. In practice, however, the search is constrained not only by the cell volume but also by the maximum allowable phase error, set here by the tolerance $\eta$ phase bins. Under extreme temporal shear (e.g., propagating from $q=0$), an initially compact orthogonal parameter cell is stretched into a thin elongated hyper-parallelepiped. While its parameter-space volume is strictly conserved, the extremal points can lie far from the template centre in phase-mismatch space. To preserve the $\eta$ bound, the sheared cell must be subdivided along its elongated axes. The resulting template count therefore approaches the same scaling as a conservative AABB cover. Consequently, tracking exact geometric shear provides no practical computational advantage over bounding box methods when a rigid $\eta$ constraint is enforced.

Given the fundamental tension between phase-space covariance and strict $\eta$ error bounds, we identify two tractable strategies for high-order hierarchical phase tracking, depending on the acceptable sensitivity tolerances:
\begin{enumerate}
    \item Aggressive Moving Grid (cost-optimized): If the primary objective is bounded and uniform computational cost across all anchor segments $q$, a moving reference frame with aggressive tiling is preferred. This minimizes the branching factor and can be implemented in either the Taylor or Chebyshev basis, with the latter providing a more isotropic starting geometry. The inevitable sensitivity gaps must then be controlled empirically, for example by tightening the search tolerance $\eta$.
    
    \item Quadrature Fixed Grid (sensitivity-optimized): If full signal coverage is required, repeated quadrature tiling in a moving frame incurs prohibitive geometric overhead. In that case, a fixed grid anchored to the accumulation start is the more stable alternative. Although this introduces a substantial penalty (up to factors of $\sim10^3$ in the representative example of Figure~\ref{fig:moving_reference}), it avoids compounding transport distortions across successive stages. In this regime, a Taylor representation is preferable because the Chebyshev coefficients inherit an additional domain-scaling overhead under non-aggressive bounding.
\end{enumerate}

Figure~\ref{fig:moving_reference} compares the cumulative grid growth produced by these basis and reference-frame choices. Aggressive moving-grid strategies achieve the lowest branching cost, whereas quadrature schemes incur a severe overhead through repeated geometric inflation. True conservative schemes are omitted from the main trend as their extreme inflation renders them entirely unfeasible. A fixed Taylor grid anchored at the observation start is costly, but remains a competitive sensitivity-preserving baseline when strict coverage is required. Furthermore, these highly inflated quadrature schemes are unusable in the overall EP framework, as excessive volume expansion increases the risk of actual signal template being pruned from the search tree due to parameter degeneracy in early stages.

Crucially, the Chebyshev basis yields roughly an order-of-magnitude reduction in required templates compared to the pure Taylor basis, driven entirely by the geometric isotropy of orthogonal gridding. For this comparison, we deliberately omitted the Chebyshev coarsening factor of $2^{k-1}$ applied to the Taylor derivatives, described in Section~\ref{sec:search_grid_design}, to isolate the native coordinate geometry gains. Applying the coarsening factor brings the Taylor branching pattern down to a similar scaling as the relevant Chebyshev profiles in Figure~\ref{fig:moving_reference}, further demonstrating its utility as a practical mechanism to mirror orthogonal basis gains.

We conclude that enforcing rigid, axis-aligned step sizes $\Delta d_k$ over a covariant parameter space is the primary geometric limitation of hierarchical polynomial tracking. A complete solution likely requires replacing fixed coordinate spacings with a local metric-based mismatch criterion. Further work is required to quantify the sensitivity loss and compare basis strategies. In this implementation, we adopt the aggressive tiling scheme in the Taylor basis as the operational default due to its strict computational efficiency, leaving the alternative propagation methods available as configurable parameters. In the remainder of this paper, aggressive tiling in the Taylor basis is assumed unless stated otherwise.

\subsection{Phase Alignment and Coherent Accumulation}
As in the P-FFA, coherent accumulation in the EP framework requires precise phase alignment between the accumulated candidate profile and the pre-computed profile from the newly added segment. The EP implementation differs from the symmetric P-FFA merge because the reference epoch of the candidate evolves through successive \func{Transform} operations, while the segment states remain fixed in their original folding frames.

Let $\mathcal{U}_s$ denote a candidate at stage $s$ with parameters $\vlambda_{\mathrm{leaf}}$ defined at the current accumulator epoch $t_{\mathrm{acc}}=t_{C,s-1}$. We combine this with the profile state $\mathcal{V}_j$ from the newly added segment $j$, whose local reference epoch is fixed at the segment midpoint $t_{\mathrm{seg}}=t_{C,j}$. The \textsc{Resolve} step performs two operations: it identifies the appropriate base-grid point in $\mathcal{G}_0$ for the incoming segment, and computes the residual phase shift $\Delta\phi_j$ accounting for both geometric path difference and cumulative phase rotations deferred from previous stages.

We first project the candidate parameters from the accumulator frame to the local segment frame using the transformation operator $\mathbf{T}(\Delta t)$,
\begin{equation}
    \vlambda_{\mathrm{proj}} = \mathbf{T}(t_{\mathrm{seg}} - t_{\mathrm{acc}}) \, \vlambda_{\mathrm{leaf}}.
\end{equation}
An analogous transformation applies in the Chebyshev case (see Appendix~\ref{app:chebyshev_basis}). Because base segments are short, the corresponding grid $\mathcal{G}_0$ remains low-dimensional (e.g., $f_0, d_2$). We therefore resolve $\vlambda_{\mathrm{proj}}$ onto the nearest grid point in $\mathcal{G}_0$ using only the active components at the base timescale, and retrieve the associated segment state $\mathcal{V}_j$.

A coordinate subtlety arises because the \func{Transform} operation updates the parameter coordinates without applying the corresponding phase rotation to the accumulated profile at each stage. For computational efficiency, the accumulated state retains an implicit phase origin at the initial anchor epoch, $t_{\mathrm{init}} = t_{C,q}$, even as the parameter vector is re-expressed at successive accumulator epochs, introducing a ``phase debt''. Consequently, the new segment must be aligned to this fixed phase origin rather than to the current coordinate origin.

The required phase shift is given by the difference between the model phase evaluated at the segment epoch and at the initial anchor epoch:
\begin{equation}
    \Delta\phi_j =
    \Phi_{\vlambda_{\mathrm{leaf}}}(t_{\mathrm{seg}}; t_{\mathrm{acc}})
    -
    \Phi_{\vlambda_{\mathrm{leaf}}}(t_{\mathrm{init}}; t_{\mathrm{acc}})
    \pmod 1,
\end{equation}
where $\Phi_{\vlambda}(t; t_{\mathrm{ref}})$ denotes the rotational phase (in cycles) predicted at time $t$ by model $\vlambda$ referenced to epoch $t_{\mathrm{ref}}$. In explicit form,
\begin{equation}
    \Delta \phi_{j} = f_{0,\mathrm{acc}} \left[ (t_{\mathrm{seg}} - t_{\mathrm{init}}) - \Delta \tau \right] \pmod 1,
\end{equation}
where the differential path delay is written as
\begin{equation}
    \Delta \tau = \frac{1}{c} \left[ d_0(t_{\mathrm{seg}} - t_{\mathrm{acc}}) - d_0(t_{\mathrm{init}} - t_{\mathrm{acc}}) \right].
\end{equation}
This expression represents the phase lag of segment $j$ relative to the fixed reference epoch $t_{\mathrm{init}}$, ensuring strict phase coherence with the accumulated profile.

The phase shift is applied to the retrieved segment profile state $\mathcal{V}_j$, and the coherent accumulation is performed via
\begin{equation}
    \mathcal{V}_{\mathrm{new}}(b) = \mathcal{V}_{\mathrm{acc}}(b) + \func{Shift}(\mathcal{V}_j, \Delta \phi_j)(b).
\end{equation}
The shift operation is implemented either as a cyclic bin rotation in the time domain or as an exact fractional phase ramp in the Fourier domain. This construction ensures that coherence is preserved with respect to the fixed observational data, despite the evolving parameter reference frame.

\subsubsection{Re-integration of Survivors}\label{sec:ascend}
The hierarchical EP algorithm performs coherent accumulation through a sequence of local parameter refinements and pruning operations. A crucial property of this framework is that the objective of the EP is not the preservation of the candidate score but the topological survival of the parameter volume containing the true signal. Although the pruning strategy is designed to ensure this signal volume survives successive threshold cuts, the accumulated fold profile of a surviving candidate is often not the maximum-coherence realization of that trajectory. At each intermediate stage, candidates are bound to discrete parameter cells; hence, mismatches introduced during branching, phase transport, or basis transformations can accumulate into a modest loss of coherent S/N.

To recover this sensitivity loss, EP performs a final \func{Ascend} operation upon completing the tree traversal. Rather than relying on the accumulated profile state stored within the surviving candidate, \func{Ascend} reconstructs the candidate profile directly from the original base-segment fold profiles. Let
\begin{equation}
\mathbb{V} = \left\{\mathcal{V}_{j}(\vlambda)\right\},\qquad j = 0,\dots,M-1,
\end{equation}
denote the collection of pre-computed fold states for all base segments. For a surviving candidate trajectory
\begin{equation}
\Gamma = \{\vlambda_0,\vlambda_1,\dots,\vlambda_{M-1}\},
\end{equation}
the reconstructed coherent profile is evaluated as
\begin{equation}
\mathcal{V}_{\Gamma}=\sum_{j=0}^{M-1}\func{Shift}\!\left(\mathcal{V}_{j}(\vlambda_j),\Delta\phi_j\right),
\label{eq:ascend_reintegration}
\end{equation}
where $\Delta\phi_j$ is the phase correction required to transport the $j$-th segment profile into the common reference frame of the candidate trajectory. The candidate score is then recomputed from the reconstructed profile,
\begin{equation}
\SNR_{\Gamma}=\func{Score}\!\left(\mathcal{V}_{\Gamma}\right),
\label{eq:ascend_score}
\end{equation}
which supersedes the accumulated score from the tree traversal.

This procedure offers two distinct advantages. First, it eliminates the coherent mismatch accumulated through intermediate approximation steps, thereby recovering the full sensitivity associated with the surviving parameter trajectory. Second, it serves as a final validation stage. Spurious trajectories can occasionally survive pruning by establishing temporary local correlations, either by mimicking noise fluctuations or by partially tracking a true signal's phase over localized sub-intervals before deviating. When re-integrated coherently over the entire observation span $\Tobs$, such candidate models fail to maintain phase consistency across all segments and are naturally suppressed. In this capacity, \func{Ascend} plays a role analogous to the final ``folding'' or candidate-refinement stage commonly employed in pulsar search pipelines \citep{Men:2023}.

Because all base fold states are already resident in memory ($\mathbb{V}$), the computational cost of \func{Ascend} scales strictly with the number of surviving candidates and the number of base segments $M$. Consequently, this step is substantially cheaper than re-running a full folding operation on the raw time series data.

A generalized execution strategy may invoke \func{Ascend} periodically during the EP traversal rather than solely at the terminal stage. After a specified number of merge levels, the coherent profile of each surviving candidate can be reconstructed via equation~\eqref{eq:ascend_reintegration}, updating the candidate state with an undegraded profile and score. This intermediate re-integration can reduce mismatch accumulation within deep search trees, providing a performance advantage for high-order polynomial searches or mixed-basis searches where candidates transition between polynomial and orbital parametrizations.

\subsection{Pruning Strategy and Threshold Schemes} \label{sec:pruning_thresholds}
To keep the EP algorithm computationally tractable, candidates inconsistent with the signal hypothesis are discarded as early as possible. The elimination strategy exploits the distinct evolution of the detection statistic $\SNR$ under the signal-present hypothesis ($\mathcal{H}_1$) and the noise-only hypothesis ($\mathcal{H}_0$). After stage $s$, the accumulated integration time is $T_s = (s+1)\Tseg$. Under $\mathcal{H}_1$, for a phase-coherent stable pulsar signal, the expected matched-filter $\SNR$ scales as $\sqrt{T_s}$. Conversely, under $\mathcal{H}_0$, the statistic is governed by the stochastic properties of noise. We therefore define a sequence of stage-dependent thresholds $\{\SNR_{t,s}\}_{s=0}^{M-1}$, such that if a candidate score at stage $s$ satisfies $\SNR < \SNR_{t,s}$, that node and its entire descendant subtree are pruned.

\begin{figure*}
\includegraphics[width=\textwidth]{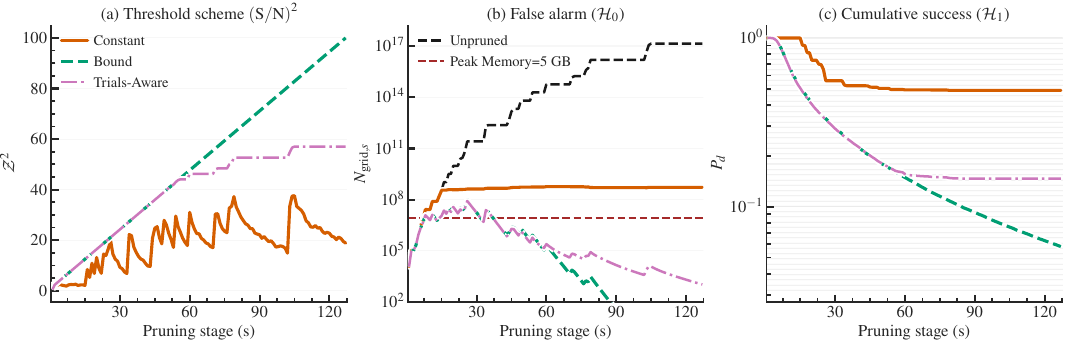}
\caption{Comparison of three heuristic threshold schemes for a circular-orbit search ($\Porb \geq \Tobs$) over an 18-min observation divided into 128 segments ($\eta = 1.0$, $N_b = 64$, target $\mathrm{S/N} = 10$). \textbf{Panel (a):} Stage-wise $(\mathrm{S/N})^{2}$ thresholds. The Bound scheme (green) increases thresholds linearly with stage. The Trials-aware scheme (magenta) adapts to per-stage false-alarm probability in later stages. The Constant Load scheme (red) caps candidate growth. \textbf{Panel (b):} Surviving $\mathcal{H}_0$ candidates $N_{\mathrm{grid}, s}$; the gray dashed curve is the unpruned total. \textbf{Panel (c):} Cumulative $\mathcal{H}_1$ detection probability~$P_d$.
\label{fig:threshold_circular_pre}}
\end{figure*}

The design of the threshold scheme constitutes a constrained optimization problem: minimize total computational work while preserving a target global detection probability. Let $\vlambda^*$ denote the true signal parameters expressed in the stage-$s$ coordinate frame. The cumulative detection probability, $P_d$, is the probability that this signal path survives all pruning decisions,
\begin{equation}
    P_{d}(\{ \SNR_{t,s} \}) = P\left( \bigcap_{s=0}^{M-1} \left\{ \SNR(\vlambda^*_s) \ge \SNR_{t,s} \right\} \;\Bigg|\; \mathcal{H}_1 \right).
    \label{eq:success_prob_pruning}
\end{equation}
The computational work is proportional to the total number of candidate evaluations performed across all stages. Let $N_{\mathrm{grid}, s}$ denote the expected number of candidates that survive to stage $s$. If $\alpha_s = P(\SNR \ge \SNR_{t,s} \mid \mathcal{H}_0)$ is the null-survival probability at stage $s$, then the expected candidate volume obeys
\begin{equation} \label{eq:complexity_rec}
    N_{\mathrm{grid}, s} \approx N_{\mathrm{grid}, s-1} \, B(s) \, \alpha_s(\SNR_{t,s}),
\end{equation}
where $B(s)$ is the stage-dependent branching factor from equation~\eqref{eq:branching_factor}. The total work is then $C_{\mathrm{total}} \propto \sum_{s} N_{\mathrm{grid}, s}$. The challenge lies in the fact that lowering the thresholds $\SNR_{t,s}$ increases $P_d$, but also drives rapid growth in  $C_{\mathrm{total}}$ through equation~\eqref{eq:complexity_rec}. Conversely, raising the thresholds suppresses cost, but increases the risk of eliminating the true signal before it accumulates enough significance to be separate from noise. The objective is therefore to determine the optimal threshold sequence $\{\SNR_{t,s}\}$ that either maximizes $P_d$ for a fixed computational budget $C_{\mathrm{max}}$ or, equivalently, minimizes cost subject to a target detection probability.

\subsubsection{Heuristic Threshold Schemes}\label{sec:heuristic_thresholds}
Before deriving optimal threshold strategies, we examine several heuristic schemes to illustrate the complexity--sensitivity trade-off inherent to the EP algorithm. Understanding these baseline strategies provides intuition for the optimization problem and offers practical baselines for comparison.

We evaluate the performance of a threshold scheme using a Monte Carlo framework that directly simulates the pruning dynamics. The evaluation requires specifying the target signal characteristics; namely, the pulse shape (typically Gaussian), minimum duty cycle, target threshold $\SNR_t$, and the branching pattern $B(s)$. We generate folded profile states for both signal-plus-noise ($\mathcal{H}_1$) and noise-only ($\mathcal{H}_0$) realizations across $M$ segments. Under $\mathcal{H}_1$, the signal power $\SNR_{t}$ is uniformly distributed across segments to simulate time-domain accumulation. At each stage, the accumulated profiles are filtered and scored exactly as in the EP algorithm, ensuring that the score distributions properly account for the look-elsewhere effect arising from maximization over trial pulse widths and phase bins. 

To maintain adequate Monte Carlo statistics at all stages, we replenish pruned states by duplicating surviving candidates (i.e., a simple resampling step), whenever pruning would otherwise reduce the ensemble size below $N_{\mathrm{MC,trials}}$. Because the newly added segment noise is independent of the existing accumulated state, this yields unbiased estimates of the one-step transition statistics used to estimate $\alpha_s$ and $P_d$. However, the replicated trajectories are no longer independent, so this procedure is used to estimate mean survival rates and expected cost rather than the full higher-order population statistics. By tracking the survival rates of these synthetic populations, we map any given threshold scheme $\{\SNR_{t,s}\}$ to the corresponding performance pair $(C_{\mathrm{total}}, P_d)$ at each stage.

We consider three representative heuristic schemes:
\paragraph*{Bound Scheme (Linear Power Growth)}
This scheme assumes that the expected signal power, $\SNR^2$, grows linearly with coherent duration, as introduced in Section~\ref{sec:pruning_overview}. The threshold at stage $s$ is scaled relative to the final desired threshold $\SNR_t$:
\begin{equation}\label{eq:snr_bound_scheme}
    \SNR_{t,s}^{2} = \SNR_{t}^{2}\,\frac{s+1}{M}.
\end{equation}
This produces a smooth, monotonically increasing threshold ramp that tracks the expected signal growth. However, it does not account for the actual tails of the score distribution under either $\mathcal{H}_0$ or $\mathcal{H}_1$, nor does it adapt to the growth of the tree through $B(s)$. It is therefore the most aggressive of the heuristic schemes, as thresholds precisely follow expected signal evolution without margin for statistical fluctuations.

\paragraph*{Trials-Aware Scheme (False Alarm Control)}
This scheme adaptively controls the per-trial false alarm rate by setting the threshold to yield a survival probability of $1/N_{\mathrm{grid}, s}$ under $\mathcal{H}_0$. This scheme is effective in later stages, where trial volume is large, but can be overly aggressive early on, discarding weak but genuine signals before they accumulate sufficient S/N. We therefore implement a hybrid approach:
\begin{equation}\label{eq:snr_trials_scheme}
    \SNR_{t,s} = \min\left( \sqrt{ \SNR_t^2 \cdot \frac{s+1}{M} }, \; \mathcal{Q}^{-1}\left(1 - \frac{1}{N_{\mathrm{grid}, s}}\right) \right).
\end{equation}
where $\mathcal{Q}^{-1}$ is the quantile function of the standard normal distribution. This enforces a conservative threshold following the linear bound in the early stages, where pruning carries high risk, while transitioning to a trials-aware criterion as tree depth and candidate volume increase. The scheme ensures the expected number of null survivors near unity.

\paragraph*{Constant Load Scheme (Bounded Complexity)}
A fundamental practical constraint of hierarchical searches is the memory capacity required to store candidate states $\mathcal{U}$ between stages. To enforce a strict storage footprint, we consider a scheme that explicitly caps the expected candidate load at each stage. The threshold is dynamically adjusted such that the survival probability under $\mathcal{H}_0$ satisfies:
\begin{equation}
    P(\SNR > \SNR_{t,s} \mid \mathcal{H}_0) = 
    \begin{cases}
        1, & \text{if } s \le s_0, \\[4pt]
        \dfrac{1}{B(s)}, & \text{if } s > s_0.
    \end{cases}
\end{equation}
The initial stages ($s \leq s_0$) allow full branching to populate the candidate buffer to capacity. Subsequent stages enforce a per-node survival rate of $\alpha_s \approx 1/B(s)$, ensuring that, on average, only one branch survives per parent node. The specific S/N thresholds required to realize these survival probabilities are obtained empirically from the Monte Carlo framework described above.

Figure~\ref{fig:threshold_circular_pre} compares the three schemes for a circular orbit search with $\Porb^{\min} = \Tobs = 18$\,min. In this example, the unpruned search space grows as $T_{s}^{10}$. While pruning reduces the effective search volume by factors of $10^7$–$10^{8}$, it also lowers the detection probability to the range $0.05$–$0.5$. This illustrates the core EP trade-off: achieving a detection probability of unity requires exploring the entire tree, whereas any pruning inevitably introduces signal loss. Each scheme balances this trade-off differently, as seen in their distinct threshold evolution and candidate survival curves.

The Bound scheme employs the most aggressive elimination by setting thresholds that exactly track the expected signal power growth at every stage. This minimizes the final trial count but incurs a severe sensitivity penalty, yielding $P_d \approx 5\%$ after $M=127$ stages. This loss originates from the stage-to-stage memory inherent in the test statistic. Because the threshold perfectly offsets the deterministic signal growth, candidate survival dictates that the running sum of noise increments must remain strictly non-negative across all $M$ stages. Mathematically, this condition maps to the boundary-crossing problem for a symmetric one-dimensional random walk. According to the Sparre Andersen theorem \citep{SparreAndersen:1954}, the survival probability for a zero-drift random walk over $M$ steps scales asymptotically as $1/\sqrt{\pi M}$. For $M=127$, it explains the empirical Monte Carlo results and shows that the Bound scheme's sensitivity loss is fundamentally governed by random-walk statistics.

The Trials-Aware scheme partially mitigates this sensitivity loss by using more conservative thresholds in later stages, reaching $P_d \approx 15\%$ at comparable total cost. The Constant Load scheme offers a particularly compelling trade-off: by deliberately accepting a controlled reduction in completeness (e.g., $\sim 50\%$ success), it actively controls the branching factor and reduces the computational load by roughly seven orders of magnitude relative to the brute-force baseline. Although tunable according to the available budget to yield higher or lower $P_d$, these heuristic schemes are not strictly optimal. 

This naturally motivates the next question: can one determine the threshold scheme that minimizes computational cost for a specific target $P_d$? In principle, such an optimal scheme should exist. There is a unique path through the parameter tree corresponding to the true signal under $\mathcal{H}_1$, and our goal is to ensure this path survives with the desired probability while minimizing exploration of spurious branches. An optimal scheme would likely distribute the risk of signal loss non-uniformly across stages, allocating pruning budget strategically based on the expected size of the search tree and the statistical separability of $\mathcal{H}_0$ and $\mathcal{H}_1$ at each intermediate stage.

\begin{figure}
\includegraphics[width=\linewidth]{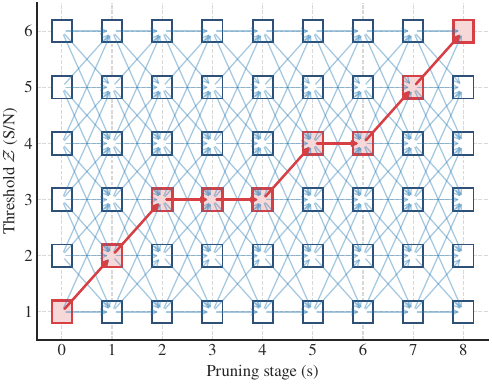}
\caption{Schematic representation of the threshold-optimization problem. Blue curves indicate the family of possible threshold paths through stage-wise S/N space. The objective is to identify the optimal sequence $\{\SNR_{t,s}\}$ (red path) that minimizes computational cost while achieving the desired detection probability. The full search space grows exponentially with the number of stages; eight stages are shown here for illustration.
\label{fig:viterbi_model}}
\end{figure}

\subsubsection{Optimal Thresholds via Dynamic Programming}\label{sec:mc_thresholds}
We formally define the pruning strategy as a sequence of thresholds $\boldsymbol{\gamma} = \{ \SNR_{t,s} \}_{s=0}^{M-1}$. The goal is to find the optimal sequence $\boldsymbol{\gamma}^*$ that minimizes total computational cost $C(\boldsymbol{\gamma})$ subject to a target detection probability constraint $P_d^{\mathrm{target}}$:
\begin{equation}
    \boldsymbol{\gamma}^* = \argmin_{\boldsymbol{\gamma}} C(\boldsymbol{\gamma}) \qquad \text{s.t.} \qquad P_d(\boldsymbol{\gamma}) \geq P_d^\mathrm{target}.
\end{equation}
The optimization landscape is inherently bi-objective and non-linear. Lowering the thresholds increases $P_d$ but drives rapid growth in $C$, whereas raising them controls cost at the expense of a higher probability of prematurely pruning the true path. Finding the optimal balance requires solving for the entire threshold sequence jointly.

Analytically deriving $\boldsymbol{\gamma}^*$ is intractable due to several factors. First, the candidate population at stage $s$ depends on all prior cuts $\{\SNR_{t,s'}\}_{s'<s}$, precluding stage-wise independent optimization. Second, the stage-wise detection statistics $\SNR_s$ are strongly correlated because the coherent score is cumulative. Finally, the branching factor $B(s)$ varies across stages, creating stage-dependent trade-offs.

A brute-force search over all threshold combinations is prohibitive. Discretizing the S/N axis into $D$ levels yields a solution space of size $D^M$. For representative values such as $D\sim100$ and $M=128$, exhaustive enumeration is impossible. Figure~\ref{fig:viterbi_model} illustrates this exponentially growing search space. We therefore recast the problem as a sequential path-finding optimization and solve it with a Viterbi-style dynamic programming framework.

\paragraph*{Viterbi-Style DP with Detection Probability Tracking}
We define a three-dimensional state space with coordinates $(s, z, r)$, where $s$ is the EP stage, $z \in [0, D-1]$ is the discretized S/N threshold index (representing a threshold value $\gamma_z$), and $r\in[0,R-1]$ is the logarithmically binned cumulative detection probability. A state $(s, z, r)$ represents partial threshold sequences ending at stage $s$ with index $z$ and cumulative detection probability in bin $r$.

Unlike the classical Viterbi algorithm \citep{Viterbi:1967}, which tracks only the minimum-cost path to each state, our method requires maintaining $P_d$ as an explicit state dimension because it is a cumulative, path-dependent quantity. By discretizing $P_d$ into bins indexed by $r$, we effectively compute the Pareto frontier of the cost-sensitivity trade-off, keeping multiple hypothesis tracks at each probability level. For each state $(s,z,r)$ we therefore store the minimum accumulated computational cost over all partial threshold sequences reaching that cell:
\begin{equation}\label{eq:dp_cost}
    C^*(s,z,r) = \min_{\boldsymbol{\gamma}\,\to\,(s,z,r)} C(\boldsymbol{\gamma}).
\end{equation}
The computational complexity $C(\boldsymbol{\gamma})$ accumulates additively across stages and can therefore be optimized recursively using dynamic programming \citep{Bellman:1957}. If multiple paths arrive at the same state cell, only the one with the lowest accumulated cost is retained, along with its predecessor pointer for backtracking.

The algorithm proceeds iteratively, stage by stage, following the same structure as the EP search. For each predecessor state $(s-1,z',r')$ and every admissible threshold index $z$ at stage $s$, we evaluate the stage transition using the Monte Carlo framework from Section~\ref{sec:heuristic_thresholds}. We first simulate the folded profile states for the next stage \emph{once} from the surviving candidate of the predecessor state: under $\mathcal{H}_0$ we add only a fresh noise realization, while under $\mathcal{H}_1$ we add the same signal injection plus noise. Applying the candidate thresholds $\gamma_z$ to this shared realization then yields the incremental expected cost, the updated null-survival factor, and the updated cumulative detection probability (which is mapped to the corresponding bin $r$). Evaluating all thresholds on identical noise ensures a fair comparison among competing choices at stage $s$.

At the final stage, the retained states $\{(M-1,z,r)\}$ provide a discrete sampling of the complexity--sensitivity frontier $C(P_d)$: for each attainable detection-probability level, the minimum achievable computational cost. In the multi-pass setting of Section~\ref{sec:ep_multi}, the relevant figure of merit is the cost-to-sensitivity ratio
\begin{equation}\label{eq:cost_ratio}
    L(\boldsymbol{\gamma}) \equiv \frac{C(\boldsymbol{\gamma})}{P_d(\boldsymbol{\gamma})},
\end{equation}
derived in Appendix~\ref{app:cost_ratio}. We select the terminal state $(M-1, z^*, r^*)$ that minimizes this ratio among all states satisfying $P_d(r^*) \ge P_d^{\mathrm{target}}$. Backtracking via the stored predecessor pointers then recovers the optimal threshold sequence $\boldsymbol{\gamma}^*$. The full procedure is summarised in Algorithm~\ref{alg:viterbi2d}.

A key practical constraint in this sequential simulation is the need to duplicate surviving candidate profiles (analogous to resampling in sequential Monte Carlo methods) in order to maintain a constant trial size $N$ as pruning proceeds. This duplication is required to avoid running out of active trials with increasing search depth, but it introduces correlations that slightly increases the variance of the final $P_d$ estimates. In practice, we use a sufficiently large number of trials $N \geq 10^{3}$ to keep this variance acceptably low.

\begin{figure*}
\includegraphics[width=\textwidth]{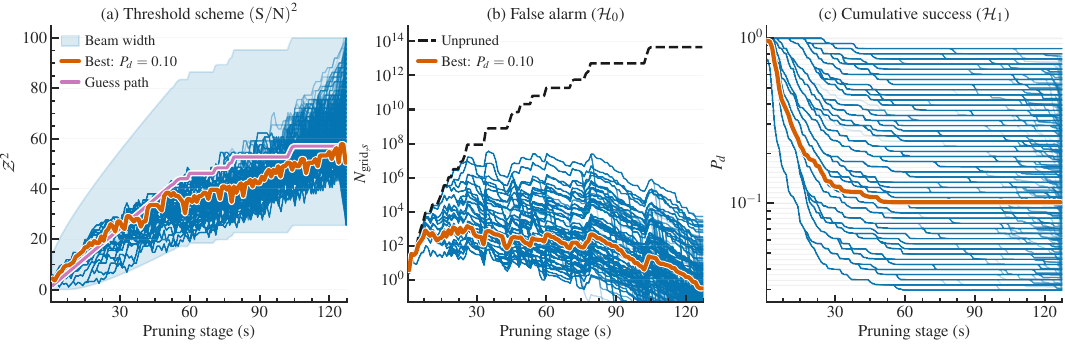}
\caption{Viterbi-optimized threshold schemes for the same configuration as Figure~\ref{fig:threshold_circular_pre}. The Trials-Aware heuristic is used as the initial guess path (orange), and the shaded band in panel (a) shows the beam-search window. Optimal schemes for various target detection probabilities are shown, with the $P_d = 0.1$ optimum highlighted. Note how optimized paths strategically set thresholds to suppress candidate explosion at critical branching stages. \textbf{Panel (a):} Stage-wise $(\mathrm{S/N})^{2}$ thresholds. \textbf{Panel (b):} Surviving $\mathcal{H}_0$ candidates $N_{\mathrm{grid}, s}$ for a single stage-0 node; the gray dashed line denotes the corresponding unpruned total. \textbf{Panel (c):} Cumulative detection probability $P_d$ under $\mathcal{H}_1$; horizontal guides mark the discrete probability levels used in the optimization.
\label{fig:threshold_circular_mc}}
\end{figure*}

This dynamic programming approach reduces the search complexity from $D^M$ to $\mathcal{O}(M D^2 R)$. In practice, we find that $D\sim100$ threshold levels and $R\sim50$ probability bins provide sufficient resolution, making the method tractable even for $M \sim 128$ stages. Not all threshold combinations are equally relevant; extremely high thresholds yield near-zero $P_d$, while very low thresholds produce maximal cost. Optimal paths therefore occupy only a narrow corridor in the full threshold space. To exploit this structure, we employ a \emph{beam-search} strategy centred on a heuristic ``guess path" $\hat{\boldsymbol{\gamma}}$. At each stage $s$, we restrict the search to a window of width $\pm \delta_{\SNR}$ around $\hat{\boldsymbol{\gamma}}_s$, reducing the per-stage search space from $D$ to $B \ll D$ active thresholds, yielding an effective complexity of $\mathcal{O}(M B^2 R)$. For the initial guess path, we use the Trials-Aware heuristic scheme from equation~\eqref{eq:snr_trials_scheme}, which provides a reasonable asymptotic path through the S/N space. The beam width $\delta_{\SNR}$ is held fixed across stages, chosen to comfortably encompass the expected S/N growth of the optimal path about the guess. Figure~\ref{fig:threshold_circular_mc} (panel a) illustrates the beam region (shaded band) together with the optimized paths recovered for different target detection probabilities.

\begin{algorithm}[H]
\caption{Viterbi-style DP for threshold scheme optimization}
\label{alg:viterbi2d}
\begin{algorithmic}[1]

\Require Threshold grid $\{\gamma_z\}_{z=0}^{D-1}$,
         detection-probability bins $\{p_r\}_{r=0}^{R-1}$ (log-spaced),
         branching pattern $\{B_s\}_{s=0}^{M-1}$,
         target $P_d^{\mathrm{target}}$,
         beam half-width $\delta_\gamma$,
         heuristic guide path $\hat{\boldsymbol{\gamma}}$,
         number of Monte-Carlo trials $N$.
\Ensure Optimal threshold sequence $\boldsymbol{\gamma}^*$
\State $\mathcal{V}[s][z][r] \gets (C=\infty) \quad \forall\, s,z,r$

\State \textbf{Stage 0 initialization:}
\State $\mathcal{T}_0, \mathcal{T}_1 \gets \func{Simulate}(\mathcal{P}_{H_0}^{\mathrm{init}}, \mathcal{P}_{H_1}^{\mathrm{init}})$ \Comment{Simulate profiles}
\For{each $z$ in $\mathrm{beam}(\hat\gamma_0,\,\delta_\gamma)$}
    \State $(\alpha_0, \alpha_1) \gets \func{Prune}(\mathcal{T}_0, \mathcal{T}_1, \gamma_z)$
    \State $C \gets B_0$;\enspace $n \gets B_0\,\alpha_0$;\enspace $P_d \gets \alpha_1$;\enspace $r \gets \mathrm{bin}(P_d)$
    \If{$C < \mathcal{V}[0][z][r].C$}
        \State $\mathcal{V}[0][z][r] \gets (C,\; n,\; P_d)$
        \State $\mathcal{P}_{H_0}^{(z,r)}, \mathcal{P}_{H_1}^{(z,r)} \gets \func{Survivors}(\mathcal{T}_0, \mathcal{T}_1, \gamma_z)$
    \EndIf
\EndFor

\State \textbf{Forward pass:}
\For{$s = 1$ to $M-1$}
    \For{each $z'$ in $\mathrm{beam}(\hat\gamma_{s-1},\delta_\gamma)$}  \Comment{previous beam}
        \For{each occupied bin $r'$ at state $(s{-}1,z',r')$}
            \State $C^{(s-1)}, n^{(s-1)}, P_d^{(s-1)} \gets \mathcal{V}[s-1][z'][r']$
            \State $\mathcal{P}_0 \gets \mathcal{P}_{H_0}^{(z',r')}$;\enspace
                   $\mathcal{P}_1 \gets \mathcal{P}_{H_1}^{(z',r')}$
            \State $\mathcal{T}_0, \mathcal{T}_1 \gets \func{Simulate}(\mathcal{P}_0, \mathcal{P}_1)$
            \For{each $z$ in $\mathrm{beam}(\hat\gamma_s,\,\delta_\gamma)$} 
                \State $(\alpha_0, \alpha_1) \gets \func{Prune}(\mathcal{T}_0, \mathcal{T}_1, \gamma_z)$
                \State $C^{s} \gets C^{(s-1)} + n^{(s-1)} \cdot B_s$;\enspace
                       $n \gets n^{(s-1)} \cdot B_s \cdot \alpha_0$
                \State $P_d^{s} \gets P_d^{(s-1)} \cdot \alpha_1$;\enspace
                       $r \gets \mathrm{bin}(P_d)$
                \If{$C^{s} < \mathcal{V}[s][z][r].C$}
                    \State $\mathcal{V}[s][z][r] \gets (C^{s},\; n^{s},\; P_d^{s},\; \gamma_z,\; (z',r'))$
                    \State $\mathcal{P}_{H_0}^{(z,r)}, \mathcal{P}_{H_1}^{(z,r)} \gets \func{Survivors}(\mathcal{T}_0, \mathcal{T}_1, \gamma_z)$
                \EndIf
            \EndFor
        \EndFor
    \EndFor
\EndFor

\State \textbf{Terminal selection:} \Comment{cost-to-sensitivity ratio}
\State $(z^*, r^*) \gets \arg\min_{z,r}\; \mathcal{V}[M-1][z][r].L$
       \enspace s.t.\enspace $P_d(r^*) \ge P_d^{\mathrm{target}}$

\State \textbf{Backtrack:}
\For{$s = M-1$ down to $0$}
    \State $\gamma^*_s \gets \gamma_{z^*}$;\enspace 
           $(z^*, r^*) \gets \mathcal{V}[s][z^*][r^*].\mathrm{ptr}$
\EndFor
\State \Return $\boldsymbol{\gamma}^* = \{\gamma^*_s\}_{s=0}^{M-1}$

\end{algorithmic}
\end{algorithm}

Although the Viterbi optimization is computationally intensive, it is a one-time pre-processing step. For a fixed survey design, the optimal threshold sequence  depends only on the search configuration: polynomial order $\kmax$, $\Tobs$, $\Tseg$, the target threshold $\SNR_t$, and signal template characteristics. A library of optimized schemes can therefore be precomputed for a range of target detection probabilities $P_d^{\mathrm{target}}$ and stored as lookup tables for operational use. During practical searches, the appropriate scheme is selected based on the desired sensitivity-cost trade-off.

\begin{figure*}
\includegraphics[width=\textwidth]{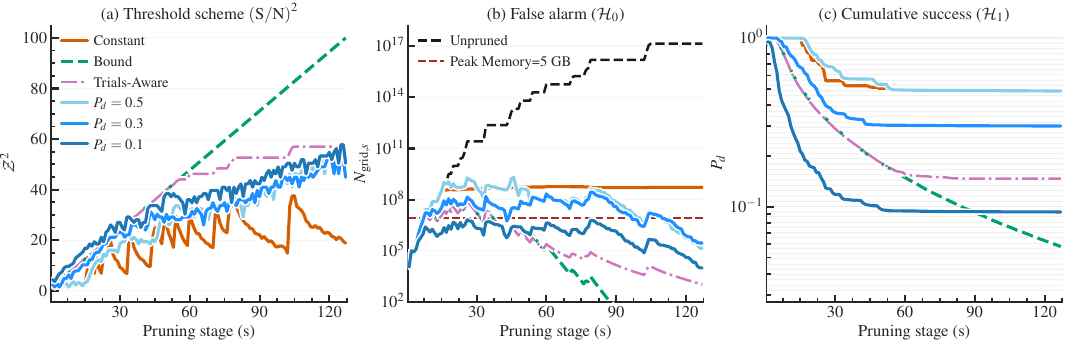}
\caption{Comparison of Viterbi-optimized schemes (solid blue tones) with the heuristic schemes of Figure~\ref{fig:threshold_circular_pre} (same panels). Optimized paths are shown for target detection probabilities $P_d = 0.1$, $0.3$, and $0.5$. Dynamic programming yields an additional order-of-magnitude reduction in pruning cost relative to the best heuristic, and roughly $9$ orders of magnitude relative to the unpruned baseline in panel~(b).
\label{fig:threshold_circular_post}}
\end{figure*}

Figure~\ref{fig:threshold_circular_mc} shows the Viterbi optimization results for the circular orbit search configuration analysed in Section~\ref{sec:heuristic_thresholds}. The dynamic programming search identifies threshold sequences across the cost-sensitivity landscape. The interplay between polynomial grid expansion and exponential pruning becomes evident near stage 30, beyond which pruning dominates and the candidate load declines even for the most conservative scheme ($P_d\sim 0.9$). This behaviour also validates the bounded complexity estimate in Section~\ref{sec:pruning_overview}.

\begin{figure}
\includegraphics[width=\columnwidth]{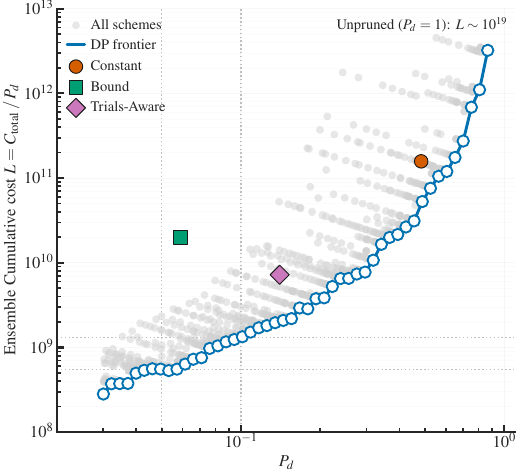}
\caption{Ensemble pruning cost $L = C_{\mathrm{total}}/P_d$ versus cumulative detection probability $P_d$ for the same search configuration as Figure~\ref{fig:threshold_circular_pre}, from the Viterbi-style DP optimization (Section~\ref{sec:mc_thresholds}). Grey points show every complete scheme at the final stage; the blue curve traces the optimized cost-efficiency frontier, connecting the minimum-$L$ scheme in each $P_d$ bin. Symbols mark three heuristic schemes (Constant, Bound, Trials-Aware). Dotted lines indicate $P_d = 0.05$ and~$0.1$. The unpruned brute-force cost at $P_d = 1$ ($L \sim 10^{19}$, off scale) is noted for reference. The cost increases steeply as $P_d \to 1$ but remains modest at low $P_d$; this strong convexity is precisely what the multi-pass ensemble strategy in Section~\ref{sec:ep_multi} exploits. }
\label{fig:cost_frontier}
\end{figure}

Figure~\ref{fig:threshold_circular_post} compares three Viterbi-optimized schemes (targeting $P_d = 0.1$, $0.3$, and $0.5$) against the heuristic baselines from Figure~\ref{fig:threshold_circular_pre}. The optimized schemes demonstrate substantial gains: for a fixed detection probability of $10\%$, the optimised path reduces computational cost by roughly one order of magnitude relative to the best heuristic (Trials-Aware) and more than nine orders of magnitude relative to the unpruned brute-force baseline. Figure~\ref{fig:cost_frontier} displays the cost-efficiency frontier recovered by the DP procedure, showing the minimum cost-to-sensitivity ratio $L = C/P_d$ for each attainable $P_d$. Its steep convex growth as $P_d\to 1$ provides the quantitative basis for the multi-pass strategy developed in the following section.

While the Viterbi framework yields statistically optimal threshold sequences, practical deployment faces two key constraints. First, system memory limits the peak candidate volume sustainable at intermediate stages. Optimal schemes for higher target $P_d$ (e.g., $P_d > 0.5$) delay aggressive pruning until later stages, leading to higher peak candidate volumes. For the search configuration analysed here, each candidate requires roughly $\sim 0.5$--1\,KB of storage (see Section~\ref{sec:ep_complexity}), restricting feasible peak volumes to $N_{\mathrm{grid}, s}\lesssim 10^7$ (roughly 5--10\,GB RAM) on typical machines. This effectively caps practical target detection probabilities to $P_d \lesssim 0.3$ for memory-constrained systems. The second constraint concerns search completeness: a $P_d$ of only 10\% is insufficient for pulsar surveys, where near-unity coherent sensitivity is desired. Nevertheless, the structure of the optimized $P_d$ curves (Figure~\ref{fig:threshold_circular_mc}, panel c) reveals latent pruning redundancy that can be exploited to overcome this limitation.

\subsection{Extreme Pruning via Multi-Pass Search}\label{sec:ep_multi}
The main limitation of single-pass pruning is that the risk of signal loss is heavily concentrated within the earliest stages of the hierarchical search. As demonstrated in Figure~\ref{fig:threshold_circular_post}, the cumulative detection probability $P_d$ for an optimized threshold scheme exhibits a highly non-uniform decay. For the $P_d = 0.1$ scheme, the survival curve displays a characteristic ``staircase'' profile: a steep drop to $\approx 20\%$ by stage 15 (out of 127), a more gradual decline to $\approx 12\%$ by stage 30, and a plateau beyond stage 50 where $P_d$ flattens at its asymptotic value of $10\%$. This structure reveals a critical operational insight: most signal loss occurs during the first $\sim 10\%$ of the accumulation stages. Once the $\mathcal{H}_0$ and $\mathcal{H}_1$ distributions become well-separated, subsequent pruning introduce negligible additional signal loss. This implies that \emph{if a true signal candidate survives the critical high-risk early window, it is statistically guaranteed to be detected}. Consequently, accumulation beyond the plateau stage no longer influences pruning decisions, suggesting a clear strategy to decouple sensitivity from memory complexity.

\begin{figure}
\includegraphics[width=\columnwidth]{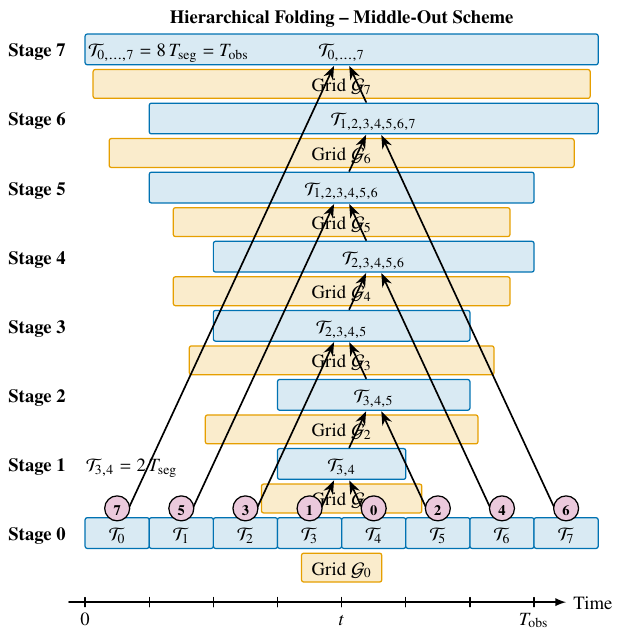}
\caption{Schematic of the middle-out integration path for $M=8$ anchored at $q=3$. Compared to the edge-forward scheme of Fig.~\ref{fig:folding_edge_forward}, the sequence of partial sums is changed, and with it the sequence of pruning decisions.
\label{fig:folding_middle_out}}
\end{figure}

In the EP algorithm, the final detection statistic is a coherent integration over $M$ segments. Although the final score is invariant to the order of segment integration (coherent addition is commutative), the intermediate \emph{pruning process} is inherently non-linear and path-dependent. A candidate trajectory is eliminated if its partial accumulated score falls below a threshold at any intermediate stage $s$. Changing the anchor segment $q$ within a middle-out traversal changes the order in which segments contribute to these partial scores. Different anchor positions therefore produce distinct pruning trajectories through the underlying parameter space, even though the final unpruned coherent statistic for any surviving candidate remains identical. Figure~\ref{fig:folding_middle_out} illustrates the merging process for a representative middle-out traversal with anchor segment $q$.

Consider two search runs with well-separated anchors, for example $q_1 = 0$ and $q_2 = 24$ for $M = 128$ segments. In the first run, the critical early-stage pruning decisions (stages $0$--$15$) are driven by the specific noise realizations present in segments $\{0, \ldots, 15\}$; in the second run, the corresponding stages process segments $\{16, \ldots, 31\}$. Because stochastic noise is temporally uncorrelated across disjoint segments, the event of pruning a true signal becomes a quasi-independent random trial in each run. Rather than performing a single high-$P_d$ search, one can instead execute many inexpensive low-$P_d$ searches with varied anchors, provided their early pruning phases probe independent subsets of the data.

\begin{figure*}
\includegraphics[width=\textwidth]{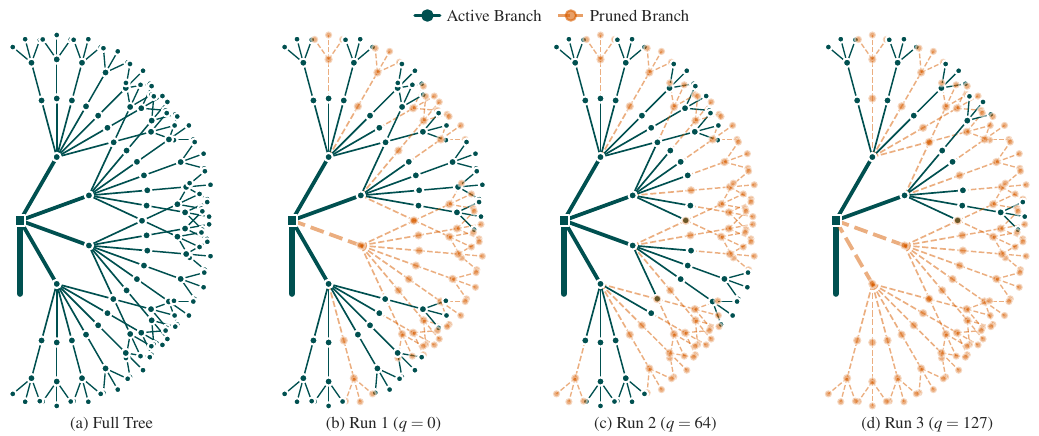}
\caption{Schematic illustration of multi-pass pruning redundancy. The figure shows a subset of the search tree with initial branching pattern $[4, 7, 1, 3]$ (additional branches omitted for clarity). Orange dashed lines indicate pruned nodes; green lines show survivors. \textbf{Panel A:} Full unpruned tree. \textbf{Panels B--D:} Three pruning runs anchored at segments $q = 0$, $64$, and $127$. Each run traverses a distinct path through parameter space during critical early stages due to different noise realizations in the integrated segments. A signal pruned early in Run 1 (e.g., due to unfavourable noise in segment 0) may survive in Run 2 (starting with segment 64), eventually converging to the correct terminal node in one of the runs. This structural redundancy allows us to recover high sensitivity from an ensemble of lossy individual runs.}
\label{fig:pruning_runs_tree}
\end{figure*}

We exploit this statistical leverage by executing an ensemble of $\nrun$ pruning passes, each anchored at a unique segment $q_i$ separated by a uniform stride $\Delta q = M / \nrun$. This geometric arrangement ensures that the high-risk early stages of each pass utilize maximally disjoint subsets of the time series. Assuming statistically independent runs, valid when the stride length exceeds the correlation length of the detection probability decay, each pass can be treated as an independent Bernoulli trial. The probability that a true signal survives in \emph{at least one} of the $\nrun$ passes is then governed by the binomial distribution:
\begin{equation}\label{eq:binomial_prob}
    P_{\mathrm{ensemble}}(\nrun, P_d) = 1 - (1 - P_d)^{\nrun}.
\end{equation}
Figure~\ref{fig:pruning_runs_tree} provides a schematic illustration of how multiple pruning passes can recover a signal missed in any individual run.

The computational advantage of this strategy follows from the strongly non-linear (convex) relationship between per-pass detection probability and search cost, as mapped in Figure~\ref{fig:cost_frontier}. Because $C(P_d)$ rises rapidly as $P_d \to 1$, the collective cost of $\nrun$ aggressively pruned searches is orders of magnitude lower than a single search achieving the same ensemble sensitivity:
\begin{equation}
    \nrun\,C(P_d) \ll C\!\left[1-(1-P_d)^{\nrun}\right].
\end{equation}
For instance, achieving $P_{\mathrm{ensemble}}\approx0.81$ via a single pass requires setting $P_d = 0.81$, incurring $C_{\mathrm{total}}\sim 10^{12}$ cumulative enumerations with peak memory demands far beyond practical limits. In contrast, an aggressive scheme with $P_d = 0.1$ requires $C_{\mathrm{total}}\sim 10^8$ per run. Executing an ensemble of $\nrun = 16$ independent passes costs only $1.6 \times 10^{9}$ candidate evaluations in total while achieving $P_{\mathrm{ensemble}} \approx 0.81$ via equation~\eqref{eq:binomial_prob}. This represents a three-order-of-magnitude cost reduction relative to the single-pass alternative for equivalent sensitivity while remaining within practical memory limits. This massive leverage, trading a linear increase in the number of passes for an exponential reduction in per-pass search space, is the defining characteristic of \emph{Extreme Pruning}. As shown in Appendix~\ref{app:cost_ratio}, minimizing the multi-pass complexity is mathematically equivalent to minimizing the single-pass cost-to-sensitivity ratio $L(\boldsymbol{\gamma})$, which justifies the objective function used in Section~\ref{sec:mc_thresholds}. Overall, an EP scheme with $P_d=0.1$ achieves a nine-order-of-magnitude reduction in search cost relative to the unpruned baseline.

The statistical independence assumption in equation~\eqref{eq:binomial_prob} represents an idealized limit that gradually breaks down as $\nrun$ increases. Two mechanisms drive this behaviour. First, \emph{segment overlap}: as runs progress to later stages, the sets of integrated segments inevitably overlap, making their pruning decisions correlated. Second, \emph{non-uniform risk distribution}:  once the early high-risk window is exhausted, subsequent runs increasingly reuse the same low-risk data spanning the late-stage plateau (see Figure~\ref{fig:threshold_circular_post}, panel c), yielding diminishing returns in independent statistical information. Although the per-pass probability $P_d$ could be reduced further and compensated by increasing $\nrun$ up to the maximum possible value, $\nrun^{\max}=M$, increasing inter-run correlations cause the ensemble sensitivity to saturate rapidly, so the optimal ensemble size is the point at which the marginal gain in sensitivity no longer justifies the linear increase in computational cost.

\begin{figure*}
\includegraphics[width=\textwidth]{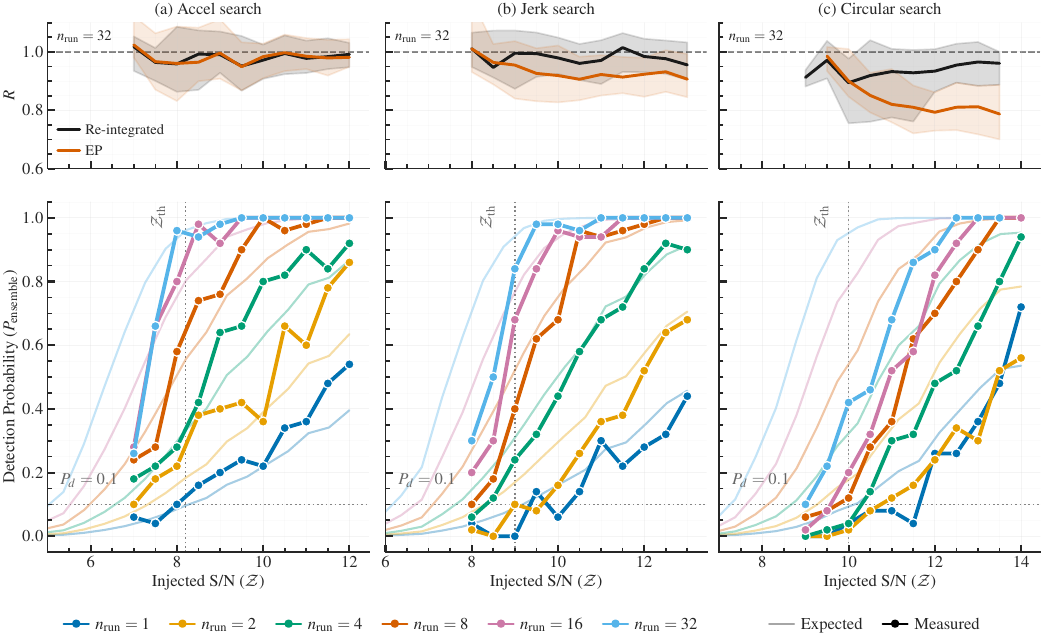}
\caption{Comprehensive validation of the EP framework ensembles for constant-acceleration (left), constant-jerk (centre), and full circular-orbit (right) searches as a function of injected signal significance $\mathcal{Z}$.
\textbf{Bottom panels:} Measured ensemble detection probability for $\nrun=\{1,\dots,32\}$ pruning passes, obtained from 50 independent signal injection per configuration. Faint curves show the prediction of the independent-trial model in equation~\eqref{eq:binomial_prob}, while the dotted vertical line marks the target design threshold $\mathcal{Z}_{\rm th}$ corresponding to the optimized per-pass scheme ($P_d=0.1$). The close agreement for the constant-acceleration and constant-jerk searches confirms that well-separated anchor positions behave as statistically quasi-independent pruning trials. For the higher-dimensional circular-orbit search, modest deviations appear only near threshold, reflecting cumulative discretization losses and inter-run correlations at the current configuration ($\eta = 1.0$, $N_b = 64$).
\textbf{Top panels:} Recovery fidelity for the representative ensemble size $n_{\rm run}=32$, quantified by the recovered significance ratio $R = (\mathcal{Z}_{\mathrm{detected}}/\mathcal{Z}_{\mathrm{injected}})^2$. Solid curves shows the median recovered significance over successful trials, with shaded regions indicating the $16^{\rm th}$--$84^{\rm th}$ percentile interval. Two recovery paths are compared: the score tracked at the end of the ensemble pass (EP; orange) and the value obtained after the final \func{Ascend} re-integration (black). While the EP score exhibits significant degradation due to pruning losses, re-integration largely restores the signal, confirming that the true parameter trajectory survives the pruning hierarchy.
\label{fig:pruning_probs}}
\end{figure*}

To empirically validate the independent-trial approximation, we conduct signal-injection experiments across three representative search regimes: constant acceleration, constant jerk, and full Keplerian circular-orbit searches. Figure~\ref{fig:pruning_probs} summarizes the validation of the EP framework along two complementary dimensions. The lower panels show the empirical ensemble detection probability as a function of the injected signal significance $\mathcal{Z}$ for ensemble sizes $\nrun \in [1, 32]$, directly verifying the independent-trial binomial model of equation~\eqref{eq:binomial_prob}. The upper panels quantify coherent signal recovery, measuring the fraction of the ideal significance retained after hierarchical traversal and subsequently restored via the terminal \textsc{Ascend} re-integration. Together, these measurements characterize both the statistical sensitivity and coherent reconstruction fidelity of the algorithm.

The simulated search space assumes an observation span $\Tobs = 18$\,min, a maximum companion mass $m_{c,\max}=10\,M_\odot$, a minimum pulsar mass $m_{p,\min}=1.2\,M_\odot$, folding resolution $N_b = 64$, phase tolerance $\eta = 1.0$, and an intrinsic pulsar spin period of $7$\,ms. To provide sufficient grid coverage, the minimum orbital period is set to $\Porb^{\min} = 10\Tobs$, $5\Tobs$, and $\Tobs$ for the constant-acceleration, constant-jerk, and full circular-orbit searches, respectively. These limits correspond to orbital coverages of approximately $10\%$, $20\%$, and $100\%$ (see Section~\ref{subsec:circ_orbit_coverage}). Each data point represents the empirical success rate from 50 independent randomized signal injections.

The lower panels of Figure~\ref{fig:pruning_probs} confirm the central prediction of the multi-pass framework. For both the constant-acceleration and constant-jerk searches, the measured recovery probability closely matches the independent-trial prediction of equation~\eqref{eq:binomial_prob} over the full range of ensemble sizes up to $\nrun = 32$. This agreement demonstrates that, for well-separated anchor positions, the dominant early-stage pruning decisions behave as statistically quasi-independent trials, with any residual correlations too weak to measurably affect the ensemble detection probability.

The dynamic thresholding scheme is optimized to achieve $P_d=0.1$ for signals near the statistical noise floor, corresponding to target detection thresholds in the range $\SNR_t\simeq8.2$--$10$. This threshold range is determined by the total enumeration volume and entropy of the search space, scaled to bound the global false-alarm expectation at unity. Accordingly, the recovery curves exhibit the expected sharp transition near the target threshold before rapidly saturating toward unity. The threshold itself is not fundamental to the algorithm. By selecting a different operating point, the EP framework can be tuned for higher target significances, trading additional pruning and lower computational cost for a corresponding rightward shift of the recovery curves while preserving their overall shape.

For the aggressive per-pass scheme adopted throughout this work ($P_d=0.1$), an ensemble of $\nrun=32$ passes achieves $P_{\mathrm{ensemble}}>0.95$ above the target threshold $\SNR_t$. Specifically, signals with injected significance $\SNR\gtrsim9$ for the constant-acceleration search and $\SNR\gtrsim10$ for the constant-jerk search are recovered with nearly $100\%$ probability. Thus, although each pruning pass intentionally discards approximately $90\%$ of detectable signals, the ensemble recovers essentially the full sensitivity while retaining the substantial computational savings of aggressive pruning.

Statistical survival, however, is only one measure of the algorithm's performance; surviving trajectories must also retain their full coherent signal strength. Since EP performs repeated coordinate transformations, discrete grid branching, and hierarchical accumulation, residual mismatches can reduce the recovered significance even when the true candidate survives the pruning cuts. The upper panels of Figure~\ref{fig:pruning_probs} tracks the recovered significance ratio, 
\begin{equation}
R=\left(\frac{\mathcal{Z}_{\rm detected}} {\mathcal{Z}_{\rm injected}}\right)^2,
\end{equation} 
evaluated for the representative ensemble size $\nrun=32$. Two recovery metrics are compared: the significance carried by the surviving candidate, and the significance obtained after applying the final \func{Ascend} re-integration described in Section~\ref{sec:ascend}. While the hierarchical EP score exhibits moderate degradation from accumulated mismatch, \func{Ascend} reconstructs the candidate directly from the stored base fold profiles, restoring nearly the full coherent significance. Across the polynomial searches, the re-integrated significance consistently exceeds $95\%$ of the injected value, confirming that the correct parameter trajectory is preserved throughout the pruning process.

The full circular-orbit search provides the most stringent test of the framework. Unlike the lower-dimensional searches, the empirical detection probability exhibits a rightward shift relative to the idealized independent-trial prediction near the nominal threshold $\SNR_t=10$. The corresponding recovery diagnostics identify the origin of this discrepancy. Although \func{Ascend} substantially restores the coherent signal for surviving candidates, a fraction of true signals are eliminated before the final reconstruction stage owing to accumulated discretization effects arising from finite phase tolerance ($\eta$), residual phase transport errors, and higher-dimensional tiling losses. Together, these effects introduce an effective threshold penalty, shifting the completeness curve slightly above its nominal design value.

Importantly, this discrepancy is confined to the immediate threshold region and does not represent a breakdown of the EP framework. As the injected significance increases modestly above the formal threshold, the recovery probability converges to unity. For the circular-orbit search presented here, a threshold scheme formally optimized for $\SNR_t=10$ achieves complete recovery for signals with $\SNR\gtrsim12$. The remaining near-threshold sensitivity loss therefore reflects implementation-level discretization effects rather than a fundamental limitation of the pruning strategy, providing a clear target for future improvements. Even in its present implementation, the ability to execute a fully coherent search over the complete circular-orbit parameter space while maintaining near-unity recovery for astrophysically relevant signals represents a substantial advancement over previously intractable search pipelines.

\subsection{Computational Complexity}\label{sec:ep_complexity}
The computational cost of the EP algorithm is determined by the total number of candidate evaluations across the $M-1$ hierarchical accumulation stages. The baseline complexity of the unpruned linear traversal, in which every candidate generated by the stage-wise grid refinement (Section~\ref{sec:unpruned_growth}) is propagated to completion without thresholding, is
\begin{equation}\label{eq:c_ep_unpruned_sum}
    C_{\mathrm{EP,unpruned}} = \sum_{s=1}^{M-1} N_{\mathrm{grid}}(T_s) \, \mathcal{O}\!\bigl(N_b(1+N_w)\bigr),
\end{equation}
plus the lower-order initialization cost of generating the base segment states with the partial P-FFA (see equation~\eqref{eq:ffa_cost_cpu}). Here, the operational factor $N_b(1+N_w)$ accounts for phase shifting, profile accumulation, and matched filtering over $N_w$ boxcar widths.

Assuming the active parameter space has stabilized to its maximum dimensionality (i.e., $\kappa_{s-1} \approx \kappa$ and the discrete expansion factors $\delta_k(s) = 1$), approximating the summation in equation~\eqref{eq:c_ep_unpruned_sum} yields
\begin{equation}\label{eq:c_ep_unpruned_final}
    C_{\mathrm{EP,unpruned}} \approx \frac{M}{\kappa + 1} N_{\mathrm{grid}}(\Tobs) \, \mathcal{O}\!\bigl(N_b(1+N_w)\bigr).
\end{equation}
Even this unpruned baseline benefits from substantial data reuse over brute-force coherent folding. Whereas brute-force folding revisits the full time series of $N_s$ samples for every trial template, EP updates candidates using only pre-computed segment-level profiles of length $N_b$. Ignoring the matched-filter scoring common to both approaches, this yields a constant-factor speed-up of order $\sim N_s(\kappa + 1)/(M N_b)$. For representative configurations ($N_s=2^{26}$ samples, $M=128$ segments, $N_b=128$ bins and $\kappa=10$), this reuse alone provides a gain of order $\sim 10^4$ over repeated time-series folding.

The actual cost of EP is lower by orders of magnitude, determined entirely by the efficacy of the chosen threshold scheme in suppressing the stage-wise candidate counts. Replacing the unpruned grid size in equation~\eqref{eq:c_ep_unpruned_sum} with the surviving candidate count $N_{\mathrm{grid},s}$ and accounting for the multi-pass strategy of Section~\ref{sec:ep_multi} gives
\begin{equation}
    C_{\mathrm{EP,total}} = \nrun \, \sum_{s=1}^{M-1} N_{\mathrm{grid}, s} \, \mathcal{O}\!\bigl(N_b(1+N_w)\bigr).
\end{equation}
Because the Viterbi-optimized threshold schemes of Section~\ref{sec:mc_thresholds} enforce aggressive early eviction, $N_{\mathrm{grid}, s} \ll N_{\mathrm{grid}}(T_s)$ throughout most of the traversal. For the representative configurations considered here, pruning reduces the total processing cost by factors of $\sim10^{9}$ relative to the unpruned baseline while maintaining an aggregate detection probability of $\sim80$--$95\%$ near the target threshold.

Memory usage is managed with a \emph{ring buffer} architecture that stores only the active candidates for the current and next stages. After each batch, surviving candidates (those exceeding $\SNR_{t,s}$) spawn child nodes, and the parent nodes are immediately overwritten. Under the explicit capacity limit \(\mathcal{C}_{\max}\), the peak memory footprint is bounded by
\begin{equation}
    M_{\mathrm{EP,peak}} \lesssim \mathcal{C}_{\max}\,\mathrm{sizeof}(\mathcal{U}),
\end{equation}
where \(\mathcal{U}\) is the full candidate state. The dominant contribution is the accumulated profile state, requiring $2 N_b$ floating-point values ($\approx 0.5$--1\,KB for the configurations considered here). Consequently, the resident memory scales linearly with the number of active candidates; for example, $\mathcal{C}_{\max}=10^7$ corresponds to a footprint of $\sim5$\,GB. The threshold scheme must therefore be ideally chosen such that $\max_s N_{\mathrm{grid},s}<\mathcal{C}_{\max}$, as discussed in Section~\ref{sec:mc_thresholds}. In practice, hardware memory places a stringent upper bound on the resident search tree, limiting feasible single-pass targets to roughly $P_d\lesssim0.3$. This also motivates the multi-pass ensemble strategy, which trades a linear increase in runtime for a substantially smaller per-pass resident tree.

\subsection{RFI Handling and Dynamic Range}\label{sec:ep_rfi}
The efficiency and theoretical complexity limits of the EP algorithm rely on the statistical assumption that the input time-series is approximately stationary. The threshold scheme $\SNR_{t,s}$ is calibrated to suppress Gaussian noise fluctuations while retaining faint signals near the detection limit. However, non-Gaussian outliers, such as bright Radio Frequency Interference (RFI) or exceptionally strong pulsar signals, violates this assumption and create a dynamic-range problem.

If a signal or RFI instance lies far above the nominal threshold ($\SNR \gg \SNR_{t,s}$), the $H_1$ hypothesis is satisfied not only by the true parameter vector but also by a large volume of adjacent parameters and their harmonic aliases. In that regime, the realized candidate load no longer follows the nominal null-survival factor $\alpha_s$. Instead, many branches remain populated simultaneously and the stage occupancy $B(s)$ can approach the unpruned branching limit, rapidly exhausting the available buffer capacity during the early stages of traversal.

The primary defence against RFI contamination is robust time-domain pre-processing (e.g., zero-DM filtering and frequency masking), which is standard in pulsar search pipelines \citep[e.g.,][]{Ransom:2002, Morello:2020}. To ensure algorithmic stability against residual contamination, EP also incorporates several defensive mechanisms. First, the \func{PruneOverload} function serves as an absolute fail-safe by enforcing a strict buffer capacity ($\mathcal{C}_{\max}$). If the number of surviving candidates exceeds $\mathcal{C}_{\max}$ because of a bright signal or RFI, the algorithm dynamically raises the detection threshold for that stage, effectively truncating the lower percentiles (e.g., the median) of the score distribution. Although this adaptive culling locally blinds the search to faint signals, it guarantees deterministic memory usage while naturally prioritizing the brightest sources.

Second, searches targeting environments containing multiple bright sources, such as globular clusters, are susceptible to candidate-tree explosion at can mask fainter signals. In such cases, we employ a two-pass exclusion strategy. An initial, low-cost coarse P-FFA search identifies and parametrizes the dominant pulsars. These detections are converted into exclusion windows in parameter space (e.g., frequency and, where appropriate, local orbital derivatives). The resulting \emph{pulsar mask} is applied at two points in EP: during \func{Seed}, seed states whose base-grid parameters fall inside masked windows are omitted; during \func{Validate}, leaf candidates mapping to masked regions are rejected before entering the next-stage buffer. This prevents known bright sources from monopolizing the candidate tree while leaving the remainder of the search space unchanged.
 
Finally, we implement an \emph{early harvesting} protocol to prevent candidate saturation by previously unknown bright sources. Rather than allowing a high-S/N candidate to spawn thousands of degenerate branches, we define a stage-dependent upper significance threshold (e.g., $\SNR \geq 10$). If a candidate exceeds this threshold at an intermediate stage (e.g., $s=20$), it is immediately recorded as a high-confidence detection. Its parameter state is serialized to disk for downstream vetting, and its coordinate domain is appended to the pulsar mask for the remainder of the search. This prevents a single dominant source from consuming most of the available branching budget. Because this procedure deliberately truncates refinement of a bright branch, it should be used conservatively and only when the candidate lies well above the survey detection threshold.

More generally, the hierarchical structure of EP provides diagnostics unavailable in a single-shot fold. Genuine astrophysical signals should persist coherently across stages and anchor choices in the multi-pass search, whereas impulsive or poorly localized RFI tends to generate broad, unstable, or harmonically repetitive structures in the candidate tree. Incorporating such stage-persistence tests into the pruning logic is a natural extension of the present framework.

\section{Circular Orbit Searches}\label{sec:pruning_circular}
Polynomial-based searches truncated at constant acceleration ($\kmax=1$) or constant jerk ($\kmax=2$) are the standard approach for detecting pulsars in binary systems \citep{Johnston:1991, Andersen:2018}. However, these methods suffer a substantial loss in sensitivity because a finite-order polynomial remains coherent over only a restricted fraction of the orbit. The computational efficiency of the EP algorithm instead enables fully coherent searches over circular orbit, maintaining phase coherence over much longer time spans, in practice, up to an entire orbital period, thereby dramatically expanding the accessible parameter space.

\subsection{Phase Model for Circular Orbits}
\begin{figure}
\includegraphics[width=\linewidth]{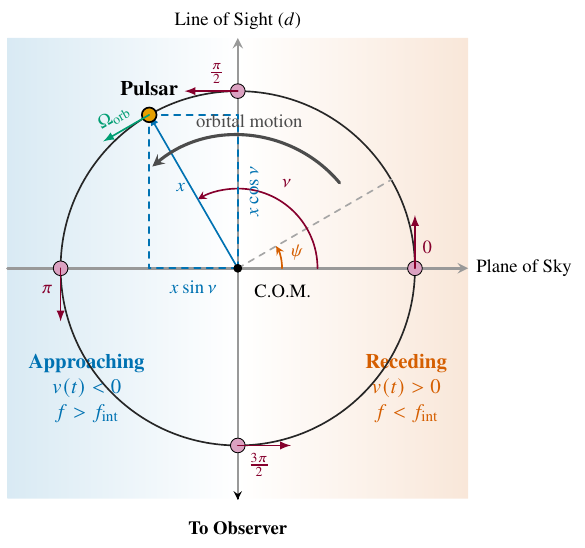}
\caption{Schematic of a pulsar in a circular binary orbit projected onto the plane of the sky. The blue-shaded region corresponds to orbital phases where the pulsar approaches the observer (apparent frequency increase); the red-shaded region corresponds to receding motion (apparent frequency decrease). The projected pulsar position is marked together with the Cartesian components of the projected semi-major axis, $x\cos\nu$ and $x\sin\nu$.}
\label{fig:circular_orbit_schematic}
\end{figure}

For a pulsar in a non-relativistic circular Keplerian orbit, the line-of-sight displacement is
\begin{equation}\label{eq:disp_circular}
d(t) = \bar{d} + a \sin(i) \, \sin\!\left(\Omegaorb t + \psi\right),
\end{equation}
where $\bar{d}$ is the distance to the binary system barycentre, $a$ is the pulsar semi-major axis about the barycentre, $i$ is the orbital inclination relative to the plane of the sky, $\Omegaorb = 2\pi/\Porb$ is the orbital angular frequency for orbital period $\Porb$, and $\psi$ is the orbital phase at $t=0$. We define the projected semi-major axis in light-seconds as $x \equiv a \sin i / c$ and the instantaneous orbital phase as $\nu \equiv \Omegaorb t + \psi$. Figure~\ref{fig:circular_orbit_schematic} illustrates the orbital geometry and these quantities.

Kepler's third law gives
\begin{equation}\label{eq:xorb}
x = \frac{\sin i}{c} \frac{G^{1/3} m_c}{(m_p + m_c)^{2/3}} \Omegaorb^{-2/3},
\end{equation}
where $m_p$ and $m_c$ are the pulsar and companion masses, respectively. Substituting equation~\eqref{eq:disp_circular} into the generic phase model in equation~\eqref{eq:phase_generic} yields the circular-orbit phase model $\Phi(t; \Lc)$. The parameter vector $\Lc = \{f_{\mathrm{int}}, x, \Omegaorb, \psi\}$ defines a four-dimensional search space for circular binaries \citep{Allen:2013}. A fully coherent search over this space requires a prohibitively dense template bank, motivating approximate or partially incoherent search strategies \citep{Knispel:2013, Balakrishnan:2022}.

\subsubsection{Orbital Coverage Limitations from Polynomial Truncation}\label{subsec:circ_orbit_coverage}
The validity of a polynomial phase model depends on both the coverage fraction $\Tobs/\Porb$ and the orbital phase at which the observation is centred. For a truncation at order $\kmax$, coherence is lost once the first neglected term becomes significant. For circular binaries, this implies that the coverage of a constant-acceleration search is inherently phase-dependent.

This distinction has significant consequences for survey completeness. The widely quoted rule-of-thumb that constant-acceleration searches ($\kmax=1$) remain valid over $\sim\!10\%$ of an orbit is not a generic limit: it is a \emph{best-case} value, achieved only at discrete orbital phases where the leading neglected term (the jerk) vanishes at the observation midpoint \citep{Johnston:1991, Ransom:2003}. At the worst-case phase, where the jerk is maximal, the coherent coverage shrinks to $\lesssim 4\%$ for the same system and S/N tolerance (see Appendix~\ref{app:circular_orbit_poly_coverage}). Adopting the $\sim\!10\%$ figure as a universal proxy therefore systematically overestimates the true orbital coverage.

For the most conservative representative configuration considered in Appendix~\ref{app:circular_orbit_poly_coverage}, extending the polynomial model to constant snap ($\kmax=3$) improves the coherent orbital coverage to approximately
\begin{equation}
\frac{\Tobs}{\Porb} \lesssim
\begin{cases}
0.17 & \text{(worst case)},\\
0.26 & \text{(best case)}.
\end{cases}
\end{equation}
More importantly, the snap-order model is the lowest-order polynomial model that contains sufficient information to recover the circular-orbit parameters uniquely. The kinematic derivatives can be inverted to obtain the physical parameters \citep{Joshi:1997}:
\begin{align}\label{eq:nu_recover_main}
    \Omegaorb &= \sqrt{-\frac{d_4}{d_2}}, \\ \nonumber
    \nu &= \arctan\bigbracket{\frac{d_2}{d_3}\sqrt{-\frac{d_4}{d_2}}}, \\ \nonumber
    x &= \frac{d_2^2}{d_4\,c\sin(\nu)},
\end{align}
where $\nu = \Omegaorb t + \psi$ is the orbital phase at the reference epoch (derived in Appendix~\ref{app:circular_orbit}).

Because higher-order derivatives for circular orbits are not independent, once $\{d_2, d_3, d_4\}$ are known, all subsequent derivatives follow from the recurrence relation derived in Appendix~\ref{app:circular_orbit}. This allows us to implicitly account for arbitrarily high-order terms without increasing the search dimensionality beyond the four parameters of a constant-snap search. Consequently, phase coherence and sensitivity can be maintained for integration times extending significantly beyond the $\sim\!26\%$ limit, approaching or exceeding $\Porb$. This motivates a coherent circular-orbit search strategy, in which the EP algorithm explores the $\{f_0,d_2,d_3,d_4\}$ space while exact circular propagation supplies the higher-order phase evolution.

\subsection{EP Algorithm Application to Circular Orbits}
In the EP algorithm, the search is carried out in the polynomial derivative basis $\Ld = \{f_0, d_2, d_3, d_4\}$. Although the underlying signal follows a circular Keplerian orbit,  the grid itself is initialized as a hyper-rectangle in Taylor coefficient space. By extending the search to $\kmax = 3$, we target the entire regime $\Porb \geq \Tobs$, where the signal exhibits significant higher-order derivatives.

\subsubsection{Search Grid Initialization}
The boundaries of the Taylor hyper-rectangle are set by the most extreme physical parameters in the target population. At fixed orbital period, the projected semi-major axis $x$ in equation~\eqref{eq:xorb} is maximized by the smallest allowed pulsar mass, largest allowed companion mass, and an edge-on orbit ($\sin i=1$). We therefore parametrize the search extent using three inputs: minimum orbital period $\Porb^{\min}$, maximum companion mass $m_{c,\max}$, and minimum pulsar mass $m_{p,\min}$. 

For a circular orbit, the amplitude of the $k$-th time derivative of the line-of-sight displacement is $|d_k| = c\,x\,\Omegaorb^k$. For derivative orders $k \geq 1$, this amplitude is maximized at the maximum orbital frequency, $\Omega_{\max} = 2\pi/\Porb^{\min}$. The corresponding search bounds are therefore set by the most compact, most massive systems:
\begin{equation}
    |d_k|_{\max} = K_{\mathrm{mass}} \, \Omega_{\max}^{k - 2/3}, \qquad k \geq 1,
\end{equation}
where the mass-dependent constant is:
\begin{equation}
    K_{\mathrm{mass}} = \frac{G^{1/3} m_{c,\max}}{(m_{p,\min} + m_{c,\max})^{2/3}}.
\end{equation}
The initial search domain is then the hyper-rectangle $\mathcal{G} = \prod_k \left[-|d_k|_{\max},\, |d_k|_{\max}\right]$, which conservatively encloses all circular orbits with $\Porb \geq \Porb^{\min}$.

\subsubsection{Exact Circular Orbit Resolution}\label{subsec:circular_resolution}
Once the accumulated span $T_s$ reaches a substantial fraction of the orbital period ($T_s \gtrsim 0.2\,\Porb$), the snap derivative $d_4$ can be measured with sufficient precision to enforce the circular-orbit constraints listed in equation~\eqref{eq:nu_recover_main}. In this regime, repeated finite-order Taylor transport ceases to be an appropriate propagation mechanism.

A truncated Taylor transformation $\mathbf{T}(\Delta t)$ re-centres a local polynomial approximation but does not preserve the exact circular manifold. This limitation reflects the fundamental mismatch between the intrinsically sinusoidal structure of circular motion and its finite-order polynomial representation. As a result, a derivative tuple $\{d_k\}$ that lies exactly on the circular manifold at one epoch is generically mapped off that manifold under finite-order Taylor transport. While this discrepancy is negligible for short time spans, it becomes significant as $T_s$ approaches a non-negligible fraction of $\Porb$, introducing systematic phase errors that grow with $\Delta t$ and accumulate across successive stages. Maintaining accuracy would therefore require progressively higher-order derivatives ($\gg 4$), which is both computationally inefficient and still formally inexact.

A more robust alternative is to propagate candidates directly in the circular orbit basis, where time evolution is exact. Given the derivative tuple $\{d_2, d_4\}$ at epoch $t_i$, we recover the orbital frequency $\Omegaorb$, advance the phase by $\Delta\phi = \Omegaorb\,\Delta t$ and evaluate the required derivatives $\{d_k(t_j)\}$ at the new epoch $t_j = t_i + \Delta t$. The explicit transformation is given in Appendix~\ref{app:circular_orbit_transform}. By construction, this transformation preserves the sinusoidal structure exactly (up to floating-point precision). Its computational cost is constant per candidate, involving a fixed number of trigonometric and arithmetic operations, and is negligible compared to the profile accumulation and scoring. Critically, the propagation remains exact even for $T_s > \Porb^{\min}$, enabling coherent integration across one or more complete orbital cycles.

The search grid itself nevertheless remains explicitly in the Taylor basis. Candidates are stored and branched as $\Ld = \{f_0, d_2, d_3, d_4\}$, which defines the natural grid for EP. The exact circular transformation is used only within the \func{Resolve} and \func{Transform} operations of Algorithm~\ref{alg:pruning_seq}: candidate grid centres are temporarily mapped to circular parameters, propagated exactly in time, and then projected back into Taylor coordinates. This hybrid strategy preserves the computational convenience of a polynomial grid while retaining the physical exactness of the circular orbit model.

\subsubsection{Singularity Handling and Basis Augmentation}\label{subsec:circular_hole}
Recovering circular orbit parameters from $\{d_2,d_3,d_4\}$ is well behaved over most of the orbit, but becomes numerically unstable near the nodal phases ($\nu \approx 0, \pi$). At those phases the even derivatives (sine-dependent terms) vanish ($d_2, d_4 \rightarrow 0$), rendering the standard frequency estimator $\Omegaorb = \sqrt{-d_4/d_2}$ ill-conditioned. In a hierarchical search, where the grid is transformed, the signal trajectory must eventually cross these nodal regions when performing a full circular orbit search ($T_s \sim \Porb^{\min}$). Ignoring them would therefore introduce local numerical singularities, or \emph{grid holes}, in the propagation. 

\begin{figure}
\includegraphics[width=\linewidth]{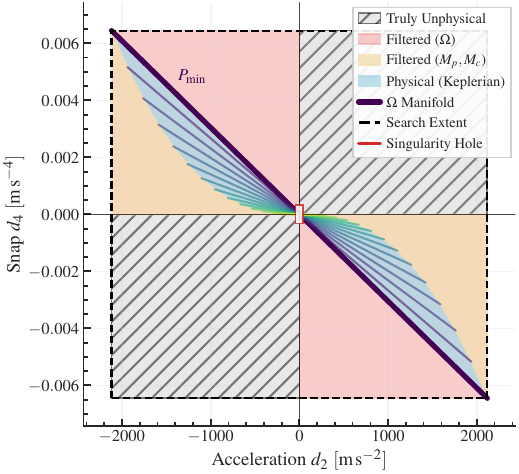}
\caption{Physical constraints on the polynomial search space in the snap ($d_4$) versus acceleration ($d_2$) plane for circular Keplerian orbits. The dashed box shows the full hyper-rectangular search region derived from $\Porb^{\min} = 1$\,h, $m_{c,\max} = 10\,M_\odot$, and $m_{p,\min} = 1.2\,M_\odot$. The gray hatched quadrants (Q1 and Q3) are strictly unphysical, corresponding to unbound orbits. The pink and orange shaded regions are excluded by the maximum orbital-frequency and mass--$\Omega_{\rm orb}$ constraints, respectively. Only the blue curved locus corresponds to physically admissible orbital configurations. Diagonal lines denote constant-$\Omegaorb$ contours, illustrating the contraction of the allowed derivative range toward longer orbital periods. The small red box at the origin marks the nodal singularity region where the snap--acceleration pair becomes numerically ill-conditioned for recovering $\Omega_{\rm orb}$.}
\label{fig:snap_accel_circular_orbit}
\end{figure}

To maintain stable propagation across the full orbit, we augment the search basis with the fifth derivative $d_5$ (crackle). Circular motion places the even- and odd-derivative families in quadrature: when the sine-driven pair $(d_2,d_4)$ vanishes, the cosine-driven pair $(d_3,d_5)$ is maximal. This provides an alternative recovery relation,
\begin{equation}
    \Omegaorb = \sqrt{-\frac{d_5}{d_3}}.
\end{equation}

We therefore implement a dual-gate classification scheme that dynamically switches between the two estimators:
\paragraph*{Snap-dominated region }
When both $|d_2|$ and $|d_4|$ exceed a significance threshold ($\sim2\sigma$) i.e., are numerically well resolved, we use the standard even-derivative estimator $\Omegaorb = \sqrt{-d_4/d_2}$. In this regime, the crackle $d_5$ is precisely determined by the circular recurrence relation and requires no additional grid refinement.

\paragraph*{Crackle-dominated region (nodal hole)}
When $|d_2|$ and $|d_4|$ fall below the significance threshold, we switch to the odd-derivative estimator based on jerk and crackle. Only in these nodal regions do we permit branching along the $d_5$ dimension. Because these holes occupy a negligible fraction of the total search volume, the additional computational cost remains minimal.

This dual-basis approach eliminates the nodal instability while preserving exact circular propagation over the full orbit.

\subsubsection{Physical Validation}
The initial hyper-rectangular grid $\mathcal{G}$ is algorithmically convenient but physically highly redundant as most cells do not correspond to viable circular Keplerian orbits. Once the snap derivative $d_4$ becomes measurable (typically when $T_s \gtrsim 0.2\,\Porb$) and inversion to circular parameters is reliable, we apply strict physical constraints to aggressively reject unphysical grid cells. This is performed via the \func{Validate} operation in Algorithm~\ref{alg:pruning_seq} at every stage using two criteria:

\paragraph*{Orbital Frequency Constraint}
A viable circular candidate must satisfy
\begin{equation}
    0 < -\frac{d_4}{d_2} \leq \Omega_{\max}^2.
\end{equation}
This condition immediately removes the first and third quadrants of the $\{d_2, d_4\}$ plane, where both derivatives have the same sign and therefore imply unbound exponential rather than oscillatory motion. It also enforces chosen minimum orbital period.

\paragraph*{Mass--$\Omega$ Constraint}
A valid orbital frequency does not by itself guarantee a physically admissible binary. For a given cell frequency $\Omega_{\mathrm{cell}}$, the orbital size $x$ is bounded by the assumed mass range. From equation~\eqref{eq:xorb}, the maximum allowed amplitude of the $k$th derivative is
\begin{equation}
    |d_k| \leq K_{\mathrm{mass}}\,\Omega_{\mathrm{cell}}^{k-2/3}.
\end{equation}
In practice, we enforce this bound only on $d_2$. This effectively removes candidates that formally satisfy the circular orbit recurrence relation $d_4 = -\Omegaorb^2 d_2$ but imply an orbital separation (energy) inconsistent with the binary mass limits. 

The resulting allowed region forms the curved locus shown in Figure~\ref{fig:snap_accel_circular_orbit}. For searches targeting strictly circular systems, these validation steps yield substantial computational savings. For broader searches that aim to retain sensitivity to mildly eccentric or otherwise non-circular binaries, the validation can be disabled or relaxed via a configurable option.

\subsubsection{Anchor-Segment Bias and Statistical Independence}\label{subsec:anchor_bias}
Because the EP framework relies on sequential data thresholding, the order in which data blocks are processed introduces a directional path dependency. A linear traversal of the $M = 128$ segments starting from an arbitrary anchor segment $q$ executes a unique sequence of pruning decisions; a candidate rejected early in one traversal path might survive if the data blocks were encountered in a modified sequence. In Section~\ref{sec:ep_multi}, we established that for standard polynomial searches, re-traversing the same dataset multiple times using maximally separated anchors mitigates this path dependence. This strategy yields a massive computational reduction by pairing a low per-pass detection probability ($P_d \sim 0.10$) with an ensemble recovery rate exceeding $90\%$. However, applying this multi-pass architecture to circular Keplerian orbits introduces phase-dependent orbital dynamics. It is therefore necessary to verify whether all anchor segments are equally viable as path origins, or if localized orbital phase boundaries introduce systematic performance biases.

\begin{figure*}
\includegraphics[width=\linewidth]{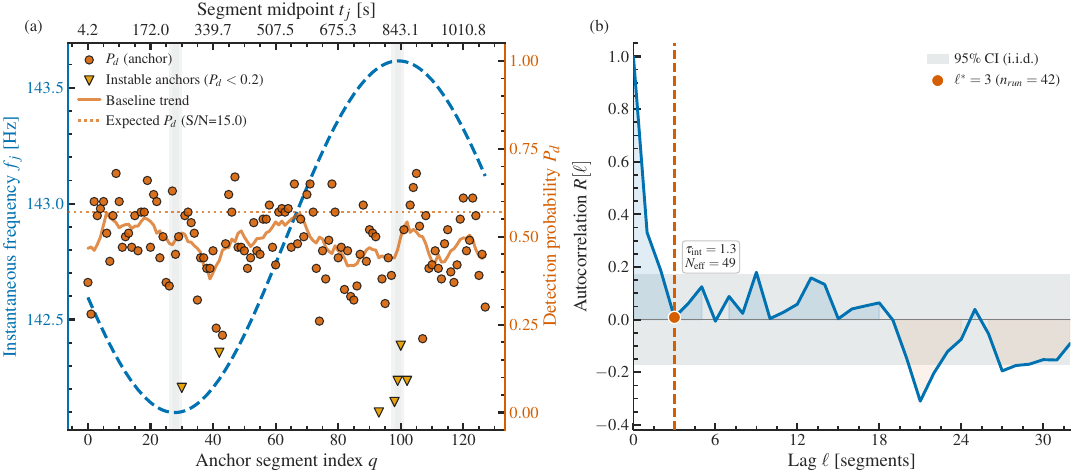}
\caption{Dependence of detection probability and pruning-path autocorrelation on the anchor segment in the hierarchical EP search for a tuned threshold scheme targeting $\mathrm{S/N}=10$.
\textbf{Panel~(a):} Injection recovery performance for a circular-orbit search ($\Porb \geq \Tobs$) over an 18-min observation divided into $M=128$ segments ($\eta = 1.0$, $N_b = 64$). A signal with $\mathrm{S/N}=15$ was injected into $n_{\rm trials} = 100$ independent noise realizations for each anchor segment $q$. The measured detection probability $P_d[q]$ (crimson points, right axis) is shown together with a localized baseline trend (solid red curve) obtained after masking numerical dropouts ($P_d < 0.2$). For reference, the instantaneous frequency $f_j$ at each segment midpoint $t_j$ is overlaid (dashed blue, left axis). While $P_d$ remains close to the expected value, $P_d^{\rm th}=0.57$, across most of the orbit, it collapses around narrow phase intervals centred on the epochs where the instantaneous orbital acceleration vanishes ($\dot{f} = 0$, shaded gray bands).
\textbf{Panel~(b):} Autocorrelation function (ACF) of the $P_d[q]$ sequence. The ACF drops below the 95\% confidence interval for an independent and identically distributed process (gray band, $\pm 1.96/\sqrt{M}$) at a decoupling lag of $\ell^* = 3$ segments. This supports treating pruning runs separated by at least $\ell^*$ segments as statistically independent, allowing up to $\nrun=\lfloor M/\ell^* \rfloor = 42$ effectively independent Bernoulli trials to be combined.}
\label{fig:ep_circular_seg_bias}
\end{figure*}

To characterize this directional dependency, we perform a Monte Carlo injection simulation using a circular binary signal whose orbital period is comparable to the total observation time ($P_{\rm orb} \sim T_{\rm obs}$). The signal is injected at a baseline $\mathrm{S/N} = 15$ across 100 independent noise realizations per anchor segment. The EP search is performed on each realization using a threshold scheme tuned for a detection threshold of $\mathrm{S/N} = 10$. Figure~\ref{fig:ep_circular_seg_bias}(a) displays the empirical detection probability $P_d[q]$ as a function of the starting anchor index $q$.  Crucially, across the vast majority of starting positions, $P_d[q]$ remains uniform and closely tracks the expected value $P_d^{\rm th} = 0.57$. This uniform behaviour demonstrates that most anchor segments are structurally safe and viable for EP traversal, confirming the general stability of the circular orbit propagation model.

Having established this baseline uniformity, we can invert the question to determine the statistical independence of neighbouring paths in a multi-pass architecture: at what spatial separation do two traversal tracks decouple? Figure~\ref{fig:ep_circular_seg_bias}(b) displays the autocorrelation function (ACF) of the $P_d[q]$ sequence. The ACF exhibits a rapid decay, plunging cleanly into the $95\%$ confidence interval for uncorrelated white noise ($\pm 1.96/\sqrt{M}$) at a critical decoupling lag of $\ell^* = 3$ segments. The short integrated autocorrelation time ($\tau_{\rm int} = 1.32$) confirms that shifting the anchor segment effectively randomizes the downstream pruning choices. Consequently, the 128-segment dataset can host up to $\nrun = \lfloor M/\ell^* \rfloor = 42$ completely independent, uncorrelated parallel search passes. This high number of independent trials permits us to tune the single-pass threshold down to a more aggressive $P_d \sim 0.05$ per run, securing massive additional complexity savings while ensuring the combined ensemble detection probability comfortably exceeds $1 - (1 - 0.05)^{42} \approx 89\%$.

Despite this overall stability, Figure~\ref{fig:ep_circular_seg_bias}(a) reveals narrow, severe performance dropouts, where $P_d$ plunges close to zero. This exposes an additional localized sensitivity loss, or phase traps, restricted to less than $5\%$ of the circular orbital phases. These deterministic phase traps coincide exactly with the epochs where the instantaneous acceleration vanishes ($\dot{f} = 0$). When the EP integration track is anchored directly inside one of these turning points, the snap--acceleration ($s\text{--}a$) relation is ill-conditioned from the outset, forcing the tree to rely exclusively on the higher-order crackle--jerk ($c\text{--}j$) pair before the tree grid has accumulated sufficient data to resolve the underlying snap. We isolate the structural mechanics of these dropouts further in Appendix~\ref{app:anchor_bias_diagnostic}.

\subsection{Extending Search to Long Durations
  \texorpdfstring{($\Tobs > \Porb^{\min}$)}{(Tobs > Porbmin)}
}
\label{sec:param_transition}
The sequential EP search framework is formulated in a local polynomial phase basis. This is well matched to the regime $\Tobs \lesssim \Porb $, where the orbital motion is observed only as a short arc and a low-order Taylor expansion remains efficient. For observation spans $\Tobs$ approaching or exceeding the minimum target orbital period $\Porb^{\min}$, the polynomial approximation becomes increasingly inefficient. A finite-order polynomial is inherently divergent with time, whereas the phase evolution for circular orbits is strictly bounded and periodic. Attempting to model periodic orbital motion with a Taylor expansion over multiple cycles requires determining high-order derivatives with extreme precision, an increasingly inefficient exercise as the observation extends beyond one orbital period.

A natural extension of the EP framework is therefore to allow a dynamic transition of the search grid basis from the local polynomial parameterization $\Ld = \{f_0, d_2, d_3, d_4\}$ to the global circular-orbit parameterization $\Lcart = \{f_0, \Omegaorb, x_{\cos\nu}, x_{\sin\nu}\}$ defined in Appendix~\ref{app:circular_orbit_cartesian}. The Cartesian basis is intrinsically bounded for circular motion and avoids the continued refinement of derivative amplitudes once the orbit has been sufficiently resolved.

As shown in Appendix~\ref{app:circular_orbit_cartesian_grid}, the required Cartesian grid spacing scales asymptotically as
\begin{align}\label{eq:circ_scaling}
    \Delta f_0, \Delta \Omegaorb &\propto T_s^{-1}, \nonumber \\
    \Delta x_{\cos\nu}, \Delta x_{\sin\nu} &\propto T_s^{0}
    \qquad (T_s \gtrsim \Porb).
\end{align}
This is the key advantage of the physical basis. In the polynomial representation, all active derivatives must continue to refine as powers of $T_s$ in order to control cumulative phase error. In the Cartesian circular basis, by contrast, the projected semi-major-axis components are physical invariants. Once the data span is long enough to resolve the orbital amplitude, further integration requires refinement only in the frequency-like coordinates $f_0$ and $\Omegaorb$.

Consequently, the grid volume for transitioned branches grows only quadratically with coherent span,
\begin{equation}\label{eq:n_grid_circ}
    N_{\mathrm{grid,circ}}(T_s) \propto T_s^2,
\end{equation}
rather than as the much steeper $T_s^{10}$ growth associated with the $\kmax=3$ polynomial representation. This does not by itself demonstrate a practical long-baseline implementation, but it does show that a basis transition is the natural route to extending EP beyond the single-orbit regime, making deep searches over many orbital periods computationally feasible.

\subsubsection{Physical Basis Transition Criterion}
A branch should transition from $\Ld$ to $\Lcart$ only once the circular parameters are constrained more finely than the physical Cartesian grid itself. We define the transition time $\Ttrans$ for a given grid cell as the epoch when the measurement uncertainties in the physical parameters ($\sigma_{\Omegaorb}, \sigma_{x_{\cos\nu}}, \sigma_{x_{\sin\nu}}$), propagated from the current polynomial cell, become smaller than the optimal physical grid spacing required in the Cartesian basis (see Appendix~\ref{app:circular_orbit_cartesian_grid}). Formally, the transition criterion is when
\begin{align}\label{eq:transition_criterion}
    \sigma_{\Omegaorb}(\Ttrans) &\le \Delta \Omegaorb(\Ttrans), \nonumber\\
    \sigma_{x_{\cos\nu}}(\Ttrans) &\le \Delta x_{\cos\nu}, \\
    \sigma_{x_{\sin\nu}}(\Ttrans) &\le \Delta x_{\sin\nu}. \nonumber
\end{align}

Since $\sigma_x$ decreases with integration time while $\Delta x$ remains constant (equation~\eqref{eq:circ_scaling}), this inequality defines a data-driven resolution horizon. Before $\Ttrans$ the orbit is observed as a local arc, so the Cartesian basis $\Lcart$ is poorly constrained and inefficient. After $\Ttrans$, continuing in the polynomial basis $\Ld$ becomes increasingly redundant as the physical amplitude coordinates have already saturated.

The detailed scaling analysis in Appendix~\ref{app:circular_grid_transition} suggests that this transition should occur only after nearly one full orbit has been sampled. In representative phase configurations the limiting condition occurs at a coverage of $\sim0.9\,\Porb$. Because this estimate depends on the uncertainty model and on the phase at which the branch is sampled, we do not treat it as a sharp universal threshold. Instead, we adopt the simpler and more conservative prescription
\begin{equation}\label{eq:transition_time}
    \Ttrans = \Porb,
\end{equation}
ensuring that all orbital phases have been sampled and that both Cartesian amplitude coordinates are well constrained.

\subsubsection{Implementation in the EP Algorithm}
This basis transition has not yet been implemented in the current EP pipeline and is therefore presented here as a proposed extension rather than a validated operating mode. The transition is required only when the search targets systems with $\Porb^{\min} < \Tobs$. For searches satisfying $\Porb^{\min} \geq \Tobs$, the polynomial basis remains efficient throughout the observation, and no transition is necessary. A straightforward implementation would modify Algorithm~\ref{alg:pruning_seq} such that, at the end of each stage $s$, every surviving grid cell is processed as follows:
\begin{enumerate}
    \item Basis Conversion: If the transition criterion is satisfied ($T_s \geq 2\pi/\Omega_{\mathrm{cell}}$), re-parametrize the cell from $\Ld$ to $\Lcart$.
    \item Heterogeneous Branching: During stage $s+1$, cells remaining in $\Ld$ continue refining the polynomial derivatives $\{d_2, d_3, d_4\}$, whereas cells that have transitioned to $\Lcart$ refine only in $\{\Omegaorb, f_0\}$.
\end{enumerate}

The resulting search tree would be heterogeneous, with different branches evolve in different parametrizations according to their local information content. The principal benefit is that transitioned branches no longer incur the cost of high-order polynomial refinement, instead following the milder $\mathcal{O}(T_s^2)$ scaling of equation~\eqref{eq:n_grid_circ}. The corresponding trade-off is increased algorithmic complexity, as candidate propagation, validation, and threshold calibration must operate consistently across a mixed-basis tree.

A complete threshold analysis for this heterogeneous regime remains future work. An initial implementation could simply retain the existing threshold scheme, which is  calibrated for the worst-case polynomial branching factor prior to the transition. Since the basis transition occurs well after the peak pruning stage, by which point the surviving search tree has already been reduced substantially, any resulting mismatch in the threshold calibration is expected to have only a minor impact on overall sensitivity. Such a scheme would preserve robust control of the null survival rate while providing a practical first implementation. A threshold scheme that explicitly accounts for the mixed-basis branching statistics should yield further improvements, but requires dedicated investigation.

\section{Algorithm implementation}\label{sec:alg_implement}
The preceding sections established the mathematical foundations of the P-FFA and EP algorithms for a localized parameter space $\mathbf{\Lambda}_d$. Controlled by the maximum derivative order $\kmax$ and the chosen parametrization, the framework supports a configurable hierarchy of searches, from constant spin frequency through successively higher spin derivatives (including acceleration, jerk, and snap) to fully coherent circular-orbit integration. We now detail the practical strategies required to scale these algorithms to full observation time series and survey-wide, multi-dimensional search spaces.

\subsection{Searching across spin frequency}\label{sec:freq_search}
Pulsars span many orders of magnitude in spin period, from slow rotators ($f \lesssim 1\,\mathrm{Hz}$) to millisecond periods ($f \sim 10^2$--$10^3\,\mathrm{Hz}$), including the fastest known pulsar at $716\,\mathrm{Hz}$ \citep{Hessels:2006}. In standard FFT-based pipelines such as \soft{PRESTO} and \soft{Peasoup}, this range is naturally covered by a single transform with uniform Nyquist sampling \citep{Ransom:2011, Barr:2020}. By contrast, both the P-FFA and EP algorithms operate most efficiently over narrow ranges of trial spin frequencies, within which folded profiles are computed using a fixed number of phase bins, $N_b$. Extending a single fixed folding resolution over several orders of magnitude in spin period is highly suboptimal: a globally large $N_b$ is computationally prohibitive, whereas a globally small $N_b$ degrades duty-cycle resolution unacceptably at long periods \citep{Morello:2020}.

Standard FFA implementations address this challenge by iteratively downsampling the time series and searching successive octaves in period, ensuring the number of phase bins remains bounded while the effective sampling time increases \citep[e.g.,][]{Kondratiev:2009, Cameron:2017, Morello:2020}. We adopt an alternative, mathematically equivalent strategy tailored to our brute-force initialization and dynamic programming framework. Rather than downsampling the input data, we partition the period range into contiguous regions, each searched with a tailored folding resolution that maintains approximately constant physical duty-cycle sensitivity.

For a minimum search period $P_{\min}$ and a user-specified minimum number of folding bins $b_{\min}$, we define a reference physical bin width
\begin{equation}
    t_{W} \equiv \max\left(\frac{P_{\min}}{b_{\min}}, t_s\right).
\end{equation}
This establishes the finest physical resolution employed anywhere in the pipeline. Sensitivity to narrow pulses is dictated by $t_{W}$, while the computational memory footprint is governed primarily by $N_b$.

\begin{figure}
\includegraphics[width=\linewidth]{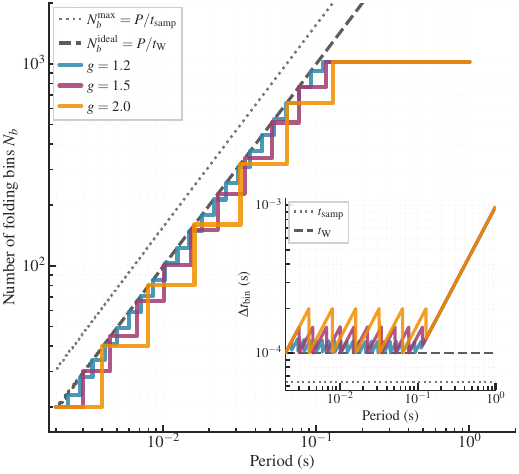}
\caption{
Frequency chunking and folding-bin allocation used in P-FFA and EP searches. The main panel shows folding bins $N_b$ as a function of trial period. The dotted curve indicates the maximum limit $b_{\max}$, while the dashed curve shows the ideal continuous scaling corresponding to a constant physical bin width $t_W$. Solid curves show the discrete binning schemes adopted for different geometric growth factors $g$. The inset shows the resulting physical bin width $\Delta t_{\rm bin}(P) = P/N_b(P)$.}
\label{fig:ffa_chunk}
\end{figure}

We divide the full period range $[P_{\min}, P_{\max}]$ into a sequence of contiguous regions indexed by $k$. Within each region, the number of folding bins $N_{b,k}$ is strictly fixed. Between successive regions, $N_{b,k}$ increases by a constant geometric growth factor $g>1$ (e.g., $g = 2$ for octave spacing):
\begin{equation}
    N_{b,k+1} = g\,N_{b,k}.
\end{equation}
Each region therefore spans the period interval:
\begin{equation}
    P \in \left[ N_{b,k} t_{W},\; g\,N_{b,k} t_{W} \right].
\end{equation}
Under this scheme, the physical bin width at the lower boundary of every region is exactly $t_{W}$, and increases by at most a factor of $g$ at the upper boundary. This geometric partitioning ensures that the duty-cycle resolution degrades smoothly, with bin widths increasing by at most a factor of $g$ within each region, while the computational cost of the P-FFA grows strictly logarithmically with the period range $P_{\max}/P_{\min}$. The allocation of folding bins and the resulting physical bin widths are illustrated in Figure~\ref{fig:ffa_chunk}.

The density of the trial spin frequency grid is governed by the ratio $\eta/N_b$, which defines the maximum allowed phase mismatch in units of phase bins. Holding the tolerance $\eta$ fixed while increasing $N_b$ would redundantly over-sample the frequency grid at long periods. To prevent this, we enforce a constant fractional duty-cycle resolution across all regions:
\begin{equation}
    \rho \equiv \frac{\eta}{N_b}.
\end{equation}
Given a user-specified tolerance $\eta_0$ evaluated at the minimum bin count $b_{\min}$, the tolerance within any region $k$ scales as:
\begin{equation}
    \eta_k = \rho\,N_{b,k} = \eta_0 \frac{N_{b,k}}{b_{\min}}.
\end{equation}
This prescription yields a frequency grid with uniform spacing in frequency space, naturally providing the dense period-space sampling required for millisecond pulsar searches while relaxing appropriately for slow rotators.

The geometric growth of $N_b$ is capped at a threshold $b_{\max}$ to limit memory usage at the longest periods. Once this limit is reached, all subsequent trial periods are searched with $N_b = b_{\max}$, allowing the physical bin width to increase linearly with period. This reflects the astrophysical expectation that progressively finer duty-cycle resolution yields diminishing returns in the long-period regime, where pulse profiles are typically broader and red-noise systematics dominate. Any resolution discontinuities introduced at region boundaries are bounded by $g$ and remain negligible compared to the intrinsic discretization imposed by finite binning.

This region-based strategy circumvents the need for explicit time-domain downsampling. When the physical bin width exceeds the native sampling interval $t_s$, the folding operation inherently performs the exact mathematical equivalent of phase-resolved averaging. Unlike conventional octave-based downsampling schemes, our approach operates directly on the original time series at native resolution, avoiding the noise-variance corrections required for non-integer downsampling factors \citep{Morello:2020}.

\subsection{Memory constraints and Minimum Coherent Volume}\label{sec:min_coh_vol}
Although frequency chunking optimizes the computational scaling of the folding resolution, practical deployments of the P-FFA and EP algorithms are ultimately constrained by available memory. Both algorithms construct a dynamic programming tree over the trial parameter grid, whose memory footprint grows rapidly with search dimensionality and coherent integration time. To enforce a prescribed memory budget, each frequency region is subdivided into contiguous frequency blocks that are processed sequentially. Each block spans a nominal frequency interval $\Delta f^{\rm nom}$, chosen such that the peak memory requirement, $M_{\rm peak}$ of its parameter tree remains within the user-specified limit. This subdivision is purely an implementation detail  and does not alter either the search configuration or the phase resolution of the parent region.

In contrast, the search space cannot be partitioned independently along non-frequency dimensions (e.g., acceleration, jerk, or higher-order Keplerian parameters). Maintaining phase coherence requires that all coupled kinematic parameters be evaluated simultaneously within a single tree. This defines the \emph{minimum coherent volume}: the smallest joint region of frequency and kinematic parameter space that can be searched as an indivisible unit. As the search order $\kmax$ or observation duration $\Tobs$ increases, this coherent volume expands, imposing a hard lower bound on the required memory that cannot be reduced through finer frequency partitioning.

For searches with $\kmax \ge 1$, each nominal frequency block must also be extended by an overlap $\delta f$ to accommodate the maximum expected frequency drift over the observation. Thus, coherently searching a target interval $[f_{\rm min}, f_{\rm max}]$ requires an active search span of $[f_{\rm min}-\delta f,\,f_{\rm max}+\delta f]$. Since the maximum drift scales with the intrinsic spin frequency ($\delta f \propto f$), the overlap fraction increases with spin frequency. Consequently, the minimum memory footprint scales as
\begin{equation}
    M_{\rm min}(f) \propto f^{\kmax} \cdot N_b.
\end{equation}

Our implementation determines the block boundaries automatically using a dynamic binary search. Starting from the high-frequency edge of each region, where the memory demand is greatest, the algorithm identifies the widest drift-padded block that satisfies the available memory budget. After processing the block, the frequency frontier is advanced and the procedure repeated until the entire region has been searched. This strategy enables fully coherent P-FFA and EP searches to operate efficiently at the available memory limit, maximizing throughput while avoiding both out-of-memory failures and coverage gaps.

\begin{figure*}
\centering
\includegraphics[width=\textwidth]{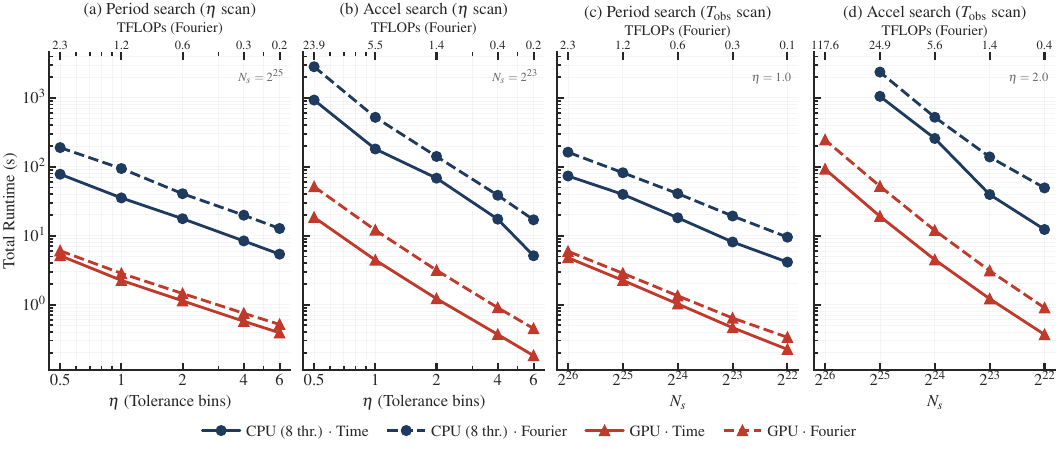}
\caption{Wall-clock runtime of the P-FFA search pipeline as a function of the tolerance parameter~$\eta$ (panels~a--b) and the number of samples $N_{s}$ (panels~c--d), for constant-period (panels~a,~c) and constant-acceleration (panels~b,~d) searches per DM trial. Panels~(a) and~(b) correspond to SKA1-mid-like configurations with $N_{s} = 2^{25}$ and $2^{23}$, respectively. Panels~(c) and~(d) show scaling with $N_s$ at fixed $\eta = 1.0$ and $2.0$, respectively. CPU results (8-thread OpenMP) are shown as circles and GPU (CUDA) as triangles; solid and dashed lines denote time- and Fourier-domain modes. The top axis indicates the analytical floating-point operation count in units of TFLOPs ($10^{12}$~FLOPs) for the Fourier-domain mode. Time resolution is fixed at $64\,\upmu$s, corresponding to observation lengths of 5--72\,min across the $N_s$ range. Folded-profile bin counts span $b_{\min}=32$ to $b_{\max}=1024$ with geometric growth factor $g=1.5$ (See Sec~\ref{sec:freq_search} for definitions).}
\label{fig:ffa_benchmark}
\end{figure*}

\subsection{Code Implementation}\label{sec:code_implementation}
The P-FFA and EP algorithms are implemented in \textsc{LOKI}, a C++20 package, publicly available on GitHub\footnote{\url{https://github.com/pravirkr/loki}}. The current release operates directly on dedispersed time series and supports configurable searches over user-defined, multi-dimensional parameter spaces, ranging from reduced-coherence modes that prioritize execution speed to fully coherent integration for maximum sensitivity.

\textsc{LOKI} provides distinct CPU and GPU execution backends, with Python bindings that expose the full public API for custom pipeline construction. The CPU backend uses OpenMP parallelism and is compiled with aggressive optimization, relying primarily on compiler auto-vectorization supplemented by explicit SIMD implementations in performance-critical kernels. The GPU backend provides an end-to-end CUDA implementation of both algorithms, with the dominant computational stages executed entirely on-device and higher-level operations managed through NVIDIA's CCCL library for parallel primitives. Coarse-grained parallelism, such as distributing DM trials across cluster nodes, can be orchestrated externally, as is standard in pulsar-search pipelines \citep{Ransom:2011}.

Performance profiling shows that the dominant runtime contribution ($\gtrsim 70\%$ in the current implementation) arises from the shift-add and \func{Score} operations, i.e.\ matched filtering over the bank of boxcar widths. Both stages exhibit intrinsically low arithmetic intensity. The time-domain shift-add kernel performs approximately $1/12$ \,FLOP\,byte$^{-1}$, increasing to only $\sim 1/3$\,FLOP\,byte$^{-1}$ for the Fourier-domain variant. The scoring kernel achieves an arithmetic intensity of $N_w/2$\,FLOP\,byte$^{-1}$ ($N_w \ll N_b$), where $N_w$ is the number of boxcar width trials. These values remain well below the roofline ridge point of contemporary GPUs ($\sim 10$--$100$\,FLOP\,byte$^{-1}$ for FP32), indicating that performance is primarily limited by global memory throughput rather than compute capacity \citep{Williams:2009}. Consequently, the CUDA implementation emphasizes parallelism across independent search grid leaves and phase bins to improve occupancy, hide memory latency, and sustain high memory throughput, rather than attempting to increase arithmetic intensity through further algorithmic restructuring. 

Given these memory-bandwidth constraints, candidates are processed in fixed-size batches to reduce per-candidate overhead. On CPU, we typically adopt $N_{\mathrm{batch}} = 2^{10}$, which improves vectorization and cache reuse during phase shifting, profile accumulation, and scoring, thereby reducing the amortized cost of candidate evaluation. On GPU, we use a larger batch size of $N_{\mathrm{batch}} = 2^{16}$ to expose sufficient parallelism to saturate memory bandwidth and maximize sustained throughput.

We also maintain a pure-Python reference implementation, \textsc{PyLOKI}\footnote{\url{https://github.com/pravirkr/pyloki}}. This modular package mirrors the C++ logic and uses \texttt{Numba} JIT compilation to optimize the underlying numerical kernels. While not intended for large-scale searches, it provides a transparent and accessible platform for algorithmic development and validation.

\subsection{P-FFA Algorithm Benchmarks}\label{sec:pffa_benchmarks}
The P-FFA functions as a complete, standalone coherent search package and constitutes a novel search algorithm in its own right. We benchmark the two regimes most relevant for practical deployment: constant-period searches ($\kmax=0$) and constant-acceleration searches ($\kmax=1$). Higher-order searches are excluded here due to their steep computational complexity, which places them in the operational regime targeted by the EP algorithm (Section~\ref{sec:ep_benchmarks}). The goal here is not a micro-architectural analysis, but to demonstrate that the P-FFA can process modern survey-scale data volumes on contemporary hardware. The reported wall-clock times correspond to the end-to-end search cost per DM trial, including coherent time-series folding, multi-width boxcar matched filtering and result serialization to disk.

Benchmarks are carried out using both time-domain folding and exact Fourier-domain folding. The baseline configuration adopts $b_{\min} = 32$, $b_{\max} = 1024$, $g=1.5$, and a frequency range of $f \in [1, 500]$\,Hz. Runtimes are measured across a range of $\eta$ values spanning practical duty-cycle resolutions. CPU benchmarks are executed on an Intel Xeon Gold~6348H system (2.30\,GHz; 96 physical cores) using 8 OpenMP threads, while GPU benchmarks are performed on an NVIDIA L40S using CUDA~13.0. Each measurement records the contiguous wall-clock time required to process the full frequency range end-to-end.

To probe performance across survey-relevant regimes, we evaluate a grid of time-series lengths \mbox{$N_s=2^{22}$--$2^{26}$} at a fixed sampling interval of $t_s=64\,\upmu\mathrm{s}$, corresponding to observation lengths $\Tobs \approx 5$--72\,min. For acceleration searches, the maximum trial acceleration is scaled inversely with observation length, yielding ranges of $\pm(700,\,350,\,175,\,87.5,\,43.75)$\Accelunits\,across this grid. This choice is conservative relative to the canonical $\Tobs^{-4/3}$ scaling expected from preserving sensitivity to a fixed orbital phase fraction, and therefore retains a broader acceleration window at longer integration times.

Figure~\ref{fig:ffa_benchmark} shows the measured runtimes. Two representative SKA1-mid--like configurations are highlighted explicitly: a constant-period search with $N_s = 2^{25}$ ($\Tobs \approx 36$\,min) and a constant-acceleration search with $N_s = 2^{23}$ ($\Tobs \approx 9$\,min) over $\pm 350$\Accelunits, corresponding to panels~(a) and~(b), respectively \citep{Keane:2025}. In the constant-period case (panels~a,~c), runtime scales approximately as $\eta^{-1}$ and $N_s \log N_s$, consistent with the hierarchical P-FFA structure and with the expected reduction in search-grid volume as $\eta$ increases. The close agreement between time- and Fourier-domain GPU runtimes, despite their differing FLOP counts, confirms that this regime is predominantly memory-bandwidth bound. 

Including acceleration trials (panels~b,~d) substantially increases the computational workload, yielding an approximate $\eta^{-2}$ scaling as expected from the 2D expansion of the search grid volume. In this regime, the per-FLOP cost of the complex Fourier-domain implementation becomes distinctly apparent on GPUs, indicating a transition toward a mixed memory- and compute-bound regime. The GPU backend consistently outperforms the 8-thread CPU baseline across all tested configurations, delivering speedups of $\sim$15--30$\times$ for constant-period searches and up to $\sim$50$\times$ for acceleration searches. CPU runtimes become prohibitive at the largest acceleration-search configurations, whereas GPU runtimes remain tractable, demonstrating that the P-FFA scales efficiently to survey-scale workloads on modern accelerator hardware. 

These benchmarks establish \textsc{LOKI}'s P-FFA module as a computationally viable solution for fully coherent  grid searches over frequency and acceleration. The reported runtimes correspond to a single DM trial; practical pulsar surveys multiply this cost by the required number of DM trials and beams, a process that scales trivially via external parallelization.

\subsection{EP Algorithm Benchmarks}\label{sec:ep_benchmarks}
We now benchmark the EP algorithm, which constitutes the architectural core of this work. While the P-FFA pipeline addresses low-dimensional parameter spaces through structured grid evaluation, the EP framework is designed to avoid exhaustive enumeration in high-dimensional searches ($\kmax \ge 2$). For consistency with the P-FFA benchmarks, all EP benchmarks adopt the same baseline profile configuration introduced in Section~\ref{sec:pffa_benchmarks}: $b_{\min} = 32$, $b_{\max} = 1024$, and $g=1.5$. For each configuration, we measure the explicit execution time for a single representative frequency chunk centred at a spin frequency of $f = 333$\,Hz. This runtime is then scaled linearly to the full search range, $f \in [1,500]$\,Hz, using the number of sequential frequency blocks required to maintain the folded-profile bin resolution described in Section~\ref{sec:freq_search}. Since the sequential block pipeline can be further optimized through improved memory saturation and chunk-wise targetted thresholding, the scaled runtimes reported here should be regarded as conservative upper-bound estimates of the computational cost.

\begin{table}
\centering
\caption{Fixed EP benchmark configurations for Fig.~\ref{fig:ep_benchmark}. Rectangular parameter bounds at each $\Porb^{\min}$ are derived from the maximum circular-orbit derivatives corresponding to that minimum orbital period. EP gain denotes the average order-of-magnitude reduction in cumulative tree node evaluations relative to an unpruned hierarchical baseline. Common to all tiers: $m_{\mathrm{p,min}} = 1.2\,M_\odot$, $m_{\mathrm{c,max}} = 10\,M_\odot$, $\Delta t = 64\,\upmu\mathrm{s}$, and $N_{\mathrm{bins}} = 31$.\label{tab:ep_bench_config}}
\begin{tabular}{lcccccc}
\hline
Bench & $\kmax$ & $\Porb^{\min}$ & $\SNR_t$ & $N_{\mathrm{s}}$ / $\eta$ & RAM & EP Gain\\
      &         &                &          &                           & (GB)& $\mathcal{O}(10^{x})$\\
\hline
Accel    & 1 & $10\,\Tobs$ & 8.5  & $2^{25}$ / $1.0$ & 4 & 4\\
Jerk     & 2 & $5\,\Tobs$  & 9.2  & $2^{25}$ / $1.0$ & 4 & 6\\
Snap     & 3 & $3\,\Tobs$  & 9.8  & $2^{25}$ / $1.0$ & 4 & 9\\
Circular & 4 & $\Tobs$     & 10.0 & $2^{23}$ / $2.0$ & 8 & 10\\
\hline
\end{tabular}
\end{table}

To characterize performance across a broad range of workloads, we sweep time-series lengths spanning 5--72\,min together with a range of $\eta$ values, matching the P-FFA benchmarks. In practice, higher-order polynomial searches and fully coherent circular-orbit searches are parametrized by physically motivated bounds rather than arbitrary grid limits. For the constant polynomial searches, rectangular parameter bounds are derived from the maximum orbital derivatives over the orbital fraction for which the corresponding polynomial approximation remains valid. For example, the constant-jerk approximation is expected to remain accurate over approximately $\lesssim 10$--$17\%$ of the orbital period (Section~\ref{subsec:circ_orbit_coverage}); we conservatively adopt a 20\% coverage fraction, setting $\Porb^{\min} = \Tobs/0.2$, and apply analogous choices to the acceleration and snap benchmarks. The fully coherent circular-orbit benchmark instead adopts the more stringent binary survey configuration with $\Porb^{\min} = \Tobs$. Throughout, we assume a maximum companion mass $m_{c,\max} = 10\,M_\odot$ and a minimum pulsar mass $m_{p,\min} = 1.2\,M_\odot$.

\begin{figure*}
\centering
\includegraphics[width=\textwidth]{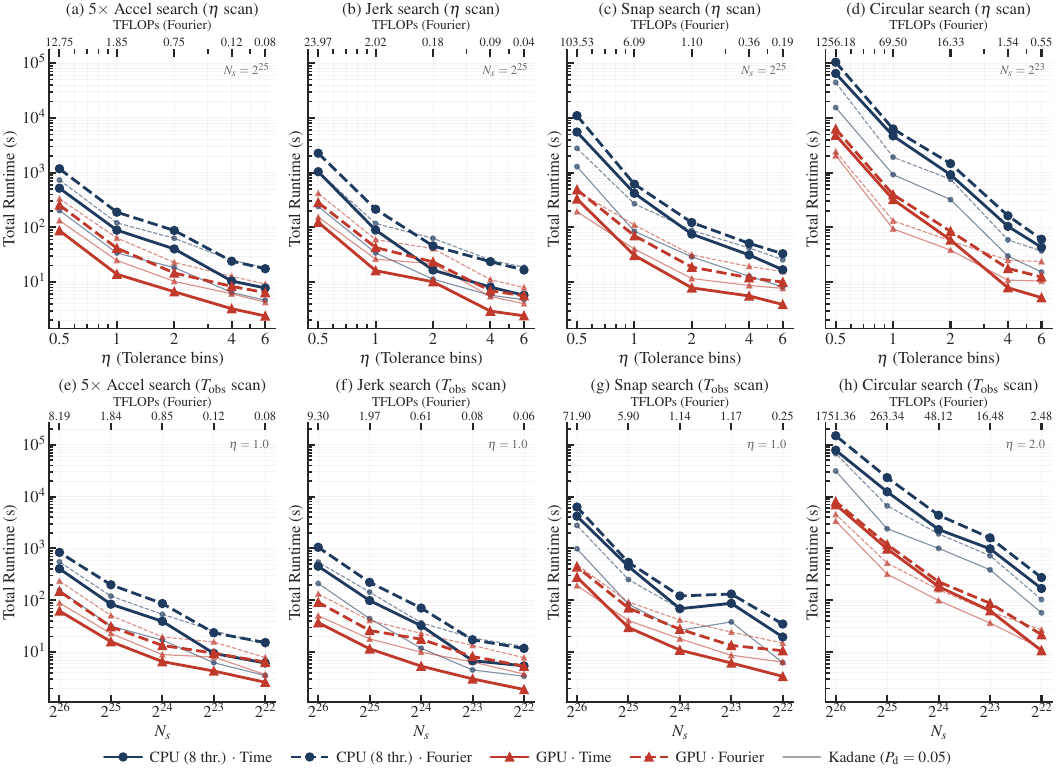}
\caption{Wall-clock runtime of the EP search pipeline for four survey configurations per DM trial.
\textbf{Top row} (panels~a--d): tolerance scan at fixed time-series length; \textbf{bottom row} (panels~e--h): scaling with the number of samples $N_{\mathrm{s}}$ at fixed tolerance.
From left to right, the columns show a $5\times$ inflated constant-acceleration search, a jerk search, a snap search, and a fully coherent circular-orbit search.
In panels~(a)--(c) and~(e)--(g), $N_{\mathrm{s}}=2^{25}$ and $\eta=1.0$, respectively; panel~(d) uses $N_{\mathrm{s}}=2^{23}$, and panel~(h) uses $\eta=2.0$ (values not held fixed are annotated in each panel).
Circles and triangles denote eight-thread CPU (OpenMP) and GPU (CUDA) executions; solid and dashed curves are time- and Fourier-domain backends, respectively.
Bold curves show the default configuration ($P_{\mathrm{d}}=0.10; \nrun=16$); faint overlays show the Kadane-based boxcar variant at $P_{\mathrm{d}}=0.05$ with $\nrun=32$.
The upper axis reports the analytical floating-point operation count in units of TFLOPs ($10^{12}$~FLOPs) for the Fourier-domain mode and $P_{\mathrm{d}}=0.10$.
All benchmarks use a sampling interval $t_s=64\,\upmu\mathrm{s}$; the remaining setup details follow Section~\ref{sec:pffa_benchmarks}.}
\label{fig:ep_benchmark}
\end{figure*}

For the constant-acceleration searches, the EP algorithm is somewhat over-engineered; pruning is sufficiently effective that the remaining workload cannot fully saturate the GPU kernels. We therefore inflate the acceleration search window by a factor of five to circumvent this low-occupancy regime and provide a more meaningful stress test. For all polynomial benchmarks, we fix $N_s=2^{25}$ for the $\eta$ sweeps and $\eta=1.0$ for the $N_s$ sweeps to maintain a consistently high computational workload. Peak memory usage is capped via the chunking mechanism described in Section~\ref{sec:freq_search}, limiting each pruning pass to 4\,GB per for the polynomial searches and 8\,GB per for the circular-orbit search. The dynamic thresholding scheme is calibrated to target detection thresholds in the range $\SNR_t = 8.5$--$10.0$, with the exact value determined by the total enumeration volume of each benchmark configuration. Relative to an unpruned hierarchical baseline, the EP algorithm reduces the cumulative number of tree-node evaluations by an average of 4--10 orders of magnitude across these tiers. This pruning efficiency is intrinsically linked to the detection threshold; the elimination component becomes increasingly effective as both the dimensionality and size of the search space grow, delivering the largest computational gains in the most demanding search configurations. The fixed benchmark configurations are summarized in Table~\ref{tab:ep_bench_config}.

Figure~\ref{fig:ep_benchmark} shows the resulting wall-clock execution profiles. The bold curves correspond to the default EP configuration, using a minimum detection probability of $P_d = 0.10$ per run and $\nrun=16$ multi-pass executions. The faint curves show an alternative configuration with $P_d=0.05$, $\nrun=32$, and the Kadane-based approximate boxcar kernel. All runtimes are reported per DM trial and include the scaled cost of the full 1--500\,Hz search range. The inflated acceleration benchmark provides a useful calibration point for the EP implementation. At $\eta=1.0$ and $N_s=2^{25}$, the GPU performance margin is already substantial ($\sim$5--6$\times$) in this deliberately constrained search space, though it is less pronounced than in higher-dimensional regimes where pruning handles a vastly larger workload. At small $N_s$ or coarse tolerance, the GPU execution curves flatten, indicating that launch latency, synchronization, and other setup overheads begin to dominate the reduced arithmetic workload.

The comparison between the inflated acceleration and jerk benchmarks yields a key algorithmic insight. Across both the $\eta$ and $N_s$ sweeps, the jerk search has nearly the same runtime as the $5\times$ inflated acceleration search. Thus, within the EP framework, moving from an artificially broadened acceleration search to a physically motivated coherent jerk search does not introduce a meaningful execution penalty. This is a central consequence of pruning: the computational cost is dictated by the surviving tree volume rather than the formal dimensionality of the original grid. The snap benchmark remains within a small constant factor of the acceleration and jerk searches on the GPU. At $\eta=1.0$ and $N_s=2^{25}$, the constant snap search is only $\sim$2--3$\times$ slower than the inflated acceleration or jerk benchmark. This behaviour demonstrates that EP makes coherent higher-order polynomial searches computationally accessible: while snap tracking is not cost-free, but its runtime remains close enough to lower-order searches to be operationally more superior option for large-scale survey processing.

The fully coherent circular-orbit benchmark represents the most demanding case in Fig.~\ref{fig:ep_benchmark}. In this regime, an exhaustive enumeration of the corresponding parameter grid would be computationally prohibitive, whereas EP reduces the search to a tractable problem. At the largest configuration $N_s=2^{26}$, the complex Fourier-domain CUDA implementation delivers an approximate $20\times$ reduction in wall-clock time relative to the 8-thread CPU baseline, highlighting the EP effectiveness on GPUs.

The relationship between compute complexity and the tolerance parameter $\eta$ matches analytical expectations, following a clean $\eta^{-4}$ trend in the circular-orbit benchmark. The middle $\eta$ values in particular should not be over-interpreted as a pure power law, because localized adjustments in the chunking layout can introduce non-monotonic efficiency variations at fixed physical search volume. Scaling with observation length $\Tobs$ remains stable and predictable because $\Porb^{\min}$ scales linearly with $\Tobs$ and is most relevant for survey planning.

The Kadane-based overlay, combined with a lower $P_d$, illustrates an additional optimization path. On the CPU backend, this configuration reduces runtime by factors of $\sim2$--5 for the larger polynomial and circular-orbit benchmarks, demonstrating that substantial algorithmic savings are available. On GPU, however, the same configuration is not consistently faster: it improves some of the hardest circular-orbit configurations, but introduces an execution overhead for the polynomial searches. This behaviour indicates that the current CUDA implementation pays additional costs from increased kernel launch frequency, synchronization, or serialization that can outweigh the arithmetic savings. Implementing regime-specific kernel optimizations is therefore likely to recover these latent performance dividends.

These benchmarks demonstrate that the EP algorithm, as implemented in \textsc{LOKI}, successfully resolves the high-dimensional scaling bottleneck, rendering deep, fully coherent binary searches computationally practical for large-scale pulsar surveys.

\section{Discussion}\label{sec:discussion}
The empirical benchmarks presented in Section~\ref{sec:alg_implement} establish that the EP algorithm changes the practical scaling of multi-dimensional pulsar searches. By constraining both the memory footprint and the operational cost of high-order phase-model evaluation, the framework makes fully coherent binary searches feasible over integration lengths relevant to modern surveys. In this section, we discuss the consequences for archival survey reprocessing, near-real-time searches with next-generation facilities, targeted globular-cluster observations, and astrophysical amplitude modulations in real pulsar data. Representative computational requirements and survey-level implications are summarized in Table~\ref{tab:survey_implications}.

\subsection{Implications for archival pulsar surveys}
Large archival datasets, including the High Time Resolution Universe South survey (HTRU-S), the Parkes Multi-beam Pulsar Survey (PMPS), and the LOFAR Tied-Array All-Sky Survey (LOTAAS), contain substantial compact-binary parameter space that has not been searched at full coherent sensitivity \citep{Manchester:2001, Keith:2010, Sanidas:2019}. The continuing yield from archival reprocessing underscores this incompleteness. For example, \citet{Sengar:2025} recently reprocessed the HTRU-S low-latitude survey using a GPU-accelerated TDAS pipeline and discovered 71 pulsars. That analysis used the full 72-min pointings, but was restricted to a constant-acceleration range of $\pm 50$\Accelunits, with a reported total cost of approximately $0.4$\,million GPU-hours. Moreover, although the Doppler correction itself is phase-coherent, the pipeline still relies on incoherent harmonic summing, which degrades sensitivity to short-period, low-duty-cycle MSPs \citep{Morello:2020}.

EP changes this trade-off by allowing the same order of compute budget to be spent on a physically richer coherent signal model rather than only on a wider acceleration grid. For a 72-min pointing ($N_s=2^{26}$ in Fig.~\ref{fig:ep_benchmark}), the benchmarked EP configuration can perform a high-resolution ($\eta=1.0$) $5\times$ expanded coherent acceleration search in roughly $2.5\times$ the computational budget of the \citet{Sengar:2025} reprocessing ($\sim 1.0$\,million GPU-hours), assuming the same number of DM trials. More importantly, this budget can alternatively be deployed to execute a fully coherent jerk search, which effectively improves the coherent orbital-phase coverage from $\lesssim 4$--$10\%$ to $\lesssim 10$--$17\%$ (depending on orbital phase) across all duty cycles. Thus, a modest increase in total compute can be traded for a substantial expansion in physically modelled parameter space.

The tolerance parameter $\eta$ provides a controlled route to reduce this cost further. The trial acceleration step size used in \soft{Peasoup} \citep{Eatough:2013, Morello:2019} and in the TDAS pipeline of \citet{Sengar:2025} corresponds approximately to an effective tolerance of $\eta \approx 4$ in the EP formalism. Operating EP at this coarser resolution reduces the acceleration-search cost by roughly a factor of $\sim \eta^2=16$ relative to the $\eta=1.0$ baseline. Consequently, EP can search a $5\times$ broader acceleration window or a fully coherent jerk space in roughly \emph{one-sixth} of the processing time required by the conventional pipeline, corresponding to $\sim 0.06$\,million GPU-hours under the same scaling assumptions. Since EP natively utilizes exact Fourier-domain phase-coherent folding, it still retains a definitive sensitivity edge across all duty-cycle regimes even when operating at identical parameter-space resolutions.

The physical importance of this capability is straightforward. The constant-acceleration approximation remains valid only when the integration time spans $\lesssim 4$--$10\%$ of the orbital period (depending on orbital phase; see Section~\ref{subsec:circ_orbit_coverage}). For 72-min HTRU-S pointings, this corresponds to full-sensitivity coverage only for wider binary systems with $\Porb \gtrsim 12$--$30$\,h. At shorter orbital periods, unmodelled orbital modulations cause acceleration searches to suffer a factor of $3$--$5\times$ degradation in minimum detectable flux density. EP directly targets this missing regime; by making coherent circular-orbit searches computationally tractable, it recovers the $\Tobs^{1/2}$ sensitivity scaling and opens a discovery window to binaries with $\Porb \gtrsim 72$\,min in the same data.

These systems are not inaccessible in an absolute sense, but they are not searched at full coherent sensitivity by acceleration-based pipelines operating on the same integrations. A blind circular-orbit EP search of the entire HTRU-S low-latitude archive at $\eta=2.0$ would require approximately 85\,million GPU-hours under the baseline configuration benchmarked here. However, utilizing a more efficient multi-run EP configuration with $\nrun=32$ passes using $P_d=0.05$ along with Kadane-based scoring already yields a $2\times$ reduction in execution time. This should be interpreted as a conservative upper bound for the current implementation; further optimisations, such as improved parameter-space gridding, optimised DM-trial placement, and restrictions to astrophysically motivated spin-frequency ranges are expected to reduce the total processing cost substantially. The appropriate conclusion is therefore not that full-orbit blind reprocessing is trivial, but that it has moved from a formally prohibitive problem to a concrete resource-allocation problem.

The PMPS provides a complementary archival case. It remains one of the most successful pulsar surveys ever conducted, with a total yield exceeding 850 pulsars \citep{Knispel:2013}. Its success is due in part to repeated re-analyses with improved search methods. \citet{Knispel:2013} performed a template-bank search and discovered 24 pulsars, although that search was structurally restricted and did not cover the full parameter space at the native data resolution. \citet{Sengar:2023} used a GPU-based acceleration search with \soft{Peasoup} and discovered 37 pulsars. These results show that the archive is not exhausted, particularly for binary systems, which remain under-represented relative to the expected population \citep{Faulkner:2004}. The native PMPS sampling interval, $t_s=250\,\upmu\mathrm{s}$, also reduces the cost of deep coherent processing relative to the 64-$\upmu$s HTRU-S low-latitude data. With EP, a full-resolution coherent circular-orbit search at $\eta=1.0$ in Fourier-domain folding mode would require approximately $0.08$\,million GPU-hours, assuming the same number of DM trials as in \citet{Sengar:2023}. This is a feasible archival campaign and would push the PMPS compact-binary search substantially closer to its instrumental sensitivity limit. In addition to enabling new discoveries, such a search would provide stronger empirical constraints for binary-pulsar population synthesis models.

The same sensitivity gap appears in other long-dwell archives. LOTAAS uses 1-hour pointings, for which the constant-acceleration approximation retains full sensitivity only for binaries with $\Porb \gtrsim 10$--$25$\,h. Because the low observing frequency and large number of tied-array beams make acceleration searches computationally expensive, the published LOTAAS processing did not include a systematic acceleration search \citep{Sanidas:2019}. The compact-binary population accessible to LOTAAS has therefore not yet been systematically explored with acceleration or orbital corrections. An EP-based acceleration or jerk search would provide a natural first reprocessing step, while a circular-orbit EP search would extend the same archive into the ultra-compact regime. 

Shorter-dwell surveys face a milder version of the same problem. The ongoing FAST Galactic Plane Pulsar Snapshot (GPPS) Survey \citep{Han:2021} uses 5-min pointings and has now discovered more than 750 pulsars, including a large population of MSPs and binary systems, while accumulating a multi-petabyte archive over the inner Galactic plane \citep{Han:2025}. For this dataset, a coherent EP acceleration search over a $5\times$ broader acceleration window can be completed in approximately $0.06$\,million GPU-hours assuming 100 DM trials, based on Fig.~\ref{fig:ep_benchmark}. More importantly, EP can execute a coherent jerk search at comparable cost, and a snap search within a small constant factor of the same budget. For short pointings, higher-order coherent searches therefore become realistic first-pass strategies rather than expensive follow-up stages, extending sensitivity to compact systems with $\Porb^{\min}$ of tens of minutes.

In all archival cases, the underlying argument remains the same: the computational cost scales with $\Tobs$, the number of DM trials, and the orbital search volume, but the sensitivity gain comes from recovering coherent integration in regimes where standard acceleration searches lose phase fidelity.

\begin{table*}
\centering
\caption{Survey-scale impact of EP for representative archival and upcoming pulsar searches. Compute estimates are scaled from Section~\ref{sec:ep_benchmarks} and should be read as order-of-magnitude processing costs. The table is intended as a compact guide to the main capabilities discussed in the text.}
\label{tab:survey_implications}
\begin{tabular}{lccccc}
\hline
Survey & $\Tobs$ & Traditional Limits & EP Search Track & EP Configuration & Projected Compute / Notes \\
\hline
\textbf{HTRU-S} & 72 min & $\pm 50$\,\Accelunits & Expanded Accel/Jerk & $\eta=1.0$, $\nrun=16$ & $1.0$\,M GPU-hours ($2.5\times$ cost) \\
                &        &                       & Expanded Accel/Jerk & $\eta\approx4.0$, $\nrun=16$ & $0.06$\,M GPU-hours ($6\times$ faster) \\
                &        &                       & Circular Orbit      & $\eta=2.0$, $\nrun=32$, Kadane & $42$\,M GPU-hours (conservative) \\
\hline
\textbf{PMPS}   & 35 min & Constant Accel        & Circular Orbit      & $\eta=1.0$, $\nrun=16$ & $0.08$\,M GPU-hours (full recovery) \\
\hline
\textbf{FAST-GPPS}& 5 min & Constant Accel       & Expanded Accel/Jerk & $\eta=1.0$, $\nrun=16$ & $0.06$\,M GPU-hours (100 DM trials) \\
\hline
\textbf{SKA1-Mid} & 10 min & $\pm 350$\,\Accelunits, 500 DM trials   & Expanded Accel & $\eta=1.0$, real-time & 120 DM trials ($18\times$ Accel range) \\
                  &        &                                         & Constant Jerk         & $\eta=1.0$, real-time & 120 DM trials (fully coherent) \\
\hline
\end{tabular}
\end{table*}

\subsection{Implications for upcoming pulsar surveys}\label{sec:upcoming_surveys}
Next-generation radio facilities, including the SKA, DSA-2000, MeerKAT and Murriyang cryoPAF systems, will increase survey speed and raw sensitivity while also increasing the data rate that must be searched, triaged, or discarded \citep{Padmanabh:2023, Dunning:2023, Keane:2025}. In this regime, the relevant question is not only whether a compact-binary search is possible offline, but whether it can be executed rapidly enough to preserve the effective sensitivity of the telescope. Search algorithms therefore become part of the observing system: insufficient compute throughput translates directly into lost sensitivity or unsearched parameter space.

The benchmarks in Section~\ref{sec:ep_benchmarks} show that coherent circular-orbit searches, previously treated as computationally prohibitive for blind processing, can be executed on modest GPU clusters for survey-relevant integration lengths. For $\Tobs=5$--$72$\,min and $\Porb^{\min}=\Tobs$, the baseline EP configuration sustains real-time throughput of $\sim\!12$ DM trials down to $\sim\!0.5$ DM trials. Acceleration and jerk searches are substantially cheaper and therefore better matched to continuous real-time survey streams.

A concrete comparison with the SKA1-Mid pulsar-search design illustrates the scale of the gain. The planned SKA pipeline uses an approximate FDAS acceleration search \citep{Ransom:2002} over $\sim100$ acceleration trials spanning $\pm 350$\Accelunits\,applied to 500 DM-corrected time series, with a maximum real-time integration time of 10\,min \citep{Levin:2025, Keane:2025}. FDAS is the dominant computational cost in this pipeline \citep{Levin:2018}. At equivalent integration time, EP in time-domain folding mode processes 120 DM trials over $18\times$ the SKA1-planned acceleration range (or a fully coherent jerk search) in real-time; in the Fourier-domain folding mode, it processes 60 DM trials over the same range. If the real-time buffer is extended to 36\,min ($N_s=2^{25}$ in Figure~\ref{fig:ep_benchmark}), EP delivers similar DM trials throughput; this scaling enables increased buffer capacity for real-time systems at the same cost, while yielding increased sensitivity with longer $\Tobs$. These figures correspond to the high resolution $\eta=1.0$ configuration and can be tuned according to the required DM spacing and candidate-recovery tolerance.

This makes EP useful in two distinct real-time roles. First, it can provide a drop-in route to wider coherent acceleration searches in survey pipelines where FDAS or TDAS currently sets the compute budget. Second, it enables a tiered strategy in which the bulk survey stream is searched with acceleration or jerk EP, while selected high-priority beams, targets, or candidate-rich regions are searched with the circular-orbit EP mode. Such a hierarchy aligns the computational effort with the astrophysical value, ensuring that increased telescope sensitivity is not lost at the search stage for the most compact binaries. 

Longer real-time buffers strengthen the case further: If a survey system can retain tens of minutes of baseband voltage data, EP can leverage that extended duration coherently rather than forcing the search into segmented acceleration approximations. The result is a direct sensitivity gain, because longer $\Tobs$ improves detectability only if the phase model remains valid over the integration span. This is precisely the regime where conventional acceleration searches saturate and where EP provides its largest return.

\subsection{Implications for globular cluster searches}\label{sec:globular_clusters}
Globular clusters (GCs) are among the most prolific sites of exotic pulsar formation in the Galaxy. Their high stellar densities drive exchange interactions and repeated binary encounters, producing compact MSP binaries with white-dwarf, neutron-star, or low-mass degenerate companions. To date, 345 radio pulsars have been confirmed in Galactic GCs, and population studies suggest that the detected sample remains incomplete by more than an order of magnitude \citep{Turk:2013, Bagchi:2025}. The limiting factor is no longer only telescope sensitivity: fast-spinning pulsars in compact, highly accelerated orbits are precisely the systems most vulnerable to coherence loss in conventional searches. 

GCs are therefore a natural application of EP. Unlike blind all-sky surveys, GC observations involve few beam pointings and a narrow DM range, typically spanning only a few pc\,cm$^{-3}$. The outer enumeration over sky position and DM trials, which dominates many survey-processing costs, is therefore greatly reduced. The computational budget can instead be directed toward the orbital parameter grid, where EP delivers its largest relative gain through hierarchical pruning of unpromising parameter-space branches. Moreover, as discussed in Section~\ref{sec:param_transition}, the EP framework extends directly to $\Tobs > \Porb^{\min}$, enabling coherent accumulation over multiple orbital cycles rather than forcing the observation to be segmented. 

Terzan~5 illustrates the scale of the opportunity. With 49 confirmed MSPs, of which 29 are in binary systems, it hosts the richest known pulsar population of any Galactic GC \citep{Padmanabh:2024}. Population estimates suggest that Terzan~5 may contain $\gtrsim 100$ pulsars in total \citep{Bagchi:2011, Chennamangalam:2013, Martsen:2022}. For example, one archival GBT dataset alone contains $\sim 206$\,h of high-time-resolution data \citep{Cadelano:2018}, and the cluster DM is sufficiently well constrained that only $\sim 20$ dispersion trials are required for a focused search. Processed in 72-min coherent chunks, a circular-orbit EP search down to $\Porb^{\min} = 72$\,min requires approximately 8000 GPU-hours for the complete dataset, using the benchmark scaling in Figure~\ref{fig:ep_benchmark}. The high time resolution of these data also permits extension of the spin-frequency axis into the sub-millisecond regime, directly testing a region of parameter space that is especially relevant for Terzan~5, where the ultra-fast pulsars ($f > 500$\,Hz) are concentrated \citep{Bagchi:2025}.

A coherent EP reprocessing campaign would therefore test two discovery spaces simultaneously: ultra-compact binaries and ultra-fast MSPs. Both are astrophysically valuable, and both are exactly the regimes in which segmented or incoherent searches lose sensitivity. The $\Tobs$ scaling in Figure~\ref{fig:ep_benchmark} further shows that several-hour coherent integrations requires no change in the underlying algorithm. Modern GPUs with tens of GB of on-board memory can accommodate the relevant search tree through the chunking strategy described in Section~\ref{sec:freq_search}, allowing coherent S/N to be recovered without segmentation loss. The combination of narrow DM range, small number of pointings, and deep integrations makes Terzan~5 an obvious first target for systematic EP reprocessing, and the same argument extends to other GCs with high-quality archival data. EP therefore provides a computationally accessible path to the ultra-compact GC binaries most relevant for strong-field gravity, dense-matter physics, and binary-evolution tests.

\subsection{Astrophysical signal modulations}\label{sec:signal_modulations}
The EP framework assumes a stationary signal whose amplitude remains constant across the observation span. Real pulsar signals are subject to intrinsic and propagation-induced modulations that violate this assumption, making it vital to examine how EP responds to each. 

The two most relevant amplitude-modulating effects for compact-binary searches are pulse nulling and radio eclipses. In pulse nulling, the radio emission ceases for intervals ranging from a few spin periods to hours before resuming \citep{Backer:1970}. Observed nulling statistics are affected by selection and sensitivity biases, but $\gtrsim8\%$ of known pulsars show nulling behaviour, with nulling fractions ranging from a few percent to nearly unity in extreme cases \citep{Sheikh:2021}. Eclipses are distinct in origin but identical in consequence: in black widow and redback systems, ionised material ablated from the companion can obscure the pulsar signal for $\gtrsim 10\%$ of the orbit at GHz frequencies, and often for larger fractions at lower frequencies \citep{Fruchter:1988, Thompson:1994}. In both cases, the effect on a whole-observation fold is the same: integrating over the inactive or eclipsed intervals accumulates noise without signal, reducing the coherent S/N in proportion to the inactive fraction.

The hierarchical structure of EP provides a natural mitigation against such intermittency. Because candidates are evaluated after each segment accumulation, a source that is visible for only part of the observation can still exceed the final detection threshold during the active segments. Once such a candidate is saved, subsequent verification can refine the search parameters using the specific subset of data in which the signal is present, instead of forcing the final statistic to integrate over inactive or eclipsed intervals. For a source visible for half of the observation, the idealised sensitivity penalty is therefore the expected factor of $\sqrt{2}$ in S/N relative to a continuously visible source, without the additional degradation caused by folding through long signal-free intervals. This selective accumulation is a consequence of the multi-pass design and requires no modifications to the core EP algorithm.

Interstellar scintillation and pulsar spectral indices introduce a separate class of effects that the present EP implementation does not address, by construction. The current pipeline operates on a single DM-corrected, frequency-averaged time series; therefore, amplitude modulations across the observing frequency band whether from scintillation \citep{Rickett:1990} or by the intrinsic radio spectrum of the pulsar, are therefore averaged before EP processing begins. Pulsar spectra are typically steep and diverse, with population studies finding mean spectral indices near $\langle \alpha \rangle \approx -1.6$ along with a tail extending to much steeper values \citep{Bates:2013, Jankowski:2018}. This is a deliberate scope limitation rather than a fundamental restriction. Extending EP to operate on sub-banded data would enable a joint search over DM and spectral index, improving sensitivity to sources that are sub-threshold in the band-averaged series but detectable under an appropriate spectral model. This is a well-defined direction for future development.

\section{Summary and Conclusions}\label{sec:conclusion}
We have presented the Extreme Pruning (EP) algorithm, a framework that addresses a long-standing bottleneck in time-domain radio astronomy: the severe computational scaling of fully coherent pulsar searches over circular binary orbits and high-order polynomial phase models. By combining a hierarchical dynamic programming structure with statistically controlled pruning, EP reserves the full coherent search only for parameter-space regions that remain consistent with a physical signal. The result is a practical route to coherent acceleration, jerk, higher-order polynomial, and circular-orbit searches using both time-domain and Fourier-domain folding modes. 

The benchmarks in Section~\ref{sec:ep_benchmarks} demonstrate that these algorithmic gains translate into practical performance. For circular-orbit searches, EP reduces the computational cost by up to ten orders of magnitude relative to exhaustive enumeration while preserving the exact coherent search statistic. The tolerance parameters provide an additional controlled trade-off between runtime and sensitivity, allowing the same framework to support both high-sensitivity searches and faster survey-scale processing. The key practical result is that coherent jerk searches can be executed at essentially the same cost as an inflated acceleration search, while snap and circular-orbit searches become accessible within well-defined computational budgets.

The scientific stakes of this algorithmic capability are commensurate with the computational investment. For decades, survey design has been forced into a compromise between short dwell times which limit raw sensitivity, or long integrations ($\Tobs$) which suffer severe coherence degradation under the constant-acceleration approximation. EP dismantles this paradigm. By recovering the optimal $\Tobs^{1/2}$ sensitivity scaling, a circular-orbit search yields a $3$--$5\times$ improvement in minimum detectable flux density for ultra-compact systems $\Porb \sim \Tobs$. For an idealised, isotropic 3D spatial distribution, this flux sensitivity translates to a detectable volume increase of $5$--$11\times$, subject to the usual caveats of survey selection effects, luminosity functions, beaming, and population incompleteness. This offers an optimal algorithmic pathway to discover the faint, ultra-compact double neutron star and neutron star--black hole binaries which are among the most valuable laboratories for strong-field gravity. 

The broader algorithmic structure is not specific to radio pulsars. EP applies whenever a search can be represented as an exhaustive enumeration over a structured, approximately linear parameter space where the detection statistic can be bounded or evaluated over data sub-segments. The pulsar-search implementation developed here is therefore both a functional astrophysical tool and a proof of principle for a wider class of coherent inference problems. For pulsar astronomy, the conclusion is straightforward: fully coherent searches over compact-binary phase models in existing and future survey data are now completely tractable; translating this capability into discoveries is now simply a question of computational resource allocation.

\section*{Acknowledgements}
BZ conceived the original algorithmic idea and developed an early prototype. PK designed and implemented the full algorithm, performed all analyses, and wrote the manuscript. Both authors contributed to interpretation and revision.

We thank Aaron Pearlman and Saif Ali for helpful comments on the manuscript. PK gratefully acknowledges the support of Maria Alessandra Papa and Bruce Allen, and the hospitality of AEI Hannover during part of this work. PK also thanks Dotan Gazith, Vivek Venkatraman Krishnan and Rahul Sengar for useful discussions.
PK and BZ are supported by the Schwartz Reisman Collaborative Science Program, which is supported by the Gerald Schwartz and Heather Reisman Foundation.
PK and BZ are supported by the Minerva Foundation with funding from the Federal German Ministry for Education and Research.
This project has been made possible in part by a grant from the SETI Institute. 
This research has made use of NASA's Astrophysics Data System Bibliographic Services and software packages, including: CUDA Toolkit, \soft{cuFFT}, \soft{NVIDIA CCCL} \citep{CCCL:2023}, 
\soft{FFTW} \citep{FFTW:2005}, \soft{NumPy} \citep{Harris:2020_numpy}, \soft{Numba} \citep{Lam:2015}, and \soft{matplotlib} \citep{Hunter:2007_matplotlib}.

\section*{Data availability}
No new data were generated or analysed in support of this research. The reference Python implementation and the production implementation of the algorithm are available at \url{https://github.com/pravirkr/pyloki} and \url{https://github.com/pravirkr/loki}, respectively. Scripts used to generate the figures are available from the corresponding author upon reasonable request.

\appendix

\section{Phase-Folding with Optimal Weights}\label{app:weighted_fold}
We derive the optimal weighting scheme for phase folding under heteroscedastic noise, assuming that the signal amplitude scales with the local mean level. We model the time-series data $\mathcal{T}_n$ as:
\begin{equation}
    \mathcal{T}_n = \mu_n (1 + s'_{\hat{b}_n}) + \epsilon_n, \quad \epsilon_n \sim \mathcal{N}(0, \sigma^2_n),
\end{equation}
where $\mu_n$ is the local mean level, $s'_{\hat{b}_n}$ is the dimensionless fractional signal strength in the phase bin $\hat{b}_n$ assigned to time $t_n$, and $\epsilon_n$ is independent Gaussian noise with time-dependent variance $\sigma^2_n$. Our goal is to estimate the fractional signal $s'_b$ in a specific bin $b$. Up to an additive constant, the log-likelihood for the subset of samples assigned to bin $b$ is:
\begin{equation}
    \ln \mathcal{L}(s'_b) = -\frac{1}{2} \sum_{n=0}^{N_s-1} \frac{(\mathcal{T}_n - \mu_n - s'_b \mu_n)^2}{\sigma^2_n} \delta_{b, \hat{b}_n},
\end{equation}
where the Kronecker delta $\delta_{b, \hat{b}_n}$ restricts the sum to samples falling in phase bin $b$. Maximizing with respect to $s'_b$ gives the maximum-likelihood estimator
\begin{equation}\label{eq:mle_fractional}
    \hat{s}'_b = \frac{\sum_{n} (\mathcal{T}_n - \mu_n) \frac{\mu_n}{\sigma^2_n} \delta_{b, \hat{b}_n}}{\sum_{n} \frac{\mu^2_n}{\sigma^2_n} \delta_{b, \hat{b}_n}} = \frac{\mathcal{P}_w(b)}{\mathcal{P}_s(b)}.
\end{equation}
Here, the numerator and denominator correspond to the folded weighted profiles $\mathcal{P}_w(b)$ and $\mathcal{P}_s(b)$, respectively. The variance of this estimator is:
\begin{equation}
    \mathrm{Var}(\hat{s}'_b) = \left( \sum_{n} \frac{\mu^2_n}{\sigma^2_n} \delta_{b, \hat{b}_n} \right)^{-1} = \frac{1}{\mathcal{P}_s(b)}.
\end{equation}
The normalized profile defined in equation~\eqref{eq:norm_profile} therefore represents the per-bin signal-to-noise ratio of the fractional-amplitude estimator:
\begin{equation}
    \mathcal{P}_{\mathrm{norm}}(b) = \frac{\mathcal{P}_w(b)}{\sqrt{\mathcal{P}_s(b)}} = \frac{\hat{s}'_b}{\sqrt{\text{Var}(\hat{s}'_b)}},
\end{equation}
so that the $\SNR_\alpha$ statistic in equation~\eqref{eq:folded_snr} is a weighted sum of these per-bin significances.

To compute the overall significance of a pulse profile matching a template shape $T(b)$, we model the expected fractional signal in each bin as $s'_b = A\,T(b)$, where $A$ is the overall amplitude and $T(b)$ is a normalized template satisfying $\sum_b T(b)^2 = 1$. The minimum-variance estimate of $A$ is obtained by combining the per-bin estimates $\hat{s}'_b$ with inverse-variance weights:
\begin{equation}
    \hat{A} = \frac{\sum_{b} \hat{s}'_b\,T(b)\,\mathcal{P}_s(b)}{\sum_{b} T(b)^2\,\mathcal{P}_s(b)} = \frac{\sum_{b} \mathcal{P}_w(b)\,T(b)}{\sum_{b} \mathcal{P}_s(b)\,T(b)^2}.
\end{equation}
The corresponding variance is
\begin{equation}
    \text{Var}(\hat{A}) = \left( \sum_{b} T(b)^2\,\mathcal{P}_s(b) \right)^{-1}.
\end{equation}
The optimal signal-to-noise ratio is therefore
\begin{equation}
    \mathrm{S/N}_{\hat{A}} = \frac{\hat{A}}{\sqrt{\mathrm{Var}(\hat{A})}} = \frac{\sum_{b} \mathcal{P}_w(b)\,T(b)}{\sqrt{\sum_{b} \mathcal{P}_s(b)\,T(b)^2}},
\end{equation}
confirming the optimal detection statistic $\SNR_\beta$ in equation~\eqref{eq:optimal_snr}. This expression is identical to the standard matched-filter statistic for a known template in Gaussian noise. The difference between $\SNR_\alpha$ and $\SNR_\beta$ becomes most important when the weight profile $\mathcal{P}_s(b)$ varies substantially across phase bins, in which case the inverse-variance weighting in $\SNR_\beta$ is required for optimal sensitivity.

\section{Taylor basis transformation}\label{app:taylor_basis}
When searching over orbital parameters in a Taylor basis, shifting the kinematic parameters $d_{k}$ from one reference time to another is frequently necessary. This operation is equivalent to re-centring the Taylor expansion. We store the coefficient vector in descending derivative order,
\begin{equation}
    \bm d =
    \begin{bmatrix}
        d_{k_{\max}} \\
        \vdots \\
        d_0
    \end{bmatrix}.
\end{equation}
To shift the reference epoch from $t_i$ to $t_j = t_i + \Delta t$, we substitute $t - t_i = (t - t_j) + \Delta t$ into the expansion. By the binomial theorem,
\begin{equation}
    (t - t_i)^k = \sum_{m=0}^{k} \binom{k}{m} (t - t_j)^m (\Delta t)^{k-m}.
\end{equation}
Collecting like powers of $(t-t_j)$ gives a linear map between the derivative vectors $\bm{d}_{j}$ and $\bm{d}_{i}$,
\begin{equation}\label{eq:transform}
    \bm{d}_{j} = \mathbf{T}(\Delta t) \bm{d}_{i},
\end{equation}
where $\mathbf{T}(\Delta t)$ is a lower-triangular $(\kmax + 1) \times (\kmax + 1)$ transformation matrix in the descending-order basis. If $a$ and $b$ denote row and column indices in this stored vector, with $a,b=0,\dots,\kmax$, then the corresponding derivative orders are $\kmax-a$ and $\kmax-b$. The matrix elements are
\begin{align}\label{eq:transform_matrix}
    T_{a,b}(\Delta t) =
    \begin{cases}
        \dfrac{(\Delta t)^{a-b}}{(a-b)!} & a \geq b,\\
        0 & a < b.
    \end{cases}
\end{align}
This transformation preserves the underlying motion $d(t)$ while changing only the reference epoch. For error (grid size) propagation under a change of reference epoch, a conservative approach accounts for the mixing of higher derivatives into lower ones, while an aggressive approach ignores the off-diagonal contributions and retains only $T_{k,k}=1$.

\section{Middle-Out Folding Scheme}\label{app:middle_out}
The middle-out folding scheme defines the family of segment accumulations order used by the EP algorithm. Given an anchor segment $q \in \{0, \dots, M-1\}$, the mapping $\mathcal{J}(s,q)$ specifies the order in which segments are processed at accumulation stage $s$.

We derive this mapping by sorting the segment index set $j \in \{0, \dots, M-1\}$ by increasing distance from the anchor, $|j-q|$, with ties broken in favour of smaller indices. Equivalently,
\begin{equation}
    \mathcal{J}(s,q) = \left(\operatorname*{sort}_{j \in \{0, \dots, M-1\}}\bigl[ (|j-q|, j) \bigr]\right)_{s},
\end{equation}
where the sorting is lexicographic in the tuple $(|j-q|,j)$ and $(\cdot)_s$ denotes the $s$-th element of the sorted sequence. This yields the corresponding middle-out ordering
\begin{equation}
    q, \; q-1, \; q+1, \; q-2, \; q+2, \; q-3, \; q+3, \; \dots,
\end{equation}
truncated to the valid index range $[0, M-1]$. Segments nearest the reference anchor are processed first, with progressively more distant segments added symmetrically. For example, with $M = 8$ segments and anchor $q = 3$,
\begin{equation}
    \mathcal{J}(s, 3) = (3, 2, 4, 1, 5, 0, 6, 7) \quad \text{for } s = 0, \dots, 7.
\end{equation}

For interior anchors ($0 < q < M-1$), this mapping initially generates an alternating expansion that symmetrically incorporates segments from both sides of the anchor. Once a boundary segment ($j=0$ or $j=M-1$) is reached, the sequence becomes a unidirectional sweep across the remaining segments. The extreme boundary configurations represent unique cases where no alternation occurs at all: $q=0$ defines a strictly monotonic \emph{edge-forward} traversal, $\mathcal{J}(s, 0) = s$, while $q=M-1$ defines a strictly monotonic \emph{edge-backward} traversal, $\mathcal{J}(s, M-1) = M-1-s$.

\subsection{Start-Epoch Gauge Bias Correction}\label{app:gauge_fix}
In the multi-run EP algorithm (Section~\ref{sec:ep_multi}), each run is initialized at a distinct anchor segment $q$ with corresponding epoch $t_{C,q}$. For implementation convenience, we adopt a \emph{run gauge} at the start epoch in which the line-of-sight velocity is set to zero:
\begin{equation}
    d_1(t_{C,q}) = 0.
\end{equation}
The frequency grid at that epoch then defines the reference frequency, $f_{q} = f(t_{C,q})$, and the search state is propagated in terms of the distance-derivative tuple $\{d_k\}_{k\geq 1}$, with the frequency grid offset represented through the velocity parameter $d_1$.

After accumulating $M$ segments, each pruning run returns a set of best-fit parameters $\{d_k^{\mathrm{final}}\}$ evaluated at the observation midpoint $t_{C,M-1} \equiv \tmid = \Tobs/2$, together with the stored reference frequency $f_q$. However, the physical observed frequency at $\tmid$ is generally different from the start-epoch frequency, so $f(\tmid) \neq f_q$. Consequently, the tuple $\{d_k^{\mathrm{final}}\}$ does not define a unique physical parameter set unless a consistent gauge is specified.

Interpreted naively, this mismatch introduces a systematic, start-epoch-dependent offset in the reported parameters, with a leading dependence tied to $(t_{C,q} - t_{C,M-1})$. This effect is not physical: it reflects the inherent degeneracy between instantaneous frequency and line-of-sight velocity. Different start epochs implicitly select different gauges, producing inconsistent reported parameters across pruning runs and complicating multi-run aggregation.

To remove this ambiguity, we transform all run outputs to a common \emph{report gauge} defined at the observation midpoint $\tmid$. By convention, we impose
\begin{equation}
    d_1^{\mathrm{report}}(\tmid) = 0,
\end{equation}
thereby fixing the gauge at a common epoch for all runs. The Doppler scale factor relating the start-epoch and midpoint frequencies is
\begin{equation}\label{eq:Sdef}
    S \equiv \frac{f_{\mathrm{obs}}(\tmid)}{f_q} = 1 - \frac{d_1^{\mathrm{final}}(\tmid)}{c}.
\end{equation}

At the common epoch $\tmid$, the observable frequency derivatives must be invariant under this transformation. Using the relation $f_k = -(f_{\mathrm{ref}}/c) d_{k+1}$, we obtain the algebraic mapping for the higher-order kinematic terms,
\begin{equation}\label{eq:map_dk}
    d_k^{\mathrm{report}}(\tmid) = \frac{d_k^{\mathrm{final}}(\tmid)}{S}, \quad k \geq 2,
\end{equation}
and for the reported frequency,
\begin{equation}\label{eq:map_f}
    f_{\mathrm{report}} = S \, f_q.
\end{equation}

Equations~\eqref{eq:Sdef}--\eqref{eq:map_f} are exact within the non-relativistic Doppler model. Although the individual quantities $\{f, d_k\}$ are gauge-dependent, the combinations entering the observable frequency derivatives remain invariant. Applying this transformation at the conclusion of each pruning run ensures that reported parameters from all runs are start-epoch invariant and can be aggregated without gauge-dependent offsets.

\section{Orthogonal polynomial basis}\label{app:chebyshev_basis}
While the monomial basis $(t - t_c)^k$ used in Section~\ref{subsec:poly_phase_model} is conceptually simple, it can lead to strong correlations between search parameters in practice. An alternative is to express the line-of-sight distance $d(t)$ in an \emph{orthogonal polynomial} basis. Two widely used families in numerical approximation theory are the Legendre and Chebyshev polynomials, both defined on the dimensionless domain $x \in [-1, 1]$. Here, we focus on the Chebyshev polynomials of the first kind, denoted $T_n(x)$, which have favourable conditioning properties. In particular, among all monic polynomials of degree $n$, $2^{1-n}T_n(x)$ has the smallest maximum absolute value on $x \in [-1, 1]$. The Chebyshev polynomials satisfy the three-term recurrence relation
\begin{align}\label{eq:cheb_recur}
T_0(x) &= 1, \\
T_1(x) &= x, \\
T_{n+1}(x) &= 2xT_n(x) - T_{n-1}(x),
\end{align}
and are orthogonal with respect to the weight function $w(x) = 1/\sqrt{1 - x^2}$ over $x \in [-1, 1]$:
\begin{equation}
\int_{-1}^{1} \frac{T_m(x)T_n(x)}{\sqrt{1-x^2}} dx = 
    \begin{cases} 
    0 & m \neq n, \\ 
    \pi/2 & m = n > 0, \\ 
    \pi & m = n = 0. 
    \end{cases}
\end{equation}
To represent the pulsar--observer distance $d(t)$ over a finite observing window, we map the time coordinate $t$ to a dimensionless variable $x \in [-1, 1]$ via:
\begin{equation}\label{eq:time_map}
    x = \frac{t - t_c}{t_s},
\end{equation}
where $t_c$ is the central time and $t_s$ is the half-span of the observation window. This maps $t \in [t_c - t_s,\, t_c + t_s]$ to $x \in [-1, 1]$. The line-of-sight distance is then expanded as:
\begin{equation}\label{eq:cheby_d_app}
    d(t) = \sum_{k=0}^{\kmax} \alpha_k T_k\bigparen{\frac{t - t_c}{t_s}},
\end{equation}
where $\{\alpha_k\}$ are the Chebyshev coefficients. Because the basis is orthogonal and well conditioned on the scaled interval, these coefficients typically exhibit reduced correlations compared to those in a monomial expansion. Higher-order terms can therefore be added with less risk of parameter degeneracy than in the monomial basis.

\subsection{Taylor--Chebyshev transformation}\label{app:cheb_taylor}
To leverage the benefits of both representations, we transform between Taylor coefficients $\{d_k\}_{k=0}^{k_{\max}}$ and Chebyshev coefficients $\{\alpha_k\}_{k=0}^{k_{\max}}$. Using the mapping $x = (t - t_c)/t_s$, we equate
\begin{equation}\label{eq:trans1}
    d(t) = \sum_{k=0}^{k_{\max}} \frac{d_k}{k!}(t - t_c)^k = \sum_{k=0}^{k_{\max}} \alpha_k T_k\bigparen{\frac{t - t_c}{t_s}}.
\end{equation}
To perform this transformation, we express powers of $x$ in terms of Chebyshev polynomials. A classical identity \citep{Mason2002} gives
\begin{equation}\label{eq:power_to_cheb}
    x^k = 2^{1-k}\sideset{}{'}\sum_{m=0}^{k}\binom{k}{\frac{k - m}{2}}T_m(x),
\end{equation}
where the prime on the summation indicates that the first term ($m=0$) carries an additional factor of $1/2$, and only terms with even $k-m \geq 0$ are included. We express this identity more systematically by introducing connection coefficients
\begin{equation}\label{eq:connection_S}
S_{k,m} =
\begin{cases}
2^{1 - k - \delta_{m0}}
\binom{k}{\tfrac{k - m}{2}},
& \substack{
(k - m)\ \text{even},\\
0 \le m \le k
} \\[4pt]
0, & \text{otherwise}.
\end{cases}
\end{equation}
where $\delta_{m0}$ is the Kronecker delta. This yields the equivalent representation
\begin{equation}
    x^k = \sum_{m=0}^{k} S_{k,m}T_m(x).
\end{equation}
Substituting into equation~\eqref{eq:trans1} and reordering summations gives
\begin{equation}
\begin{aligned}
d(t)
&= \sum_{k=0}^{k_{\max}} \frac{d_k t_s^k}{k!}
   \sum_{m=0}^{k} S_{k,m} T_m(x) \\
&= \sum_{k=0}^{k_{\max}}
   \biggl[
   \sum_{m = k}^{k_{\max}} \frac{d_m t_s^m}{m!} S_{m,k}
   \biggr] T_k(x),
\end{aligned}
\end{equation}
from which the forward transformation follows:
\begin{equation}\label{eq:final_transform_cheb}
\alpha_k = \sum_{m = k}^{k_{\max}} \frac{d_m t_s^m}{m!} S_{m,k}.
\end{equation}
For $k_{\max}=5$, the coefficients reduce to
\begin{align}
\alpha_5 &= \frac{d_5 t_s^5}{1920}, \\
\alpha_4 &= \frac{d_4 t_s^4}{192}, \\
\alpha_3 &= \frac{d_3 t_s^3}{24} + \frac{d_5 t_s^5}{384}, \\
\alpha_2 &= \frac{d_2 t_s^2}{4} + \frac{d_4 t_s^4}{48}, \\
\alpha_1 &= d_1 t_s + \frac{d_3 t_s^3}{8} + \frac{d_5 t_s^5}{192}, \\
\alpha_0 &= d_0 + \frac{d_2 t_s^2}{4} + \frac{d_4 t_s^4}{64}.
\end{align}

The inverse transformation from Chebyshev coefficients back to the Taylor coefficients is also of interest. We express Chebyshev polynomials as power series,
\begin{equation}\label{eq:cheb_to_power}
    T_k(x) = \sum_{m=0}^{k} R_{k,m} x^m,
\end{equation}
with connection coefficients \citep{Mason2002}
\begin{equation}\label{eq:connection_R}
R_{k,m} =
\begin{cases}
(-1)^{k/2},
& \substack{
m = 0,\\
k\ \text{even}
} \\[4pt]
(-1)^{\tfrac{k-m}{2}}
2^{m-1}
\frac{2k}{k+m}
\displaystyle\binom{\tfrac{k+m}{2}}{\tfrac{k-m}{2}},
& \substack{
(k-m)\ \text{even},\\
0 < m \le k
} \\[4pt]
0, & \text{otherwise}.
\end{cases}
\end{equation}
Substituting into equation~\eqref{eq:cheby_d_app} and collecting powers of $x$ gives
\begin{equation}\label{eq:final_transform_power}
    d_k = \frac{k!}{t_s^k} \sum_{m = k}^{k_{\max}} \alpha_m\,R_{m,k}.
\end{equation}
For $k_{\max}=5$, this becomes
\begin{align}
    d_0 &= \alpha_0 - \alpha_2 + \alpha_4 \nonumber,\\
    d_1 &= \frac{1}{t_s}(\alpha_1 - 3\alpha_3 + 5\alpha_5) \nonumber,\\
    d_2 &= \frac{4}{t_s^2}(\alpha_2 - 4\alpha_4) \nonumber,\\
    d_3 &= \frac{24}{t_s^3}(\alpha_3 - 5\alpha_5) \nonumber,\\
    d_4 &= \frac{192}{t_s^4}\alpha_4 \nonumber,\\
    d_5 &= \frac{1920}{t_s^5}\alpha_5.
\end{align}

Equivalently, the Taylor coefficients can be obtained by directly evaluating the $k$-th derivative of the Chebyshev series at $t = t_c$, or $x = 0$:
\begin{equation}\label{eq:derivative_method}
    d_k = \sum_{m=0}^{\kmax} \frac{\alpha_m}{t_s^k}\left.\frac{d^k T_m(x)}{dx^k}\right|_{x = 0}.
\end{equation}

\subsection{Optimal Gridding via Orthogonalization}\label{app:optimal_gridding}
A principal advantage of an orthogonal basis is that it provides a well-conditioned coordinate system for constructing a computationally efficient search grid. In the Chebyshev basis, correlations among the coefficients $\{\alpha_k\}$ are substantially reduced compared to the monomial basis. Consequently, a phase error from a mismatch $\Delta\alpha_k$ is less readily compensated by adjustments in other coefficients $\Delta\alpha_m$.

The gridding criterion from equation~\eqref{eq:grid_criteria} requires the maximum phase error to remain below the tolerance $\delta_\Phi = \eta/N_b$. In the Chebyshev representation, a mismatch $\Delta\alpha_k$ in a single coefficient produces a phase deviation $\Delta\Phi(t) = -(f_{\mathrm{int}}/c)\Delta\alpha_k T_k(x)$. Since $|T_k(x)|\le 1$ for $x\in[-1,1]$, the criterion imposes an independent bound on each Chebyshev coefficient:
\begin{equation}\label{eq:cheb_grid_alpha_app}
    |\Delta\alpha_k| \lesssim \frac{c}{f_{\mathrm{int}}}\frac{\eta}{N_b}.
\end{equation}
This defines a uniform grid spacing $\delta_\alpha$ for each Chebyshev coefficient.

The key step is mapping this bound back to the physical Taylor parameters $\{d_k\}$. The transformation in equation~\eqref{eq:final_transform_cheb} demonstrates that a variation in $d_m$ cascades into  multiple Chebyshev coefficients $\alpha_k$ for $k \le m$. To establish an optimal grid spacing, we isolate the uncompensable error component. In a top-down approach, the variation $\Delta d_k$ uniquely determines the highest-order orthogonal term $\alpha_k$, as lower-order adjustments ($d_{m<k}$) cannot absorb it. From the transformation matrix in equation~\eqref{eq:final_transform_cheb}, the direct contribution of $d_k$ to $\alpha_k$ is:
\begin{equation}
    \Delta\alpha_k = \frac{\partial \alpha_k}{\partial d_k} \Delta d_k = \frac{t_s^k}{k!}S_{k,k} \Delta d_k,
\end{equation}
where $t_s$ is the half-span of the observation. Requiring this component to satisfy the bound in equation~\eqref{eq:cheb_grid_alpha_app}, and substituting $S_{k,k} = 2^{1-k}$ for $k \ge 1$, yields the optimal grid spacing
\begin{equation}\label{eq:dk_optimal_app}
    \Delta d_k^{\rm opt} = \frac{2^{k-1} c}{f_{\mathrm{int}}}\frac{\eta}{N_b}\frac{k!}{t_s^k}, \quad k \ge 2.
\end{equation}
This grid is coarser than the naive monomial-basis spacing by a factor of $2^{k-1}$ for each derivative order $k \geq 2$. The resulting reduction in the number of required grid points is critical for making higher-order polynomial searches computationally tractable.

\subsection{Chebyshev domain transformation}\label{app:cheb_domain_transformation}
In practice, as observations accumulate over extended time spans, the polynomial representation must be updated to maintain numerical accuracy. This requires transforming Chebyshev coefficients between different temporal domains. Consider a function expanded in Chebyshev polynomials over two domains,
\begin{equation}
\begin{aligned}
t &\in [t_{c,1} - t_{s,1},\, t_{c,1} + t_{s,1}], \\
t &\in [t_{c,2} - t_{s,2},\, t_{c,2} + t_{s,2}].
\end{aligned}
\end{equation}
Each of these expansions, when re-expressed in terms of the monomial basis, can be viewed as
\begin{equation}\label{eq:matrix_form}
    d(t) = \bm{\alpha}_1^\top \mathbf{M}_1 = \bm{\alpha}_2^\top \mathbf{M}_2
\end{equation}
where $\bm{\alpha}_i$ is the $(k_{\max}+1)$-dimensional vector of Chebyshev coefficients and $\mathbf{M}_i$ is the matrix that maps Chebyshev coefficients in domain $i$ to monomial coefficients in the common basis. The corresponding coefficient transformation is
\begin{equation}\label{eq:alpha1-alpha2}
    \bm{\alpha}_2^\top = \bm{\alpha}_1^\top \mathbf{M}_1\mathbf{M}_2^{-1}.
\end{equation}
However, this matrix approach can become numerically unstable for high-order polynomials or after multiple successive transformations. We therefore use a direct transformation method.

Consider the same function expressed in two coordinate systems:
\begin{equation}\label{eq:two_expansions}
    \sum_{k=0}^{k_{\max}} \alpha_{1,k} T_k\bigparen{\frac{t-t_{c,1}}{t_{s,1}}} = 
           \sum_{k=0}^{k_{\max}} \alpha_{2,k} T_k\bigparen{\frac{t-t_{c,2}}{t_{s,2}}}
\end{equation}
The key step is to express the Chebyshev polynomials from one domain in the basis of the other. We define the scaling parameters:
\begin{equation}\label{eq:scale_params}
    p = \frac{t_{s,2}}{t_{s,1}}, \quad q = \frac{t_{c,2}-t_{c,1}}{t_{s,1}}
\end{equation}
which relate the two dimensionless coordinates through
\begin{equation}
    \frac{t-t_{c,1}}{t_{s,1}} = p\frac{t-t_{c,2}}{t_{s,2}} + q
\end{equation}
The linear transformation problem then reduces to expressing $T_n(px + q)$ in the target-domain Chebyshev basis:
\begin{equation}\label{eq:transform_basis}
    T_n(px + q) = \sum_{k=0}^n C_{n,k}(p,q)T_k(x)
\end{equation}
where $C_{n,k}(p,q)$ are the transformation coefficients. Using the power series representation of Chebyshev polynomials from Appendix~\ref{app:cheb_taylor} and the connection coefficients $R_{k,m}$ and $S_{i,k}$ defined above,
\begin{align}
    T_n(px + q) &= \sum_{m=0}^n R_{n,m}(px + q)^m \nonumber\\
    &= \sum_{m=0}^n R_{n,m}\sum_{i=0}^m \binom{m}{i}p^i q^{m-i}\sum_{k=0}^i S_{i,k}T_k(x) \nonumber\\
    &= \sum_{k=0}^n \bigbracket{\sum_{m=k}^n R_{n,m} \sum_{i=k}^m \binom{m}{i}p^i q^{m-i} S_{i,k}} T_k(x)
\end{align}
This gives the transformation coefficients
\begin{equation}\label{eq:transform_coeff}
    C_{n,k}(p,q) = \sum_{m=k}^n R_{n,m} \sum_{i=k}^m \binom{m}{i}p^i q^{m-i} S_{i,k}
\end{equation}
The final transformation between Chebyshev coefficient vectors is
\begin{equation}\label{eq:final_transform}
    \alpha_{2,k} = \sum_{m=k}^{k_{\max}} \alpha_{1,m} C_{m,k}(p,q)
\end{equation}
This direct transformation avoids the numerical instabilities associated with matrix inversion while maintaining full precision and is therefore better suited to high-order polynomial representations and repeated domain updates.

\section{The Cost-to-Sensitivity Ratio as the Multi-Pass Pruning Metric}\label{app:cost_ratio}
The dynamic programming recursion of Section~\ref{sec:mc_thresholds} minimizes the additive complexity $C$ at fixed detection probability, yielding the complexity--sensitivity frontier $C(P_d)$. Let a target ensemble detection probability $P_{\mathrm{ensemble}}$ be reached with $\nrun$ statistically independent pruning passes, each of per-pass detection probability $P_d$ and per-pass cost $C(P_d)$. Inverting the binomial probability relation in equation~\eqref{eq:binomial_prob} gives the required number of passes,
\begin{equation}\label{eq:nrun_required}
    \nrun(P_d) = \frac{\ln\!\left(1 - P_{\mathrm{ensemble}}\right)}{\ln\!\left(1 - P_d\right)}
    \;\xrightarrow[\;P_d \ll 1\;]{}\;
    \frac{\ln\!\left[\,1/(1 - P_{\mathrm{ensemble}})\,\right]}{P_d},
\end{equation}
where the limit uses $\ln(1-P_d)\simeq -P_d$, appropriate for the aggressive low-$P_d$ schemes of interest. The total computational cost of the ensemble is therefore
\begin{align}\label{eq:total_cost_ratio}
    C_{\mathrm{ensemble}}(P_d) &= \nrun(P_d)\,C(P_d)\\
    &\;\simeq\;
    \ln\!\left[\,1/(1 - P_{\mathrm{ensemble}})\,\right]\;\frac{C(P_d)}{P_d}.
\end{align}
Since the leading factor depends only on the fixed target $P_{\mathrm{ensemble}}$ and not on the threshold scheme, minimizing the total ensemble cost is \emph{equivalent} to minimizing the ratio $L = C/P_d$.

\section{Circular Orbit Kinematics}\label{app:circular_orbit}
For a circular orbit, successive derivatives of the line-of-sight displacement in equation~\eqref{eq:disp_circular} are
\begin{equation}\label{eq:dk_circular}
    d^{(k)}(t) = c\,x\,\Omegaorb^k\,\sin\!\left(\Omegaorb t + \psi + \tfrac{\pi k}{2}\right), \quad k \geq 1,
\end{equation}
where each derivative introduces a factor of $\Omegaorb$ and a phase shift of $\pi/2$. Writing the instantaneous orbital phase as $\nu \equiv \Omegaorb t + \psi$, the derivatives evaluated at a reference epoch $\tref$ satisfy
\begin{equation}\label{eq:d_relations}
    d_3 = d_2 \, \Omegaorb \cot(\nu), \qquad
    d_4 = -d_2 \, \Omegaorb^2.
\end{equation}
These relations can be inverted to recover the physical circular-orbit parameters:
\begin{align}
    \Omegaorb &= \sqrt{\!-\frac{d_4}{d_2}}, \nonumber \\
    \nu &= \arctan\left[\frac{d_2}{d_3}\sqrt{-\frac{d_4}{d_2}}\right], \label{eq:circular_recover} \\
    x &= \frac{d_2^2}{c \, d_4 \sin(\nu)}. \nonumber
\end{align}
For circular motion, the higher derivatives are not independent. Equation~\eqref{eq:dk_circular} implies the recurrence
\begin{equation}\label{eq:fnf_recurrence}
    d_{k+2} = -\Omegaorb^2 d_k = \frac{d_4}{d_2} d_k, \quad k \geq 2.
\end{equation}
Thus $\{d_2, d_3, d_4\}$ determine all higher derivatives. In closed form (for $d_2 \neq 0$),
\begin{equation}
\label{eq:dn_closed_form_app}
d_k \;=\;
\begin{cases}
   \displaystyle \left(\frac{d_4}{d_2}\right)^{(k-2)/2} d_2, & k \text{ even}, \\[1em]
   \displaystyle \left(\frac{d_4}{d_2}\right)^{(k-3)/2} d_3, & k \text{ odd}.
\end{cases}
\end{equation}
This is the key structural simplification exploited by the circular-orbit EP search: once the local circular manifold is identified, the omitted higher-order Taylor terms are no longer free parameters.

\subsection{Orbital Coverage Limitations from Polynomial Truncation}\label{app:circular_orbit_poly_coverage}
A polynomial phase model truncated at order $\kmax$ neglects higher-order contributions to the phase evolution ( equation~\eqref{eq:phase_taylor}). Over a finite observation span $\Tobs$, the dominant loss of phase coherence is set by the first neglected term. Taking the reference epoch at the observation midpoint ($\tref = t_c$), and requiring the phase error at the interval endpoints to remain below a tolerance of $\eta$ phase bins, we obtain
\begin{equation}\label{eq:trunc_condition_app}
    \left| \frac{f_{k+1}}{(k+2)!} \left(\frac{\Tobs}{2}\right)^{k+2} \right| \lesssim \frac{\eta}{N_b},
\end{equation}
where $N_b$ is the number of fold bins. Here $\eta$ serves as a proxy for the S/N-dependent detection threshold.

Substituting the circular-orbit derivatives from equation~\eqref{eq:dk_circular} into the equation~\eqref{eq:fderiv}, and approximating the intrinsic spin frequency by the search frequency ($f_{\rm int} \simeq f_0$), yields
\begin{equation}\label{eq:trunc_full}
    f_0 x \Omegaorb^{k+2} \frac{(\Tobs/2)^{k+2}}{(k+2)!} \left|\sin\!\left(\nu_c + \frac{\pi(k+2)}{2}\right)\right| \lesssim \frac{\eta}{N_b},
\end{equation}
where $\nu_c = \Omegaorb t_c + \psi$ is the orbital phase at the observation midpoint. For an optimally sampled profile, $N_b \simeq 1/(f_0 t_s)$, where $t_s$ is the sampling time, causing the spin-frequency dependence to cancel identically. Defining the orbital coverage fraction $R \equiv \Tobs/\Porb$ and utilizing $\Omegaorb = 2\pi/\Porb$, we obtain the strictly phase-dependent bound:
\begin{equation}\label{eq:tobs_general_k}
    R(\nu_c) \lesssim \frac{1}{\pi} \left[ \frac{\eta \, t_s \, (k+2)!}{x \left|\sin\!\left(\nu_c + \frac{\pi(k+2)}{2}\right)\right|} \right]^{\frac{1}{k+2}}.
\end{equation}
At fixed $k$, the limit scales as $R \propto x^{-1/(k+2)}$; astrophysical parameters enter solely through the projected semimajor axis $x$. 

Because $\nu_c$ is uniformly distributed on $[0,2\pi]$ for a generic circular orbit, $R$ can be treated probabilistically. The worst-case coverage floor, $R_{\rm w}$, occurs when the first neglected derivative is maximized ($|\sin(\cdot)| = 1$):
\begin{equation}\label{eq:tobs_worst}
    R_{\rm w} = \frac{1}{\pi} \left[\frac{\eta \, t_s \, (k+2)!}{x}\right]^{\frac{1}{k+2}}.
\end{equation}
Conversely, when the sine term vanishes the truncation error is capped by the next non-vanishing term, $f_{k+2}$, providing a finite best-case ceiling, $R_{\rm b}$:
\begin{equation}\label{eq:tobs_best}
    R_{\rm b} = \frac{1}{\pi} \left[\frac{\eta \, t_s \, (k+3)!}{x}\right]^{1/(k+3)}.
\end{equation}

For an arbitrary phase, the local bound scales as $R(\nu_c) = R_{\rm w} \left|\sin(\nu_c + \pi(k+2)/2)\right|^{-1/(k+2)}$. Given the cumulative distribution of $|\sin \nu_c|$, the probability that a randomly phased system remains coherent at a required coverage ratio $r$ is:
\begin{equation}\label{eq:coverage_survival}
    P(R \ge r) = \frac{2}{\pi} \arcsin\left[ \left(\frac{R_{\rm w}}{r}\right)^{k+2} \right], \quad \text{for } R_{\rm w} \le r \le R_{\rm b}.
\end{equation}
The median coverage fraction ($P = 0.5$) is
\begin{equation}\label{eq:coverage_median}
    R_{50} = R_{\rm w} \cdot 2^{\frac{1}{2(k+2)}}.
\end{equation}
For standard acceleration ($k=1$), jerk ($k=2$), and snap ($k=3$) searches, $R_{50}$ exceeds $R_{\rm w}$ by only $\sim\!12\%$, $\sim\!9\%$, and $\sim\!7\%$, respectively. The $1/(k+2)$ exponent inherently skews the distribution toward the worst-case limit, demonstrating that high-coverage best-case scenarios are statistically rare.

\begin{figure}
\centering
\includegraphics[width=\linewidth]{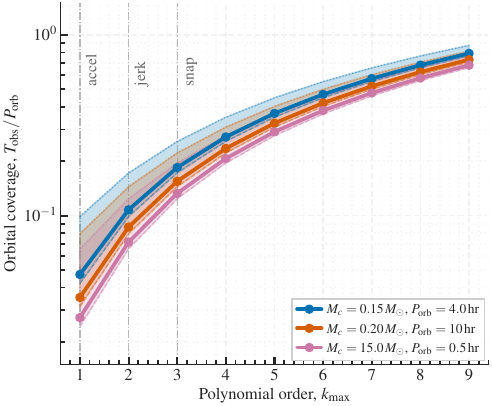}
\caption{Orbital coverage fraction $\Tobs/\Porb$ as a function of polynomial order $\kmax$ for the three representative configurations described in the text. Shaded bands span the worst-case floor $R_{\rm w}$ (dashed) and best-case ceiling $R_{\rm b}$ (dotted); solid curves show the phase-marginalized median $R_{50}$.}
\label{fig:orbital_coverage}
\end{figure}

Figure~\ref{fig:orbital_coverage} illustrates $R$ as a function of $\kmax$ for three search-relevant configurations assuming $m_p=1.4\,M_\odot$, $\eta=2$, and $t_s=64\,\mu\mathrm{s}$: a mid-period Galactic MSP binary ($m_c=0.15\,M_\odot$, $P_{\rm orb}=4\,\mathrm{hr}$; redback-like), a recycled pulsar with a He/CO white dwarf ($m_c=0.2\,M_\odot$, $P_{\rm orb}=10\,\mathrm{hr}$; MSP--WD), and a compact-orbit, high-mass-companion stress case ($m_c=15\,M_\odot$, $P_{\rm orb}=30\,\mathrm{min}$). For the most conservative configuration, equations~\eqref{eq:tobs_worst} and~\eqref{eq:tobs_best} yield $R_{\rm w}=0.04$ and $R_{\rm b}=0.10$ at $\kmax=1$, and $R_{\rm w}=0.17$ and $R_{\rm b}=0.26$ at $\kmax=3$.

The commonly quoted $\sim\!10\%$ limit for constant-acceleration searches is therefore not a generic circular-orbit coverage bound \citep{Ransom:2003}. It is \emph{only} the best-case ceiling $R_{\rm b}$, valid at favourable orbital phases, where the leading neglected term vanishes. This distinction is important in survey completeness analyses, where adopting the best-case value as a universal rule will systematically overestimate the orbital coverage of acceleration searches.

Finally, we note two conservative assumptions built into this analytical bound:
\begin{enumerate}
    \item \emph{Parameter Absorption:} A true template grid search maximizes over all polynomial coefficients $f_0, \ldots, f_k$, partially absorbing the truncation error (see Appendix~\ref{app:optimal_gridding}). By the Chebyshev minimax property, absorbing a pure $(t-t_c)^{k+2}$ phase drift into lower-order terms reduce the peak-to-peak residual by a factor of $2^{-(k+1)}$, potentially increasing the limiting coverage by $2^{(k+1)/(k+2)}$ (e.g., a factor of $\sim\!1.6$ for $k=1$). 
    \item \emph{Inclination:} The projected semi-major axis is strictly $x \propto \sin i$. An edge-on orbit ($\sin i = 1$) provides a conservative lower bound on $R$. Marginalizing over an isotropic inclination distribution (median $\sin i = \sqrt{3}/2$) would shift the median acceleration coverage upward by only an additional $\sim\!5\%$.
\end{enumerate}

\subsection{Circular Basis Transformation}\label{app:circular_orbit_transform}
A finite-order Taylor shift is exact only within the polynomial manifold. For circular orbits, an exact propagation is obtained by evolving the oscillatory derivatives directly. The key observation is that the pair $(d_2,d_3/\Omegaorb)$ transforms under a rigid phase rotation.

Let $\bm{d}_i = [d_0,\, d_1,\, d_2,\, d_3,\, d_4,\, d_5]_i^\top$ denote the derivative state at epoch $t_i$. We first infer the orbital frequency from the even-derivative pair, $\Omegaorb = \sqrt{-d_{4,i} / d_{2,i}}$ or, in the nodal regime discussed in Section~\ref{subsec:circular_hole}, from the odd-derivative pair, $\Omegaorb = \sqrt{-d_{5,i} / d_{3,i}}$. With $\Delta t = t_j - t_i$ and the accumulated phase advance $\Delta\phi=\Omegaorb \Delta t$, the exact evolution of the oscillatory subspace is
\begin{subequations}\label{eq:circ_full}
\begin{align}
    d_{2,j} &= d_{2,i}\cos\Delta\phi + \frac{d_{3,i}}{\Omegaorb}\sin\Delta\phi, \label{eq:circ_d2_full} \\
    d_{3,j} &= d_{3,i}\cos\Delta\phi - \Omegaorb d_{2,i}\sin\Delta\phi, \label{eq:circ_d3_full} \\
    d_{4,j} &= -\Omegaorb^2 d_{2,j}, \label{eq:circ_d4_full} \\
    d_{5,j} &= -\Omegaorb^2 d_{3,j}. \label{eq:circ_d5_full}
\end{align}
\end{subequations}

The lower-order derivatives $d_1$ and $d_0$ follow by integrating the oscillatory solution once and twice. It is convenient to define the two epoch-invariant integration constants
\begin{equation}
A \equiv d_{1,i} + \frac{d_{3,i}}{\Omegaorb^2},
\qquad
B \equiv d_{0,i} + \frac{d_{2,i}}{\Omegaorb^2}.
\end{equation}
The propagated lower-order terms are then
\begin{align}
    d_{1,j} &= -\frac{d_{3,j}}{\Omegaorb^2} + A, \label{eq:circ_d1_full} \\
    d_{0,j} &= -\frac{d_{2,j}}{\Omegaorb^2} + A\,\Delta t + B. \label{eq:circ_d0_full}
\end{align}
For a strictly circular orbit with no secular drift, $A=0$ and $B=\bar d$, but we retain it here to accommodate small non-circular perturbations from the initial accumulation epoch in EP algorithm.

Equations~\eqref{eq:circ_full}--\eqref{eq:circ_d0_full} give the exact propagation of a circular-orbit grid centre. The map is nevertheless non-linear as a grid operation, because $\Omegaorb$ is inferred from the candidate itself. This is harmless for propagating individual candidate centres, but it prevents the use of a single global linear transport operator for cell extents. We therefore propagate candidate centres with the exact circular map, while the extents of axis-aligned Taylor cells are still transported with the generic Taylor operator $\mathbf{T}(\Delta t)$ when aggressive or conservative bounding is required.

\subsection{Anchor segment Bias Diagnostic}\label{app:anchor_bias_diagnostic}
\begin{figure*}
\includegraphics[width=\linewidth]{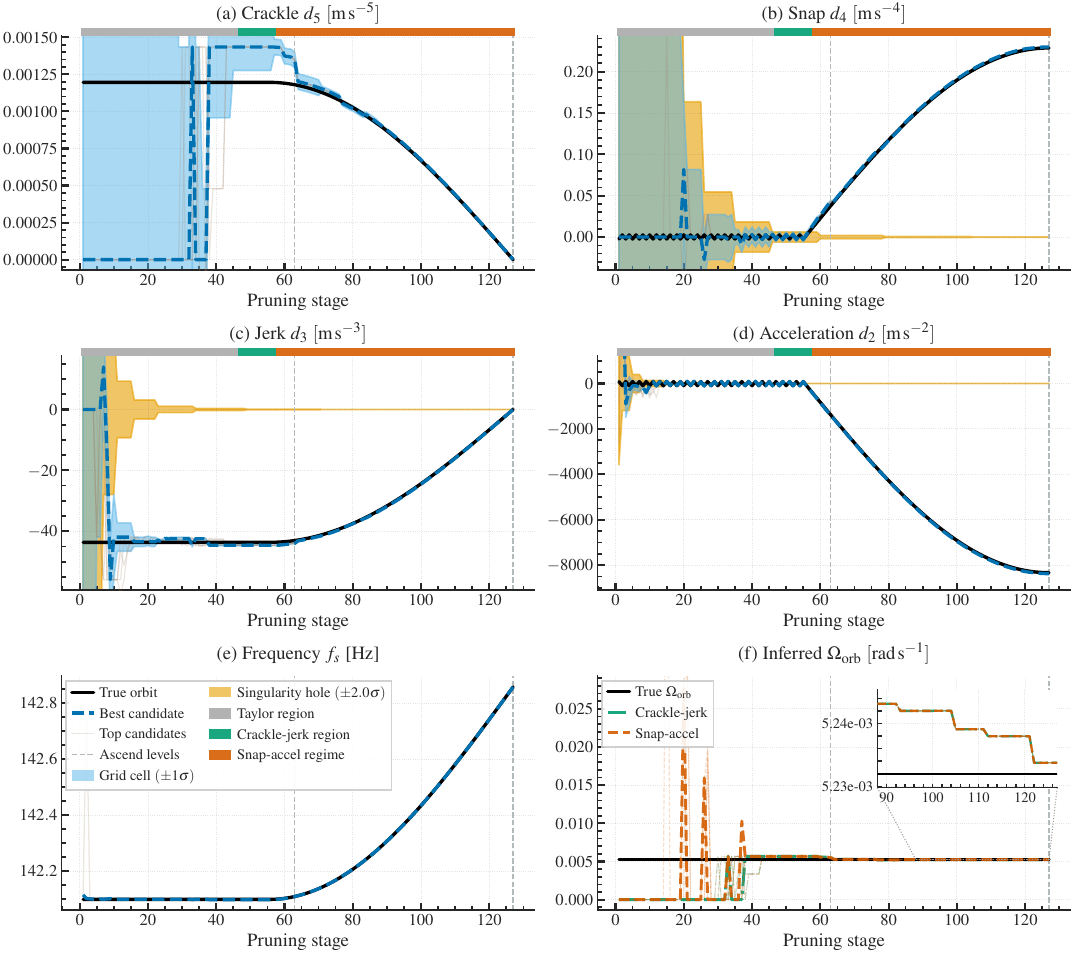}
\caption{Evolution of the five highest-ranked candidate paths through successive pruning stage for a circular-orbit injection with $\mathrm{S/N} = 30$ and anchor segment $q=28$, corresponding to one of the phase intervals exhibiting anomalously low detection probability in Fig.~\ref{fig:ep_circular_seg_bias}. The search configuration and orbital injection parameters are identical to those of Fig.~\ref{fig:ep_circular_seg_bias}. Panels~(a)--(d) show the evolution of the higher-order Taylor-kinematic parameters, panel~(e) shows the frequency $f_s$; and panel~(f) shows the inferred $\Omega_{\rm orb}$ obtained from the snap--acceleration and crackle--jerk estimator pairs. The solid black curve denotes the true Keplerian trajectory, the dashed blue curve traces the highest-ranked tree candidate, and the fainter curves show show the four sub-dominant survivors. The cyan band indicates the instantaneous $\pm 1\sigma$ grid-cell boundary surrounding the best candidate, and the amber region marks the singularity hole where the snap-based propagator is numerically ill-conditioned. Vertical dotted lines denote stages at which the \func{ascend} operation is invoked. The coloured strip above panels~(a)--(d) identifies the active $\Omega_{\rm orb}$ propagation regime of the leading candidate. Although the candidate path remains close to the true trajectory throughout the search, incomplete convergence of $\Omega_{\rm orb}$ (inset, panel~f) because of gridding inaccuracies around crackle-jerk handover produces a residual phase-model mismatch and the associated loss in recovered S/N.}
\label{fig:ep_tracking_paths}
\end{figure*}

To trace the inner workings of our dynamic basis-switching rules within the localized dropouts identified in Section~\ref{subsec:anchor_bias}, we execute a diagnostic run at an elevated $\mathrm{S/N} = 30$, anchored at one of the problematic index $q = 28$. Figure~\ref{fig:ep_tracking_paths} maps the real-time state-space trajectories of the top five surviving candidates across successive pruning stages. Panels~(a) through~(d) demonstrate how the higher derivatives act as a predictive buffer, shielding the search tree from geometric cell explosions. Within the shaded amber singularity holes where the low-order acceleration is poorly resolved, the tracking path successfully locks onto the stable crackle--jerk regime (orange indicators), keeping the primary carrier frequency $f_s$ centred on the true analytical track in panel~(e).

However, as revealed by the trajectory magnification in the panel~(f) inset, the inferred $\Omega_{\rm orb}$ does not completely converge to its true analytical value by the final stage. Because the track spends its crucial initial stages inside the numerical hole, gridding inaccuracies, discretization mismatches and pruning decisions during the high-order $c\text{--}j$ to low-order $s\text{--}a$ handover introduce subtle, irreversible coordinate losses. Consequently, the true, physical grid cell is pruned early in the tree, and the top-ranked survivor shown in panel~(f) represents a sub-dominant candidate that suffers a residual phase-model mismatch, dragging down the final recovered S/N.

Figure~\ref{fig:ep_tracking_paths} also highlights the corrective behaviour of the \func{ascend} function, which is invoked at the mid-point (stage 63) and termination (stage 127) of the traversal. This operation re-integrates the accumulated fold data across the surviving grid boundaries, forcing the diverging candidate tracks back toward the true analytical trajectory. While the current handover losses inside the $\dot{f} \approx 0$ windows are small enough to be safely neglected in the initial version of the algorithm, this structural behaviour provides a clear path for future optimization. The localized gridding mismatches can be entirely mitigated by inserting additional, intermediate \func{ascend} levels throughout the early pruning stages, providing a localized patch that stabilizes the crackle--jerk handover and recovers full sensitivity across the entire orbit.

\subsection{Cartesian Coordinate Formulation}\label{app:circular_orbit_cartesian}
To recover physical circular-orbit parameters from the local kinematic derivatives (equation~\eqref{eq:circular_recover}) while avoiding numerical singularities (e.g., when $\sin\nu \to 0$), we adopt a Cartesian parametrization that defines orthogonal projections of the projected semi-major axis at orbital phase $\nu$:
\begin{equation}
    x_{\cos\nu} \equiv x \cos\nu, \qquad x_{\sin\nu} \equiv x \sin\nu.
\end{equation}
This transformation replaces the polar form $(x, \nu)$ with bounded, smooth coordinates. The Cartesian grid is naturally compact and uniform, eliminating special treatment near singularities of $\tan\nu$, $\sin\nu$, or $\cos\nu$ in traditional formulations. Using the circular derivative relations from equation~\eqref{eq:d_relations}, the Cartesian amplitudes may be written directly in terms of the Taylor coefficients:
\begin{equation}\label{eq:nu_recover2_app}
    x_{\cos\nu} = -\frac{d_3}{c\,\Omegaorb^{3}}, \qquad
    x_{\sin\nu} = -\frac{d_2}{c\,\Omegaorb^{2}}.
\end{equation}

Assuming independent local uncertainties in the measured derivatives, standard error propagation gives
\begin{align}
\left(\frac{\sigma_{\Omegaorb}}{\Omegaorb}\right)^2 &= \frac{1}{4} \left[\left(\frac{\sigma_{d_2}}{d_2}\right)^2 + \left(\frac{\sigma_{d_4}}{d_4}\right)^2 \right], \\
\left(\frac{\sigma_{x_{\cos\nu}}}{|x_{\cos\nu}|}\right)^2 &= \left(\frac{\sigma_{d_3}}{d_3}\right)^2 + \frac{9}{4} \left[\left(\frac{\sigma_{d_2}}{d_2}\right)^2 + \left(\frac{\sigma_{d_4}}{d_4}\right)^2 \right], \\
\left(\frac{\sigma_{x_{\sin\nu}}}{|x_{\sin\nu}|}\right)^2 &= 4\left(\frac{\sigma_{d_2}}{d_2}\right)^2 + \left(\frac{\sigma_{d_4}}{d_4}\right)^2.
\end{align}
Although not part of the minimal Cartesian search basis, the instantaneous line-of-sight velocity is sometimes needed for Doppler corrections. It follows from
\begin{equation}
d_1 = -\frac{d_3}{\Omegaorb^2},
\end{equation}
with propagated uncertainty
\begin{equation}
\left(\frac{\sigma_{d_1}}{d_1}\right)^2 =  \left(\frac{\sigma_{d_3}}{d_3}\right)^2 + \left[\left(\frac{\sigma_{d_2}}{d_2}\right)^2 + \left(\frac{\sigma_{d_4}}{d_4}\right)^2 \right].
\end{equation}

\subsection{Grid Spacing in the Cartesian Circular Parametrization} \label{app:circular_orbit_cartesian_grid}
We derive optimal grid spacing for $\Lcart = \{f_0, \Omegaorb, x_{\cos\nu}, x_{\sin\nu}\}$ by enforcing the usual phase-mismatch tolerance. Substituting the Cartesian identities (equations~\eqref{eq:phase_generic}–\eqref{eq:disp_circular}) and writing $\tau=t-\tref$, the circular phase model becomes
\begin{align}\label{eq:cartesian_phase}
\Phi(t;\Lcart)
&=
\Phi_{\mathrm{ref}} + f_0 \tau \nonumber\\
&\phantom{=}
-f_0\!\left[
x_{\cos\nu}\sin(\Omegaorb \tau)
+
x_{\sin\nu}\cos(\Omegaorb \tau)
\right].
\end{align}

Using the metric mismatch criterion in equation~\eqref{eq:grid_criteria}, we determine optimal grid spacing for $\Lcart$. For the intrinsic frequency $f_0$, the phase dependence on $t$ is linear:
\begin{equation}
    \Delta f_0 = \frac{\eta}{N_b(\Tobs - \tref)}.
\end{equation}
The key point is that the amplitude coordinates $x_{\cos\nu}$ and $x_{\sin\nu}$ enter the phase linearly and with bounded derivatives:
\begin{equation}
    \left| \frac{\partial \Phi}{\partial x_{\cos\nu}} \right| \leq f_0, \qquad 
    \left| \frac{\partial \Phi}{\partial x_{\sin\nu}} \right| \leq f_0.
\end{equation}
Therefore the required spacing in these coordinates is
\begin{equation}\label{eq:grid_constant}
    \Delta x_{\cos\nu} = \Delta x_{\sin\nu} = \frac{\eta}{N_b f_0},
\end{equation}
which is independent of the coherent span. Once the orbit is resolved, the amplitude grid no longer needs to refine. The orbital frequency $\Omegaorb$ is the most sensitive parameter. Differentiating equation~\eqref{eq:cartesian_phase} with respect to $\Omegaorb$ yields a term growing linearly with time:
\begin{equation}
    \frac{\partial \Phi}{\partial \Omegaorb} \approx -f_0 x (t - \tref) \cos(\nu + \Omegaorb(t - \tref)).
\end{equation}
Evaluating at the maximum extent $t = \Tobs$, the spacing scales inversely with observation time:
\begin{equation}
    \Delta \Omegaorb = \frac{\eta}{N_b f_0 x (\Tobs - \tref)}.
\end{equation}

\subsection{Grid Transition Criterion}\label{app:circular_grid_transition}
At the grid resolution, we estimate the polynomial parameter uncertainties from the measurement precision achievable with a given observation span. While our search employs the Chebyshev-optimized grid spacing (equation~\eqref{eq:dk_optimal}) for computational efficiency, the transition criterion depends on the \emph{physical information content} of the orbital coverage, not the specific gridding strategy. We therefore estimate the local polynomial uncertainties by the Taylor-grid spacings at coherent span $T_s$ (equation~\eqref{eq:grid_params}):
\begin{align}
    \sigma_{d_2} &\approx \Delta d_2 = \frac{8c\eta}{N_b f_0 T_s^2}, \\
    \sigma_{d_3} &\approx \Delta d_3 = \frac{48c\eta}{N_b f_0 T_s^3}, \\
    \sigma_{d_4} &\approx \Delta d_4 = \frac{384c\eta}{N_b f_0 T_s^4},
\end{align}
where we have set $\tref = T_s/2$.

Using transformation relations from equation~\eqref{eq:nu_recover2_app} and error propagation from Appendix~\ref{app:circular_orbit_cartesian}, we substitute the circular orbit relations $d_2 = cx\Omegaorb^2\sin(\nu)$, $d_3 = cx\Omegaorb^3\cos(\nu)$, and $d_4 = -cx\Omegaorb^4\sin(\nu)$ to express fractional uncertainties. After algebraic manipulation and cancellation of common factors $\eta/(N_b f_0)$, the transition conditions in equation~\eqref{eq:transition_criterion} reduce to inequalities in the dimensionless phase-coverage variable $u \equiv \Omegaorb T_s = 2\pi T_s/\Porb$:

\begin{enumerate}
    \item From $x_{\sin\nu}$:
    \begin{equation}\label{eq:transition_xsin}
        \frac{256}{u^4} + \frac{147456}{u^8} < 1.
    \end{equation}
    
    \item From $x_{\cos\nu}$: Setting $\cot(\nu) = 1$ (i.e., $\nu = \pi/4$) for a phase-averaged estimate,
    \begin{equation}\label{eq:transition_xcos}
        \frac{2304}{u^6} + \frac{144}{u^4} + \frac{331776}{u^8} < 1.
    \end{equation}
    
    \item From $\Omegaorb$: Setting $\sin(\nu) = 1$ for the most favorable case,
    \begin{equation}\label{eq:transition_omega}
        \frac{4}{u^2} + \frac{9216}{u^6} < 1.
    \end{equation}
\end{enumerate}

Remarkably, all configuration-dependent parameters ($\eta$, $N_b$, $f_0$, $x$) cancel in these expressions, yielding \emph{universal} transition criteria that depend only on orbital phase coverage. The transition must satisfy the most stringent of these conditions. Solving equations~\eqref{eq:transition_xsin}--\eqref{eq:transition_omega} numerically yields $u_{\mathrm{trans}} \approx 5.83$ (from $x_{\cos\nu}$), corresponding to approximately $93\%$ of one orbital period. In practice, we adopt the conservative choice:
\begin{equation}\label{eq:transition_time_app}
    \Ttrans = \Porb,
\end{equation}
representing completion of one full orbit. This choice is motivated by several considerations. After one orbit, all orbital phases $\nu \in [0, 2\pi]$ have been sampled, ensuring both $x_{\cos\nu}$ and $x_{\sin\nu}$ are well-constrained regardless of the initial phase $\psi$. The criterion is independent of the search configuration parameters, making it universally applicable. Thus, $\Ttrans = \Porb$ serves as a robust, conservative threshold for basis transition in the EP algorithm.

\bibliographystyle{mnras}
\bibliography{references.bib}

\bsp	
\label{lastpage}
\end{document}